\documentclass[12pt]{iopart}

\usepackage{makeidx,iopams,cite} 
\usepackage{eurosym}
\usepackage{graphicx} 
\usepackage{subeqnar} 
\usepackage{multicol} 
\usepackage{cropmark} 
\makeindex

\newcommand{\be}{\begin{equation}} \newcommand{\ee}{\end{equation}}
\newcommand{\beas}{\begin{eqnarray*}}
\newcommand{\eeas}{\end{eqnarray*}}
\newcommand{\bea}{\begin{eqnarray}} 
\newcommand{\eea}{\end{eqnarray}}
\newcommand{\bsy}[1]{\boldsymbol{#1}} 
\newcommand{\req}[1]{(\ref{#1})}
\renewcommand{\pt}[1]{\left(#1\right)} 
 
\def\P{{\cal P}}

\renewcommand{\l}{\left} 
\renewcommand{\r}{\right}
\newcommand{\ii}{{\rm i}} 
\newcommand{\sign}{{\rm sign}}
\newcommand{\tra}{{\rm tr}}
\renewcommand{\det}{{\rm det}}
\newcommand{\erf}{{\rm erf}}
\newcommand{\erfc}{{\rm erfc}}
\newcommand{\prob}{{\rm Prob}}
\newcommand{\avg}[1]{\left\langle{#1}\right\rangle}
\newcommand{\davg}[1]{\left\langle\!\left\langle{#1}
        \right\rangle\!\right\rangle}

\newcommand{\ovl}[1]{\overline{#1}}

\newcommand{\argmax}{{\rm arg\,} \max}
\begin{document}

\topical{Statistical mechanics of socio-economic systems with
heterogeneous agents}

\author{Andrea De Martino\dag ~and Matteo Marsili\ddag} \address{\dag
  CNR/INFM, Dipartimento di Fisica, Universit\`a di Roma ``La
  Sapienza'', p.le A. Moro 2, 00185 Roma (Italy)} \address{\ddag The
  Abdus Salam International Centre for Theoretical Physics, Strada
  Costiera 11, 34014 Trieste (Italy)}

\ead{andrea.demartino@roma1.infn.it,marsili@ictp.trieste.it}

\begin{abstract}
  
We review the statistical mechanics approach to the study of the
emerging collective behavior of systems of heterogeneous interacting
agents.  The general framework is presented through examples is such
contexts as ecosystem dynamics and traffic modeling.  We then focus on
the analysis of the optimal properties of large random
resource-allocation problems and on Minority Games and related models
of speculative trading in financial markets, discussing a number of
extensions including multi-asset models, Majority Games and models
with asymmetric information.  Finally, we summarize the main
conclusions and outline the major open problems and limitations of the
approach.

\end{abstract}


\tableofcontents

\newpage


\section{Introduction}

Collective phenomena in economics, social sciences and ecology are
very attractive for statistical physicists, especially in view of the
empirical abundance of non-trivial fluctuation patterns and
statistical regularities -- think of returns in financial markets or
of allometric scaling in ecosystems -- which pose intriguing
theoretical challenges. On an abstract level, the problems at stake
are indeed not too different from, say, understanding how spontaneous
magnetization may arise in a magnetic system, since what one wants in
both cases is to understand how the effects of interactions at the
microscopic scale can build up to the macroscopic scale.  Clearly,
ecologies or financial markets are quite more complex systems than
magnets, being composed of units which themselves follow complex (and
far from understood) behavioral rules. Still, in many cases it may be
reasonable to assume that the collective behavior of a crowd of
individuals presents aspects of a purely statistical nature which
might be appreciated already in highly stylized models of such
systems. This is ultimately the rationale for applying statistical
mechanics to such problems.

In general, statistical physics offers a set of concepts (e.g. order
parameters and scaling laws) and tools (both analytical and numerical)
allowing for a characterization in terms of phases and phase
transitions which might be useful in shaping the way we think about
such complex systems. The considerable progress achieved in the last
decades in the statistical mechanics of non-equilibrium processes and
of disordered systems, thanks to which it is now possible to deal
effectively with fluctuations and heterogeneity (respectively) in
systems with many interacting degrees of freedom, is particularly
important for socio-economic applications. In fact, while equilibrium
and homogeneity are important in physics, non-equilibrium and
heterogeneity are the rule in economics, as each individual is
different both in his/her characteristics and in the way he/she
interacts with the environment. Deriving general macroscopic laws
taking the specific details of each and every individual's behavior
into account is a desperate task. However, as long as one is
interested in collective properties, a system with complicated
heterogeneous interactions can be reasonably well represented as one
with random couplings \cite{Wigner1}. In the limit of infinite system
size, some of the relevant macroscopic observables will be subject to
laws of large numbers, i.e. some quantities will be self-averaging,
and, if the microscopic dynamics follows sufficiently simple rules,
one may hope to be able to calculate them explicitly. It is with these
properties -- which we call {\it typical} -- that the statistical
mechanics approach is concerned.

In what follows, we shall mostly concentrate on problems arising in
economics and finance. When modeling these systems one must be aware
that their microscopic behavior is very different from that governing
particles or atoms in physics. Economic agents typically respond to
incentives and act in a selfish way. This is usually modeled assuming
that individuals strive to maximize their private utility functions,
with no regard for social welfare.  Not only agents might have
conflicting goals, as their utility functions will in general be
different, but their selfish behavior may lead to globally inefficient
outcomes -- e.g. to a coordination failure or to a lack of
cooperation. Such outcomes, called {\em Nash equilibria} in Game
Theory, are in general different from socially optimal states where
the total utility is maximized. Hence, generally, in a system of
interacting agents there is no global energy function to be minimized.

Another important difference between the dynamics of a physical
system, such as a magnetic material, and that of an economic system is
that, while in the former spins at a particular time depend at most on
the past states of the system, in the latter the agents' choices also
depend on the expectations which they harbor about the future
states. This suggests that the collective dynamics may have a
non-causal component (indeed, backward induction in time plays a big
role in the strategic reasoning of rational agents
\cite{GameTheory}). In many cases, however, it is reasonable to assume
that agents are boundedly rational or `inductive', i.e. that their
behavior as well as their expectations adjust as a result of
experience. We shall concentrate our analysis to these cases of
adaptive agents following a learning dynamics. We shall see that the
lack of a global Hamiltonian is reflected in the fact that such a
dynamics, in general, violates detailed balance.

Actually, in many cases it is realistic to assume that agents behave
as if they were interacting with a {\em system} as a whole -- be it a
market or the crowd -- rather than directly with a number of other
individuals. In economics, this is termed a {\em price-taking}
assumption, because it amounts to stating that agents act as if prices
do not depend on what they actually decide to buy or sell (i.e. they
take prices as given), and it is usually justified by saying that the
contribution of a single agent to the total demand is negligible when
the number of agents is large. The equilibria of systems where agents
behave as price-takers are called {\em competitive equilibria}.
However, prices depend on the aggregate demand and supply and hence on
the choice of each agent, and the statistical physics approach
provides a very transparent description of how price-taking behavior
modifies the global properties of a system.

This review gives a survey of some recent quantitative developments on
the statistical mechanics of systems of many interacting adaptive
agents. This is a subject that has been shaped over the past few years
around a few basic models (like the El Farol problem) and a few
analytical techniques, mostly borrowed from the mean-field theory of
spin glasses (like the replica method). The models, though highly
stylized to an economist's eyes, possess a strong physical content and
in many cases provide important indications as to whether the
phenomenology of real systems is specific of each of their particular
natures or rather it is generic of large systems of adaptive units
interacting competitively. Ultimately, it is not too unfair to say
that separating system-specific features from general features can be
seen as the main contribution statistical physics can provide to this
field (besides techniques). 

Our choice of arguments is clearly biased, and the reader may dispose
of several recent books that cover some of the important issues
(especially finance-inspired) we merely touch here
\cite{MantegnaStanley,Dacorogna,BouchaudPotters,NFJBOOK,Voit}.  Along
with a core of problems related to the emergence of non-trivial
fluctuation phenomena, cooperation and efficiency (understanding which
has been the original goal of these studies), other issues such as the
impact of different information structures or the interaction between
different multi-agent systems have just started to be analyzed and are
likely to attract a great deal of attention in the near future. On the
physical side, precisely because of the differences in the microscopic
modeling of economics and physics, these systems pose a number of
fascinating questions that open several directions for further work,
some of which will be outlined here.

The review is organized as follows. In Sec. 2 we present a general
discussion of resource allocation by complex adaptive systems and a
few exemplary models from different contexts like ecology and traffic
dynamics, including the El Farol problem. Sec. 3 is devoted to the
statistical analysis of optimal properties of large random economies,
that is, more precisely, to a survey of the macroscopic properties of
classical economic optimization problems. Most of our attention will
be on the model of competitive equilibrium for linear production
economies and on Von Neumann's model of economic growth.  In Sec. 4 we
review the basic properties of the Minority Game, a minimal and yet
highly non-trivial model of speculative trading derived from the El
Farol problem, and discuss the role of the different parameters
involved in its definition. Besides its physical richness, the
Minority Game provides a simple adaptable framework where a number of
important issues related to financial markets (such as the emergence
of `stylized facts', the role of different types of traders and the
effect of information asymmetries) can be analyzed in great
detail. Some of them are discussed in Sec. 5. Finally, some concluding
remarks are expounded in Sec. 6. The main analytical techniques
employed for these studies will be discussed in some detail only for
cases where details are not available in the published literature: the
replica technique for a model of a competitive ecosystem in Sec. 2;
the continuous-time limit approach for the El Farol problem, also in
Sec. 2; the dynamical generating functional for the canonical
multi-asset Minority Game in Section 5.


\section{Statistical mechanics of resource allocation: some examples}

We start our discussion by introducing a general class of problems
where a population of heterogeneous agents competes for the
exploitation of a number of resources. Then we will discuss a few
examples -- ranging from ecosystems to urban traffic -- where this
generic framework can be formalized in specific models where the
nature of resources and the laws governing the behavior of agents are
completely specified.

\subsection{General considerations}

In a nutshell, the models we consider address the decentralized
allocation of scarce resources by $N$ heterogeneous selfish agents
subject to public and/or private information. The word `allocation' is
to be intended here in a broad sense that includes the exchange of
resources (for example, commodities) among agents, the production of
resources by means of other resources and the consumption of
resources.  Agents take decisions on the basis of some type of
information aiming at some pre-determined goals, like maximizing a
certain utility function, and are to various degrees adaptive
entities. We shall consider cases in which they are perfect optimizers
(or `deductive') as well as cases in which their decision-making is
governed by a learning process (`inductive'). Heterogeneity may reside
in a number of factors, like the agents' initial endowments, their
learning abilities or in how differently they react to the receipt of
certain information patterns.

In general, the allocation is a stochastic dynamical process, where
the noise may be present in both the information sources and the
agent's learning process. We shall mostly be concerned with the
steady-state properties and, more than on individual performances, we
shall focus on the resulting distribution of resource loads and in
particular on
\begin{enumerate}
\item[a.] how evenly are resources exploited on average (i.e. whether
  the allocation process leads typically to over- or
  under-exploitation of some resources)
\item[b.] the fluctuations of resource loads (i.e. how large the
deviations from the average can be)
\end{enumerate} 
In such contexts as production economies, ecosystems or traffic the
meaning and the relevance of the above observables is immediately
clear. 
In toy models of financial markets, where, as we shall see, the
role of resources is played by information bits, the former quantity
plays the role of a `predictability' while the latter measures the
`volatility'.

It is implicitly assumed that optimal allocations are those where
resources are exploited as evenly as possible and where fluctuations
are minimal. In an economic setting, this corresponds to allocations
with minimal waste whereas in financial markets, optimality implies
information being correctly incorporated into prices with minimal
volatility.

In what follows, we shall denote by $\avg{\cdots}$ time averages
performed in the steady state: \be \avg{X}=\lim_{T,T_{{\rm
eq}}\to\infty}\frac{1}{T-T_{{\rm eq}}}\sum_{t=T_{{\rm eq}}}^T X(t) \ee
where $T_{{\rm eq}}$ is an equilibration time.  Moreover, we shall
label agents by the index $i\in\{1,\ldots,N\}$ and resources by the
index $\mu\in\{1,\ldots,P\}$. In the statistical mechanics approach,
the relevant limit is ultimately that where $N\to\infty$ and $P$
scales linearly with $N$, so that $\alpha=P/N$ remains finite as $N$
diverges. To give a loose name, we shall call the relative number of
information patterns $\alpha$, which will be our typical control
parameter, the `complexity' of the system.

Denoting by $Q^\mu(t)$ the load of resource $\mu$ at time $t$, which
is determined by the aggregate action of all agents (for instance,
$\mu$ may be a certain commodity and $Q^\mu(t)$ the demand for it at
time $t$), one easily understands that the relevant macroscopic
quantities are given respectively by \be H=\frac{1}{P}\sum_\mu
\avg{A^\mu}^2,~~~~~~~ A^\mu(t)=Q^\mu(t)-\ovl{\avg{Q}}, \ee
($\ovl{\avg{Q}}=(1/P)\sum_\mu\avg{Q^\mu}$) which measures the
deviation of the distribution of resource loads from uniformity (if
$H\neq 0$ at least one resource is overexploited or underexploited
with respect to the average load) and by \be
\sigma^2=\frac{1}{P}\sum_\mu\l[\avg{(A^\mu)^2}-\avg{A^\mu}^2\r]
=\frac{1}{P}\sum_\mu\l[\avg{(Q^\mu)^2}-\avg{Q^\mu}^2\r] \ee which
measures the magnitude of fluctuations. Efficient steady states have
$H=0$ and $\sigma^2$ ``small'' in a sense that will be specified from
case to case. To fix ideas, whenever fluctuations are smaller than
those which would be obtained by zero-intelligence agents who act
randomly and independently at every time step one can infer that
agents are to some degree cooperating to reduce fluctuations.

An important question we shall typically ask is how efficient are the
steady-state resource loads distributions generated by a particular
group of agents with a given information stream. Besides this, we
shall also look at the inverse problem, namely under which conditions
can a steady state satisfy criteria for efficiency. For example, what
type of information should one inject into the system in order to
facilitate the reach of a steady state in which $H$ and $\sigma^2$ are
as small as possible? Indeed the structure of the information agents
have access to may drastically affect global efficiency in many cases
(e.g. traffic models).

\subsection{A simple model of ecological resource competition}
\label{specie}

Ecosystems constitute a foremost example of the class of problems we
outlined above \cite{Til}. The following can be seen as a minimal
model of a competitive ecology with limited resources. Such a model
will be taken as a prototype to illustrate the statistical mechanics
(static) approach. The statistical mechanics approach to ecosystems
has been pioneered in \cite{Rieger} based on the generating functional
approach. The central issue is that of the May's biodiversity paradox
\cite{May}, which shows that, contrary to expectations, increases in
biodiversity in a random ecosystem enhance its instability. We shall
indeed find the same result.

\subsubsection{Definition}

Let us consider a system with $N$ species whose populations $n_i(t)$
($i\in\{1,\ldots,N\}$) are governed by Lotka-Volterra type of
equations:

\be 
\frac{\dot n_i(t)}{n_i(t)}=f_i+\sum_{\mu=1}^PQ^\mu(t)
q_i^\mu
\label{lotk}
\ee 
\noindent
$Q^\mu$ denotes the abundance of resource
$\mu\in\{1,\ldots,P\}$ (be it a mineral, a particular habitat,
water\ldots) while $q_i^\mu$ is a coefficient saying how much species
$i$ benefits from that resource. The constant $f_i$ is the
population's decay rate `in absence of resources'. To simplify things,
we mimic the complex interdependence between species and resources by
assuming that the $q_i^\mu$'s are independent, identically distributed
quenched random variables.

The abundance of resource $\mu$ depends on the population of each
species, i.e.  

\be 
Q^\mu(t)=Q^\mu_0-\sum_{j=1}^{N}q_j^\mu n_j(t) 
\ee
\noindent
where $Q^\mu_0$ is the amount of resource $\mu$ that would be present
in the system if no species fed on it. To fix ideas, let us suppose
that 

\be 
Q_0^\mu=P+s\sqrt{P}~x^\mu \label{qomu} 
\ee 
\noindent
where $s>0$ is a constant and $x^\mu$ is a quenched Gaussian random
variable with zero average and $\avg{x^\mu
x^\nu}=\delta_{\mu\nu}$. (The $P$-scaling is introduced in order to
obtain a well-defined limit $N\to\infty$, or $P\to\infty$) Loosely
speaking, the parameter $s$ is related to the variability of
resources: for small $s$, the resource level is roughly the same for
all resources, while increasing $s$ the distribution of resource
levels gets less and less uniform.  Clearly, the number of species
that survive (i.e. such that $n_i(t)>0$) in the steady state will
depend on a number of factors, like the distribution of available
resources and how similar the species are among themselves, that is on
the distribution of $q_i^\mu$'s, which we take to have first moments

\be 
\davg{q_i^\mu}=q,~~~~~~~~~~~~\davg{(q_i^\mu-q)^2}=1 
\ee
\noindent
(here and in what follows we denote averages over the quenched
disorder by $\davg{\cdots}$) Along with the questions concerning the
resulting resource loads distribution, an interesting problem to raise
is the following: what is the typical maximum number of species that
can be supported asymptotically when the number of resources $P$ is
large ($P\to\infty$) as a function of $s$?

This issue can be tackled by noting that 

\be 
H(t)=\frac{1}{2}\sum_\mu
Q^\mu(t)^2-\sum_{i=1}^{N_s} f_i n_i(t) 
\ee 
\noindent
is a Lyapunov function of the dynamics (i.e. $\dot H(t)\leq 0$; this
can be easily shown by a direct calculation). This implies that the
steady state properties are described by the minima of $H$ over
$\{n_i\geq 0\}$. Note that in the steady state \be
H\simeq\sum_\mu\l(\avg{Q^\mu}-\ovl{\avg{Q}}\r)^2,~~~~~~~~~~~
\ovl{\avg{Q}}=\frac{1}{P}\sum_\mu \avg{Q^\mu} \ee

In the rest of this section we shall first work out in detail the
minimization of $H$ and then discuss the resulting scenario.

\subsubsection{Statics (replica approach)}

The task of minimizing $H$
can be carried out by introducing a `partition sum' 

\be
Z=\Tr_{\bsy{n}} e^{-\beta H},~~~~~~~~~~~\bsy{n}=\{n_i\}
\ee and applying the replica trick: \be \fl
\lim_{N\to\infty}\davg{\min_{\bsy{n}}\frac{H}{N}}= -\lim_{\beta\to\infty}
\lim_{N\to\infty}\frac{1}{\beta N}\davg{\log Z}=
-\lim_{\beta\to\infty}\lim_{r\to 0}\lim_{N\to\infty} \frac{1}{\beta
r N}\log\davg{Z^r}\label{tricco}
\ee

\noindent
The calculations are relatively straightforward. Using the
Hubbard-Stratonovich trick we can write 

\be 
\fl 
Z=\Tr_{\bsy{n}}
\l[\prod_{\mu=1}^P e^{-\frac{\beta}{2N}\pt{Q^\mu}^2}\r]
e^{-\frac{\beta}{N}\sum_i f_i n_i} = \Tr_{\bsy{n}}
\avg{\prod_{\mu=1}^P e^{\ii\sqrt{\frac{\beta}{N}}z^\mu Q^\mu}}_z
e^{-\frac{\beta}{N}\sum_i f_i n_i} 
\ee 

\noindent
where $\avg{\ldots}_z$ is an
average over the Gaussian variables $z^\mu$ with $\avg{z^\mu}_z=0$
and $\avg{z^\mu z^\nu}_z=\delta_{\mu\nu}$. So we have 

\[
\fl
\davg{Z^r}= \Tr_{\{\bsy{n}_a\}} \avg{\prod_{\mu=1}^P
\davg{e^{\ii\sqrt{\frac{\beta}{N}}\pt{\sum_a z^\mu_a} Q^\mu_0}}
\prod_{i=1}^N \davg{e^{-\ii\sqrt{\frac{\beta}{N}}\pt{\sum_a z^\mu_a
n_{ia}} q_i^\mu}} }_z e^{-\frac{\beta}{N}\sum_i f_i \sum_a n_{ia}} 
\]

\noindent
where the index $a$ runs over replicas ($a=1,\ldots,r$). The first
disorder average is done over $Q_0^\mu$ as given by \req{qomu} (thus
more properly over $x^\mu$), while the second is done over the
$q_i^\mu$'s. The former is easily performed. As for the latter we note
that if $\beta/N\ll 1$ (which is the case since we first take the
limit $N\to\infty$ and then the limit $\beta\to\infty$), then

\[
\davg{e^{-\ii\sqrt{\frac{\beta}{N}}\pt{\sum_a z^\mu_a n_{ia}}
q_i^\mu}} \simeq e^{-\ii\sqrt{\frac{\beta}{N}}\davg{q_i^\mu}\pt{\sum_a
z^\mu_a n_{i,a}} -\frac{\beta}{2N}\pt{\sum_a z^\mu_a
n_{i,a}}^2\davg{(q_i^\mu-q)^2}}
\]

\noindent
We thus find 
\[ 
\fl \davg{Z^r}= \Tr_{\{\bsy{n_a}\}}
\avg{\prod_{\mu=1}^P e^{\ii\sqrt{\beta N}\sum_a
z^\mu_a\pt{\alpha-\frac{q}{N}\sum_i n_{ia}}}
e^{-\frac{\beta}{2}\sum_{a,b}z^\mu_a z^\mu_b \pt{\alpha
s^2+\frac{1}{N}\sum_i n_{ia}n_{ib}}} }_z e^{-\frac{\beta}{N}\sum_i f_i
\sum_a n_{ia}} 
\] 

\noindent
The leading term in the above exponential is the first one. However,
it corresponds to an undesirable super-extensive term in the free
energy unless \be \frac{1}{N}\sum_i n_{ia}= \frac{\alpha}{q} \ee If
so, the annoying term acts as a $\delta$-distribution that ensures the
above condition:

\[
\fl
e^{\ii\sqrt{\beta N}\sum_a z^\mu_a\pt{\alpha-\frac{q}{N}\sum_i
n_{ia}}} \propto \prod_a\delta\pt{N\alpha/q-\sum_i n_{ia}}\propto \int
d\bsy{w}~ e^{\beta\sum_a w_a\pt{N\alpha/q-\sum_i n_{ia}}}
\]

\noindent
Furthermore one sees that the relevant macroscopic order parameter is
the overlap

\be
G_{ab}=\frac{1}{N}\sum_i n_{ia}n_{ib} 
\ee 

\noindent
which can be introduced in the replicated partition sum with the
identities

\be
\fl 
1=\int \delta
\left( G_{ab}-\frac{1}{N} \sum_i n_{ia}n_{ib}\right)dG_{ab}\propto
\int dR_{ab}\ dG_{ab}\ e^{-\frac{N\alpha\beta^2}{2} R_{ab}\left(
G_{ab}-\frac{1}{N}\sum_i n_{ia}n_{ib}\right)} 
\ee 

\noindent
for all $a\ge b$.  Noting that when $\beta\to\infty$ only the minima
of $H$ contribute to the partition sum, it is easy to understand that
$G_{ab}$ measures how similar different minima $a$ and $b$ are to each
other. We may now factorize over resources to obtain

\[
\prod_\mu\avg{e^{-\frac{\beta}{2}\sum_{a,b}z^\mu_a z^\mu_b \pt{\alpha
s^2+G_{ab}}}}_z=e^{-\frac{P}{2}\tra\log\l[\bsy{I}+\beta\pt{\alpha s^2
+\bsy{G}}\r]}
\]

\noindent
so that, finally, factorizing over species, we arrive at 

\be
\davg{Z^r}=\int e^{-\beta rN
f(\bsy{w},\bsy{G},\bsy{R})}d\bsy{w}d\bsy{G}d\bsy{R} 
\ee 

\noindent
with
\begin{eqnarray}
\fl 
f(\bsy{w},\bsy{G},\bsy{R})=&\frac{\alpha}{2r\beta}\tra\log\l[\bsy{I}+
\beta\pt{\alpha s^2 +\bsy{G}}\r] +\frac{\alpha\beta}{2r}\sum_{a\ge b}
R_{ab}G_{ab}\nonumber\\ 
&-\frac{\alpha}{rq}\sum_a w_a
-\frac{1}{r\beta}\log\avg{\Tr_n e^{ \frac{\alpha\beta^2}{2}\sum_{a\ge
b} R_{ab} n_a n_b- \beta\sum_a w_a n_a-\frac{\beta f}{N}\sum_a n_a}}_f
\end{eqnarray}

\noindent
where now $\avg{\cdots}_f$ stands for an average over the distribution
of decay rates. By the principle of steepest descent, when
$N\to\infty$, $\davg{Z^r}$ is dominated by the saddle point values of
the order parameters $\bsy{G}$, $\bsy{R}$ and $\bsy{w}$ (which we
shall denote by a $^\star$) so 

\be
\lim_{N\to\infty}\min_{\bsy{n}}\frac{H}{N}=
\lim_{\beta\to\infty}\lim_{r\to 0}
~f(\bsy{w}^\star,\bsy{G}^\star,\bsy{R}^\star)
\ee

\noindent
To proceed further, we assume that $\bsy{G}^\star$, $\bsy{R}^\star$
and $\bsy{w}^\star$ take the replica-symmetric (RS) form\footnote{This
assumption gives the exact results in almost all the cases we shall
discuss in this review because the functions to be minimized have a
unique minimum. Should this condition fail, one must resort to more
complicated Ans\"atze known as replica-symmetry breaking.}
\be\label{replicasym}
G_{ab}^\star=g+(G-g)\delta_{ab}~~~~~~~R_{ab}^\star=2r-(r+\rho/\beta)\delta_{ab}
~~~~~~~w_a^\star=w \ee which leads, in the limit $r\to 0$, to the free
energy density
\begin{eqnarray}
\fl
f_{{\rm RS}}(g,G,r,\rho,w)=&\frac{\alpha}{2\beta}\log\l[1+\beta(G-g)\r]+
\frac{\alpha}{2}\frac{\alpha s^2+g}{1+\beta(G-g)}+ \frac{\alpha
r}{2}\beta(G-g)\nonumber\\ 
&-\frac{\alpha}{2}G\rho-\frac{\alpha}{q} w
-\frac{1}{\beta}\avg{\log\int_0^\infty dn~ e^{-\beta V(n|z,f)}}_{z,f}
\label{bobo}
\end{eqnarray}
where the ``potential'' $V$ is given by 

\be
V(n|z,f)=\frac{1}{2}\alpha\rho n^2+\pt{w+f/N-\sqrt{\alpha r} z}n 
\ee

\noindent
and the average $\avg{\cdots}_{z,f}$ is over both the unit Gaussian
variable $z$ and the decay rate $f$, whose distribution we left
unspecified up to now. It is clear that if this distribution has
finite moments and does not get broader with $N$, we can drop the term
$f/N$ above.  Now let us take the remaining limit $\beta\to\infty$,
where minima are selected, assuming that $H$ has a unique minimum. In
this case, clearly, $G\to g$ (there is only one minimum by
assumption!) and we may look for solutions with 

\be
\lim_{\beta\to\infty}\beta(G-g)=\chi 
\ee 

\noindent
finite. Moreover, the last
integral in \req{bobo} in the limit $\beta\to\infty$ is dominated by
the minimum of $V$. Therefore we end up with 
\begin{eqnarray}
\fl
\lim_{\beta\to\infty}f_{{\rm RS}}(g,G,r,\rho,w)=
\frac{\alpha}{2}\frac{\alpha s^2+G}{1+\chi}\nonumber\\+ \frac{\alpha
r}{2}\chi-\frac{\alpha}{2}G\rho-\frac{\alpha}{q} w
+\frac{1}{2}\alpha\rho\avg{n^2}_\star+w\avg{n}_\star-\sqrt{\alpha r}\avg{z
n}_\star 
\end{eqnarray}

\noindent
where $\avg{\cdots}_\star$ are averages over the normal variable
$z$, with the $n=n^\star(z)$ which minimizes $V$:
\[
n^\star(z)=\frac{\sqrt{\alpha r}}{\alpha \rho}(z-z_0)\theta(z-z_0),
~~~~~~~~~~~z_0=w/\sqrt{\alpha r}.
\] 

\noindent
Notice that this operation
corresponds to an `effective species' problem whose solution describes
the collective behavior of the original $N$-species system.  

The saddle point equations are 

\bea 
\frac{\partial f_{{\rm RS}}}{\partial w}=0&~~~~~~\Rightarrow ~~~~~~&
\avg{n}_\star=\frac{\alpha}{q}\nonumber\\ 
\frac{\partial f_{{\rm RS}}}{\partial \rho}=0&~~~~~~\Rightarrow
~~~~~~& \avg{n^2}_\star=G\nonumber\\
\frac{\partial f_{{\rm RS}}}{\partial r}=0&~~~~~~\Rightarrow ~~~~~~&
\avg{nz}_\star=\sqrt{\alpha r}\chi\\ 
\frac{\partial f_{{\rm RS}}}{\partial G}=0&~~~~~~\Rightarrow ~~~~~~& 
\rho=\frac{1}{1+\chi}\nonumber\\ 
\frac{\partial f_{{\rm RS}}}{\partial \chi}=0&~~~~~~\Rightarrow ~~~~~~&
r=\frac{\alpha s^2+G}{(1+\chi)^2}\nonumber 
\eea 

\noindent
It is easier to find a parametric solution in terms of $z_0$: let us define
\begin{eqnarray}
\fl
I_1(z_0)=\int_{z_0}^{\infty}\frac{dz}{\sqrt{2\pi}}e^{-z^2/2}(z-z_0)=
\frac{e^{-z_0^2/2}}{\sqrt{2\pi}}-\frac{z_0}{2}\erfc\pt{z_0/\sqrt{2}}\nonumber\\
\fl
I_2(z_0)=\int_{z_0}^{\infty}\frac{dz}{\sqrt{2\pi}}e^{-z^2/2}(z-z_0)^2=
\frac{1}{2}\pt{1+z_0^2}\erfc\pt{z_0/\sqrt{2}}- \frac{z_0
e^{-z_0^2/2}}{\sqrt{2\pi}}\\
\fl
I_z(z_0)=\int_{z_0}^{\infty}\frac{dz}{\sqrt{2\pi}}e^{-z^2/2}z(z-z_0)=
\frac{1}{2}\erfc\pt{z_0/\sqrt{2}} \nonumber
\end{eqnarray} 

\noindent
After some manipulations we find 
\begin{eqnarray} 
\alpha = \frac{1}{2}\l[I_2+\sqrt{I_2^2+4s^2q^2 I_1^2}\r]
\label{alpha}\nonumber\\
G = \frac{\alpha^2 I_2}{{q}^2 I_1^2}\\ 
\chi=\frac{I_z}{\alpha-I_z}\nonumber
\end{eqnarray}

\begin{figure}[t]
\centering 
\includegraphics*[angle=-90,width=10cm]{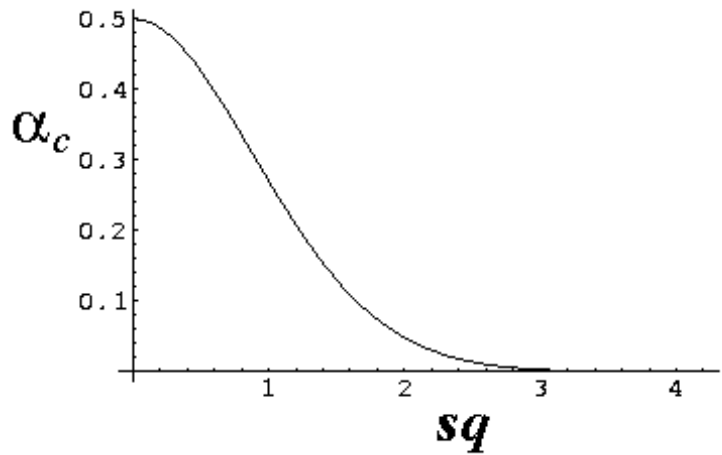}
\caption{Behavior of $\alpha_c$ as a function of $sq$.}
\label{eco2}
\end{figure}

\begin{figure}[t]
\centering 
\includegraphics*[angle=-90,width=10cm]{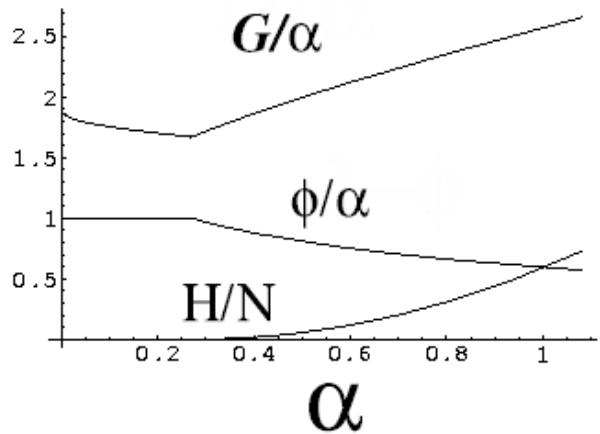}
\caption{Behavior of $H$, $G$ and of the fraction of surviving species
$\phi$ as a function of $\alpha$ for $sq=1$.}
\label{eco1}
\end{figure}

\noindent
The assumed scaling of parameters with $\beta$, and hence the 
above equations, are valid only for $z_0<z_0^\star$ where $z_0^\star$
is the solution of 
\be
I_z(z_0^\star)\l[I_z(z_0^\star)-I_2(z_0^\star)\r]=s^2 q^2 I_1^2(z_0^\star).
\ee
Indeed $\chi\to\infty$ as $z_0\to z_0^\star$. This singularity marks a
phase transition at a point $\alpha_c=I_z(z_0^\star)$ between a phase
$\alpha>\alpha_c$ which is described by the equations above, and one
where $\chi=\infty$.
The critical point
$\alpha_c$ is a decreasing function of $sq$ (from a value
$\alpha_c=1/2$ for $sq=0$) which rapidly vanishes as
$sq$ increases (it's already $10^{-5}$ for $sq =4$). It is
reported in Fig. \ref{eco2}.
At the transition, the susceptibility $\chi\sim |\alpha-\alpha_c|^{-1}$
diverges and the free energy, which as we said is proportional to the
variance of the resource loads distribution, vanishes. This means that
below $\alpha_c$ all resources are exploited to the same extent, while
above $\alpha_c$ the resource load distribution is not uniform. For
the fraction of surviving species (with $n_i>0$) we get 

\be
\phi=\avg{\theta(z-z_0)}_z=\int_{z_0}^\infty
\frac{dz}{\sqrt{2\pi}}e^{-z^2/2}=
\frac{1}{2}\erfc\pt{z_0/\sqrt{2}}=I_z(z_0) 
\ee 

\noindent
so $\phi<\alpha$ for $\alpha>\alpha_c$ and $\phi\to\alpha$ at
$\alpha_c$. This means that
below $\alpha_c$ the number of surviving species equals that of
resources while for $\alpha>\alpha_c$ there is on average less than
one species per resource (or
$\phi/\alpha <1$). The behavior of $H$, $G$ and of the fraction of
surviving species per resource $\phi/\alpha$ is
displayed in Fig. \ref{eco1} 
as a function of $\alpha$ for fixed $sq$.

\subsubsection{Stability}

\begin{figure}[t]
\centering 
\includegraphics*[angle=-90,width=10cm]{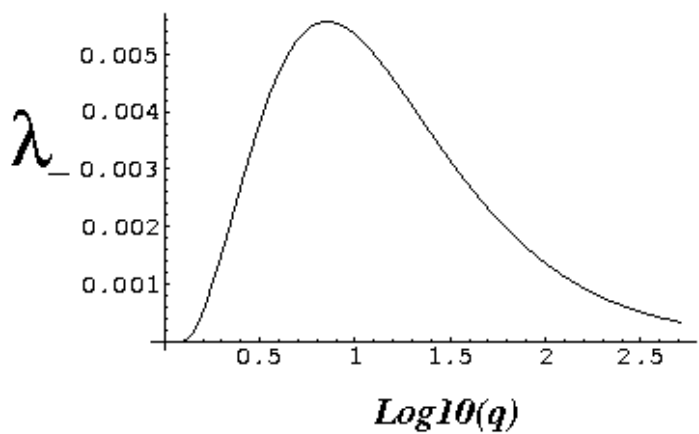}
\caption{Behavior of $\lambda_-$ as a function of $q$ for $\alpha=0.2$
and $s=1$. The ecosystem is marginally unstable when $\lambda_-=0$,
whereas maximal stability occurs when $\lambda_-$ attains a maximum.}
\label{lameno}
\end{figure}

Note that at fixed $q$ the maximal number $P/\alpha_c$ of species that
can be sustained in an ecosystem with $P$ resources is an increasing
function of $s$, so that by increasing the variability of resources
the ecosystem gets more stable. The threshold of stability also
increases if $q$ increases. Having fixed the variance of $q_i^\mu$ to
$1$, increasing $q$ means that species get more and more similar. This
seems at first sight a contradictory scenario.  To sort out this
issue, let us analyze the linear stability of the system.  Let
$n_i(t)=n_i(\infty)+\sqrt{n_i(\infty)}\eta_i(t)$ where $n_i(\infty)$
is the asymptotic value of the population of species $i$ and
$\eta_i(t)$ is a small perturbation. To leading order, the dynamics is
given by
\begin{equation}
\dot \eta_i(t)=-\sum_{j=1}^{N}\Delta_{ij}\eta_j(t)
\end{equation}
with 
\begin{equation}
\Delta_{ij}= \frac{1}{P}\sum_{\mu=1}^P
\sqrt{n_i(\infty)}\pt{q_i^\mu-q}\pt{q_j^\mu-q}\sqrt{n_j(\infty)}.
\end{equation}
The stability is related to the smallest eigenvalue $\lambda_-$ of
$\Delta_{ij}$. This can be computed explicitly \cite{mitra} and it is
given by:
\begin{equation}
\lambda_-=\frac{1}{q}\pt{\sqrt{\alpha}-\sqrt{\phi}}
\end{equation}
This shows that the phase transition point, where
$\alpha_c=\phi(\alpha_c)$, is the onset of dynamical instability of
the system. The presence of the factor $1/q$ in $\lambda_-$ causes an
interplay of the effects of increasing $s$ and increasing $q$ (since
$\lambda_-\to 0$ as $q\to\infty$), ultimately leading to a maximal
stability for intermediate values of $q$, as can be seen by the
behavior of $\lambda_-$ versus $q$, Fig. \ref{lameno}.  It is also
easy to show that $\lambda_-$ is an increasing function of $\alpha$,
for all values of $s$ and $q$. Hence the introduction of new species
always decreases the stability of the ecosystem, in agreement with
May's classical result \cite{May}.

\subsection{The El Farol problem}

The El Farol problem is the paradigm of resource allocation games with
inductive agents. It can be stated as follows \cite{arth}. $N$
customers labeled $i$ have to decide independently on each night $t$
whether to attend ($a_i(t)=1$) or not ($a_i(t)=0$) the El Farol bar,
which has a capacity of $L<N$ seats. The place is enjoyable only if
it's not overcrowded, that is only if the attendance $A(t)=\sum_i
a_i(t)$ doesn't exceed the number of seats. In order to make their
decisions, customers aim at predicting whether the bar will be crowded
or not on any given night based on the past attendances. 

In his seminal work, Arthur has pointed out the frustration inherent
in such a situation. If everybody expects that the bar will be
crowded, no one will go and the bar will be empty. Conversely all
agents may attend the bar at the same time, if they all expect it to
be empty. Hence he argued that this is a situation which forces
expectations of different agents to diverge. It is reasonable to think
that, if agents start with different expectation models and revise
them according to the history of the attendance, their expectations
will never converge and agents' heterogeneity will be preserved
forever. He then showed by computer experiments with $N=100$ and $L=60$
that inductive agents endowed with fixed `predictors' (namely look-up
tables associating to each series of past attendances a binary
decision like go/don't go) are able to self-organize so that the
attendance $A(t)$ fluctuates around the comfort level $L$.

Note that the El Farol problem can be regarded as an embryonic market
where $L$ units of an asset or a commodity must be allocated on any
given day $t$. They are offered to $N$ agents who may decide to invest
$1$\euro ~to buy it ($a_i(t)=1$) or not ($a_i(t)=0$).  The attendance
$A(t)$ is then the demand of the asset (the number of available units,
or supply, is fixed at $L$). Each unit of asset delivers a return of
$1$\euro ~to its owner at the end of the period. Imagine that the
price at which the asset is sold is determined at each period by a
market clearing condition (demand $=$ supply): $A(t)=L p(t)$. Then an
agent who invests $a_i(t)$\euro ~in the asset, receives $a_i(t)/p(t)$
units of it. These will be worth $a_i(t)/p(t)$\euro ~at the end of the
period. If $p(t)>1$, which occurs if $A(t)>L$ (crowded bar), it is not
convenient to invest (attend). If $p(t)<1$ it is instead worthwhile to
invest (attend). 

\subsubsection{Definition}

In what follows, we focus on a tractable version of the model that
differs from Arthur's original work in the form of the predictors but
preserves all the main qualitative features of the model \cite{elfa}.
In order to formalize the problem it is reasonable to assume that (i)
customers have a finite memory, that is, their analyzing power is
limited and they must base their prediction on the attendances of a
finite number (say, $m$) of past nights, and that (ii) they are
insensitive to the actual size of the attendance (perhaps simply
because they don't have access to it) and rather only know whether the
bar was overcrowded or not on a given night. This means that the
information available to customers on night $t$ is encoded in the
string \be\label{dyninfo}
\mu(t)=\{\theta\l(L-A(t-1)\r),\ldots,\theta\l(L-A(t-m)\r)\}\in\{0,1\}^m
\ee where $\theta(\cdot)$ is the Heaviside function:
$\theta(L-A(t))=1$ if the bar is enjoyable ($A(t)<L$) while
$\theta(L-A(t))=0$ if the bar is overcrowded ($A(t)>L$). The time
evolution of the string $\mu(t)$ is governed in time by the map \be
\mu(t+1)=\l[2\mu(t)+\theta(L-A(t))\r]{\rm mod}(2^m) \ee The above
equation completely defines the structure of the information available
to agents in the case in which they base themselves on the past
attendances.
\begin{figure}[t]
\centering 
\includegraphics*[width=8cm]{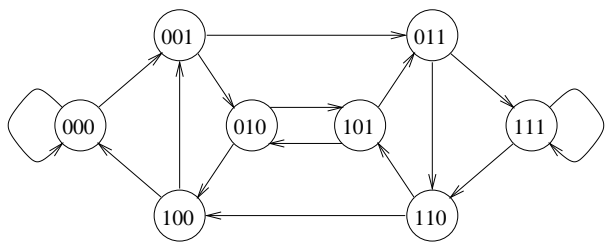}
\caption{De Bruijn graph of order 3 (from \cite{CMmem}).}
\label{debr}
\end{figure}
Graphically, the evolution of history strings is constrained to occur
on a de Bruijn graph \cite{Deb} of order $m$, Fig. \ref{debr}.

We shall consider, for comparison, another possibility, namely that
the information supplied to customers is a random binary string of
length $m$ or equivalently a random integer (`information pattern')
drawn from $\{1,\ldots,2^m\equiv P\}$ with equal probability at each
time step. We shall refer to the latter as the case of exogenous
random information, as opposed to the former of endogenous
information.  The obvious difference between the two choices is that
while in the latter case the space of informations is sampled
uniformly by construction, in the former this is in principle not
true. There is however a deeper difference that has serious
consequences on the analytical solubility of the model: in the case
of random information the dynamics is Markovian.

Having defined the information source, let us specify the agents'
behavior. Even in a simplified context, making the optimal decision
for each given string requires an unrealistic computational capacity
that should be shared by all agents. Inductive reasoning requires that
customers stick instead to simple decision rules. In particular, we
assume that they have at their disposal a small number $S$ of
$2^m$-dimensional vectors called `strategies' (analog to Arthur's
predictors) that map information strings into binary actions (go/don't
go): \be \bsy{a}_{ig}:\{0,1\}^m\ni\mu\to a_{ig}^\mu\in\{0,1\}~~~~~~~
(i=1,\ldots,N;g=1,\ldots,S) \ee In the table below one such possible
strategy is shown for $m=3$ (or $P=8$).

~

\begin{center}
\begin{tabular}{|c|c|c|}\hline
        past attendance string&pattern $\mu$&decision $a^\mu$\\
        \hline\hline 000&1&1\\ \hline 001&2&0\\ \hline 010&3&0\\
        \hline 011&4&1\\ \hline 100&5&1\\ \hline 101&6&0\\ \hline
        110&7&1\\ \hline 111&8&0\\ \hline
\end{tabular}
\end{center}

~

Customers are heterogeneous as of course different agents have
different strategies. This is modeled by assuming that each component
$a_{ig}^\mu$ of every strategy $\bsy{a}_{ig}$ is drawn independently
for all $i$, $\mu$ and $g$ with probability distribution \be
P(a)=\ovl{a} \delta(a-1)+\l(1-\ovl{a}\r) \delta(a) \ee where $\ovl{a}$
is the average attendance frequency of agents. Strategies are assigned
to agents at time $t=0$ and are kept fixed throughout the game. In
order to decide which strategy to adopt on every night, agents keep
tracks of their performance via a score function that is updated
according to the following rule: \be\label{elfarol_dyn}
U_{ig}(t+1)-U_{ig}(t)=\l(1-2 a_{ig}^{\mu(t)}\r)\l[A(t)-L\r] \ee with
the rationale that strategies suggesting not to go
($a_{ig}^{\mu(t)}=0$) are rewarded when the attendance is higher than
$L$ and punished when it is lower than $L$ (and vice versa when
$a_{ig}^{\mu(t)}=1$). Then on each night every agent selects the
strategy with the highest cumulated score: \be g_i(t)=\argmax_g
U_{ig}(t) \ee and acts accordingly: $a_i(t)=a_{i g_i(t)}^{\mu(t)}$. In
short, the model's rules can be summarized as follows:
\begin{eqnarray}
g_i(t)=\argmax_g U_{ig}(t)\nonumber\\
A(t)=\sum_i a_{i g_i(t)}^{\mu(t)}\label{elfar}\\
U_{ig}(t+1)-U_{ig}(t)=\l(1-2 a_{ig}^{\mu(t)}\r)\l[A(t)-L\r]\nonumber
\end{eqnarray}
(from top to bottom: strategy selection; aggregation; updating). It is
understood that scores are initialized at time $t=0$ at certain values
$U_{ig}(0)$.

\subsubsection{Macroscopic properties}

\begin{figure}[t]
\centering 
\includegraphics*[width=8cm]{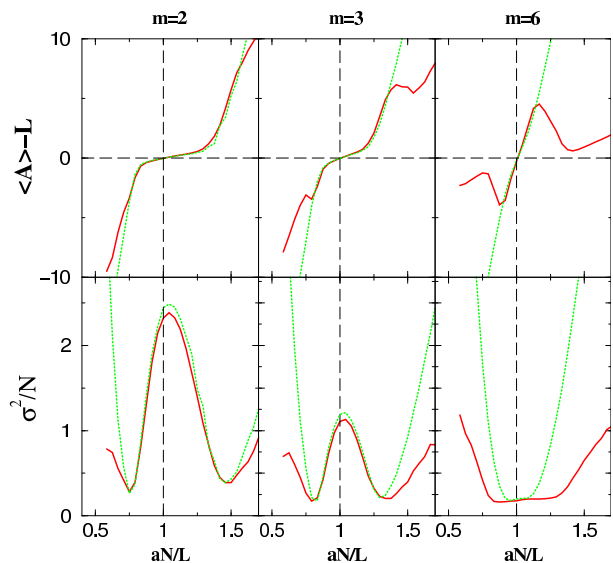}~
\caption{Average deviation of the attendance from the comfort level
(top) and fluctuations (bottom) versus $\ovl{a}N/L$ for endogenous
(solid lines) and exogenous (dashed lines) information (from
\cite{elfa}).}
\label{figelfarol}
\end{figure}

After a transient, the dynamics defined by \req{elfar} will reach a
steady state whose global efficiency can be conveniently characterized
by two parameters: the average deviation of the attendance from the
comfort level $L$, $\avg{A-L}$ and its fluctuations
$\sigma^2=\avg{(A-L)^2}$. The former measures the degree to which
agents coordinate to generate attendances around the comfort
level. The latter measures the waste of resources: the larger
$\sigma^2$ the bigger the deviations of the attendance from the
comfort level (in either direction). In a nutshell, it quantifies the
quality of the coordination. The behavior of the two quantities
(properly normalized with $N$) at fixed $L=60$, $\ovl{a}=1/2$ and
$m=2,3,6$ and varying $N$ is displayed in Fig. \ref{figelfarol} for
endogenous (solid lines) and random (dashed lines) information.
In the former case, a general feature that emerges is that the average
attendance settles at the comfort level in a window of values of
$\ovl{a}$ centered around $L/N$ whose size shrinks as $m$ increases.
Out of this window, sensible deviations may occur. In parallel,
fluctuations are maximal at $\ovl{a}=L/N$ for $m=2$ and the height of
the maximum decreases with increasing $m$ until it disappears. This
implies that the waste of resources is comparatively larger when $m$
is smaller, so that for instance the fraction of losers is larger for
small $m$. Thus one can say that global efficiency increases when $m$
increases. The behavior in the case of random information is
qualitatively similar to the previous case in the vicinity of
$\ovl{a}N/L=1$. Quantitative deviations occur outside this region. 

\begin{figure}[t]
\centering 
\includegraphics*[width=8cm]{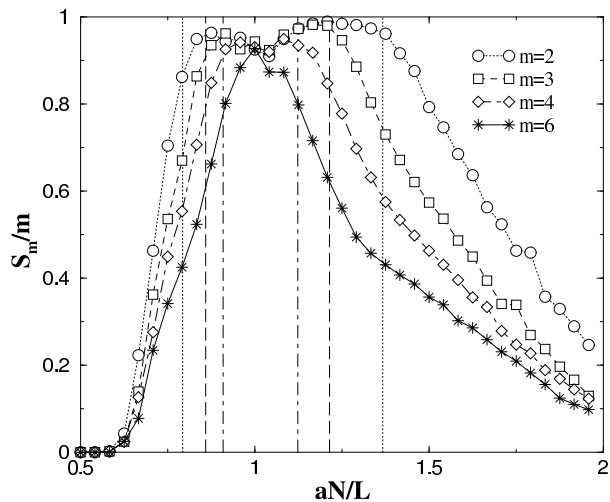}
\caption{Normalized entropy $S_m/m$ versus $\ovl{a}N/L$ for different
memory lengths $m$ (from \cite{elfa}).}
\label{entel}
\end{figure}

Based on this, one expects that with endogenous information the
information space is sampled uniformly around $\ovl{a}\simeq
L/N$. This is indeed so. To see it, one can measure the frequency with
which histories are sampled in the steady state, $\rho(\mu)$, and
calculate the entropy \be S(m)=-\sum_\mu \rho(\mu)\log_2\rho(\mu) \ee
such that $S(m)=m$ when the information space is sampled uniformly
(the `effective' number of information patterns visited by the
dynamics is $2^{S(m)}$). As shown in Fig. \ref{entel}, $S(m)/m\simeq
1$ only for when $\ovl{a}\simeq L/N$.
Outside this phase, the entropy decreases, signaling that the
information dynamics is biased.

These findings indicate that the degree to which inductive agents are
able to coordinate the exploitation of the limited resource in a way
that is collectively efficient depends on the size of the information
space they base themselves on. While the average level of activity
always settles at the resource level, fluctuations get smaller and
smaller as the information space grows. When the average attendance
frequency is close to $L/N$, then, the particular nature of the
information provided to agents doesn't affect the stationary
macroscopic properties. The relevant requirement is that all agents
possess the same information, independently of whether it's the true
attendance history or a random string.  

\subsubsection{Dynamics (continuous-time limit approach)}

The mathematical analysis of this model can be carried out in the case
of exogenous information by studying the continuous-time limit of
\req{elfar} along the lines of \cite{CTL}. A few simplifications are
necessary to this aim. First note that the dynamics \req{elfar} is
non-linear in a way that doesn't allow to write it in the form of a
gradient descent, that is physically the model is defined by a set of
$N$ globally coupled Markov processes that violate detailed balance
and it is not clear that a Lyapunov function exists. It is however
possible to regularize the dynamics by smoothing the choice rule
$g_i(t)=\argmax_g U_{ig}(t)$ to \be\label{rules1}
\prob\{g_i(t)=g\}=C(t) ~e^{\Gamma U_{ig}(t)}~~~~~~~C(t)={\rm
normalization} \ee with $\Gamma>0$ the `learning rate' of agents (the
original choice is recovered for $\Gamma\to\infty$). This modification
is not without consequences and $\Gamma$ indeed turns out to play a
rather non-trivial role in the macroscopic properties. With
\req{rules1}, it is possible to construct the continuous-time limit of
\req{elfar}. 

The crucial observation is that there is a `natural' characteristic
time scale for the dynamics given by $P$ (intuitively, agents have to
check the efficiency of their strategies against all information
patterns before they can evaluate their performance
meaningfully). This implies that if one is interested in steady state
properties, time should be re-scaled as $t\to\tau=t/P$. Iterating
\req{elfar} from time $t=P\tau$ to time $t=P(\tau+d\tau)$ and setting
$u_{ig}(\tau)=U_{ig}(P\tau)$ one obtains \be
u_{ig}(\tau+d\tau)-u_{ig}(\tau)=
\frac{1}{P}\sum_{t=P\tau}^{P(\tau+d\tau)} \l(1-2
a_{ig}^{\mu(t)}\r)\l[A(t)-L\r] \ee The arguments of the sum on the
right-hand side can be separated into a deterministic and a
fluctuating term: \be \l(1-2 a_{ig}^{\mu(t)}\r)\l[A(t)-L\r] =\ovl{(1-2
a_{ig})\avg{\l[A(t)-L\r]}_\pi}+X_{ig}(t) \ee where we used the fact
that information is exogenous and random and we denoted by
$\avg{\cdots}_\pi$ an average over the distributions \be
\pi_{is}(\tau)=\frac{1}{Pd\tau} \sum_{t=P\tau}^{P(\tau+d\tau)}
C(t)~e^{\Gamma U_{ig}(t)} \ee We have therefore \be
u_{ig}(\tau+d\tau)-u_{ig}(\tau)=\ovl{(1-2
a_{ig})\avg{\l[A(t)-L\r]}_\pi}~d\tau+dW_{ig}(\tau) \ee where
$dW_{ig}(\tau)=(1/P)\sum_{t=P\tau}^{P(\tau+d\tau)} X_{ig}(t)$ is a
noise term whose statistics (average and correlations) can be derived
by noting that $X_{ig}(t)$ are independent identically-distributed
zero-average random variables, so $\avg{dW_{ig}(\tau)}=0$ and
\begin{eqnarray}
\avg{dW_{ig}(\tau)dW_{jg'}(\tau')}&=\avg{\frac{1}{P^2}
\sum_{t=P\tau}^{P(\tau+d\tau)} \sum_{t'=P\tau'}^{P(\tau'+d\tau)}
X_{ig}(t) X_{jg'}(t')}\nonumber\\
&=\frac{\delta(\tau-\tau')}{P}\avg{X_{ig}(t)X_{jg'}(t)}_\pi d\tau
\end{eqnarray}
The remaining term, $\avg{X_{ig}(t)X_{jg'}(t)}_\pi$ can be
evaluated from the statistics of disorder and of $A(t)$. Finally,
taking the limit $d\tau\to 0$ one arrives at the following
Langevin process:
\begin{eqnarray}
\dot u_{ig}(\tau)=\ovl{(1-2
a_{ig})\avg{\l[A(t)-L\r]}_\pi}+\eta_{ig}(\tau)\nonumber\\
\avg{\eta_{ig}(\tau)}=0\label{ghjk}\\
\avg{\eta_{ig}(\tau)\eta_{jg'}(\tau')}\simeq\frac{\ovl{\avg{(A-L)^2}_\pi}}{P}
\ovl{(2 a_{ig}-1)(2 a_{jg'}-1)}\delta(\tau-\tau')\nonumber
\end{eqnarray}
where in the last relation we have factorized the average over
$\mu$'s. Note also that the averages on the right-hand side of
Eq. \req{ghjk} are taken at fixed $\pi_{ig}(\tau)=C(\tau)~\exp[\Gamma
u_{ig}(\tau)]$ so they are themselves time-dependent. Therefore
\req{ghjk} is a set of complex, non-linear stochastic differential
equations in which the noise correlation is also time-dependent. At
the same time, the probability to choose a predictor $g$,
$\pi_{ig}(\tau)$, is easily seen to satisfy, in the re-scaled time
$\Gamma\tau=\widetilde{\tau}$ the stochastic equation \be \dot
\pi_{ig}(\widetilde{\tau})=\pi_{ig}(\widetilde{\tau})F[\bsy{\pi}]+
\sqrt{\Gamma}G[\bsy{\pi},\bsy{\eta}] \ee where $F$ and $G$ are
$\Gamma$-independent functions whose form is not relevant for our
scopes. This tells us that agents' preferences are subject to
stochastic fluctuations of strength proportional to $\sqrt{\Gamma}$
around their average. The larger $\Gamma$ the longer it takes to
average fluctuations out. Moreover only in the limit $\Gamma\to 0$, in
which the dynamics if the $\pi_{ig}$'s (and consequently of the
$u_{ig}$'s) becomes deterministic, the system performs a gradient
descent with the Lyapunov function \be H=\frac{1}{P}\sum_\mu
\l(\avg{A|\mu}-L\r)^2,~~~~~~~ \avg{A|\mu}=\sum_{i,g}f_{ig}a_{ig}^\mu
\ee where $\avg{\cdots|\mu}$ denotes a time-average in the steady
state conditioned on the occurrence of pattern $\mu$:\be
\avg{X|\mu}=\lim_{T,T_{{\rm eq}}\to\infty}
\frac{1}{T_\mu}\sum_{t=T_{{\rm eq}}}^{T}X(t)\delta_{\mu(t),\mu},
~~~~~~~ T_\mu=\sum_{t=T_{{\rm eq}}}^{T}\delta_{\mu(t),\mu}
\ee
and $f_{ig}=\avg{\pi_{ig}}$.

Thus the minima of $H$ over $f_{ig}$ (subject to $\sum_g f_{ig}=1$ for
all $i$) describe the steady state.  From a physical viewpoint, $H$
measures the amount of exploitable information produced in the system,
or the `predictability': if e.g. $\avg{A|\nu}\neq L$, the signal
$\mu(t)$ carries information which is useful to predict whether one
should attend or not to the bar when $\mu(t)=\nu$. The fact that the
stationary state corresponds to minimal $H$ means that agents exploit
to their best the system's predictability. We shall term phases with
$H=0$ `unpredictable' or `symmetric', while phases with $H>0$ will be
called `predictable' or `asymmetric'.

Notice also that the noise correlations are proportional to the
volatility $\sigma^2=\avg{(A-L)^2}$ which in turn depends on the set
of all $u_{ig}$'s. Hence calculating the volatility requires solving a
much more complex self-consistent problem.

\begin{figure}[t]
\centering 
\includegraphics*[width=8cm]{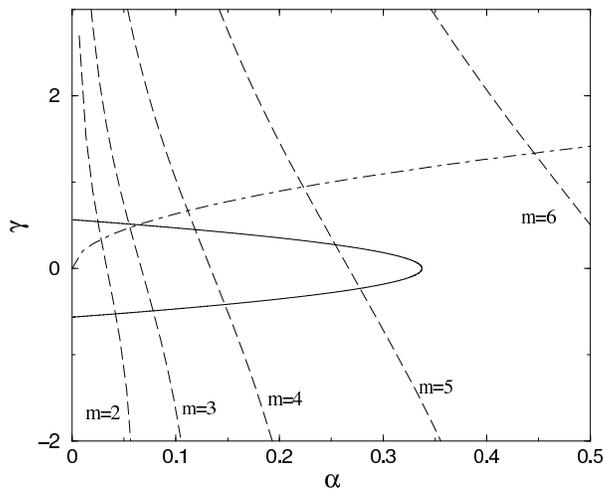}
\caption{Phase diagram of the El Farol bar problem. The solid line
    encloses the `unpredictable' phase where $H=0$. The dashed lines
    correspond to the trajectories of systems with $L=60$, $\bar
    a=1/2$ and $m=2,\ldots,6$ as the number of agents increases (from
    bottom to top). The dot-dashed line corresponds to a typical
    trajectory of a system with fixed $L,~N$ and $\bar a>L/N$ as the
    agents' memory changes (from \cite{elfa}).}
\label{pdel}
\end{figure}

The minimization can be carried out analytically resorting again to
the replica trick. The thermodynamic limit to be considered in this
case is $N\to\infty$ with $\ell=L/N$ and $\alpha=P/N$ finite. The
interesting case is that where the average attendance frequency
$\ovl{a}$ fluctuates around $\ell$ so that
$\ovl{a}-\ell=O(1/\sqrt{N})$. Indeed, if $\ovl{a}-\ell=O(1)$ then each
agent will always use the strategy that prescribes him to go more
(resp. less) often if $\ovl{a}<\ell$ (resp. $\ovl{a}>\ell$). A
convenient parametrization is given by \be
\ovl{a}-\ell=\gamma\sqrt{\frac{\ell(1-\ell)}{P}} \ee with $\gamma$
finite and independent of $N$. The resulting phase diagram in the
$(\alpha,\gamma)$ plane is reported in Fig. \ref{pdel}.
We see a region for small $\alpha$ and small $\gamma$ where $H=0$,
i.e.  $\avg{A}=L$.  In this `unpredictable' phase the average
attendance converges to the comfort level but fluctuations are
large. On the other hand, the typical attendance differs from $L$
outside this phase. Looking at the $m$-dependence, we see that as $N$
varies with $L$ and $\bar a$ and $m$ fixed, the system follows the
trajectories shown in dashed lines. For small values of $m$ these
cross the symmetric phase in the region $\bar a N\simeq
L$. 

This rich phenomenology, and specifically the non-trivial interplay
between predictability and fluctuations, is characteristic of the
complexity of many other resource-allocation models, two of which we
shall now discuss.

\subsection{Buyers and sellers in the `fish market'}

Market organization, namely the establishment of stable relationships
between buyers and sellers, is one of the basic mechanisms that
determine the efficiency of commodity markets. An important question
concerns the effects that organization has on prices and their
fluctuations. This issue has been investigated in detail in
\cite{weis} in the context of an empirical study of the Marseille fish
market. This is the sense in which this section refers to a model of a
`fish market'. Loosely speaking, one can think that a seller with
loyal buyers has an incentive to take advantage of the situation by
raising prices, thus removing the incentive of buyers to be loyal to
him. Once the relationship is broken, buyers will seek cheaper sellers
thus driving a reduction of the average price. This mechanism however
is expected to cause an increase of fluctuations (and thus a decrease
of cost certainty), since in the `disorganized' phase buyers will be
switching from one seller to another. This elementary scenario, from
which it is clear that efficiency is a two-sided concept, is worth of
a deeper investigation. A highly stylized yet non-trivial model
addressing this issue was introduced in \cite{spie}. 

One considers a system with $N$ buyers and $P$ sellers, which for
simplicity may be assumed to sell different commodities each (say each
seller supplies a different type of fish). Ultimately, the limit
$N\to\infty$ with $n=N/P$ finite shall be considered.  On each day
$t=1,2,\ldots$, every consumer $i$ has to acquire one of $S$ possible
bundles of commodities, for instance for his or her subsistence. A
bundle is a vector $\bsy{q}_{ig}=\{q_{ig}^\mu\}$ such that
$q_{ig}^\mu$ denotes the amount of goods buyer $i$ demands from seller
$\mu$ ($\mu\in\{1,\ldots,P\}$). $g\in\{1,\ldots,S\}$ labels the
different feasible bundles. We are interested to model the case in
which buyers are heterogeneous, in the sense that different buyers
have different needs and thus different possible bundles.  We
therefore assume that bundles $\bsy{q}_{ig}$ are quenched random
vectors with probability distribution\be
P(\bsy{q}_{ig})=\prod_\mu\l[(1-q)\delta(1-q_{ig}^\mu)+q\delta(q_{ig}^\mu)\r],
\ee ($0<q<1$ being the probability that any given commodity is part of
a bundle) that are assigned to consumers independently on $i$ and $g$
on day $t=0$ and are kept fixed. In this way, we introduce a further
simplification in that each seller is either visited or not by a
buyer, and the purchased quantities play no role. Moreover, we are
implicitly assuming that the different commodities are equivalent to
consumers, that is there is no commodity that all buyers will need to
buy. Coming to sellers, we assume that they set the daily price of
their commodity according to the demand they receive, denoted by
$D^\mu(t)$, so that the higher the demand the higher the price. Each
buyer, on the other hand, aims at purchasing, on each day, the bundle
he or she finds more convenient, labeled by $g_i(t)$, with the
limitation that when the choice is made the price at which the
purchase will take place is not known yet (it is determined by the
collective decision of all consumers, which form the demands). Hence
they try to learn the convenience of different bundles from experience
in order to be able to predict which bundle will have the highest
marginal utility on any given day. The events taking place on each day
$t$ can be summarized by the following scheme:
\begin{eqnarray}
g_i(t)=\argmax_g U_{ig}(t)\nonumber\\ D^\mu(t)=\sum_i q_{i
g_i(t)}^\mu\label{aggregate}\\
U_{ig}(t+1)-U_{ig}(t)=\frac{1}{P}\sum_\mu
q_{ig}^\mu\l[k-D^\mu(t)\r]\nonumber
\end{eqnarray}
At the decision stage, each buyer chooses the bundle which carries the
highest (cumulated) utility $U_{ig}(t)$. The different choices are
then aggregated and the demands are formed. Finally utilities are
updated with the following rationale: if the demand of a commodity
$\mu$ is above a certain threshold $k$, consumers perceive that
commodity as too costly and the utility of all of his feasible bundles
that include it will tend to be reduced. Similarly, if the demand has
been lower than $k$ the commodity will be seen as `cheap' and will
tend to increase the utility of the bundles that contain it. The
utility of a bundle is determined by the demands of all commodities in
it. Finally, we assume that the score updating is initialized at
values $U_{ig}(0)$.
 
It is clear that $k$ plays in this model the role of the comfort level
$L$ of the El Farol problem. Based on the discussion of the previous
section, we concentrate on the case $k=N(1-q)$. The relevant
macroscopic observables are given by
\begin{eqnarray}
H=\frac{1}{P}\sum_\mu\avg{D^\mu-N(1-q)}^2\\
\Delta=\frac{1}{P}\sum_\mu\l[\avg{(D^\mu)^2}-\avg{D^\mu}^2\r]
\end{eqnarray}
$H$ measures of how evenly buyers are distributed over sellers. Indeed
if $H=0$, each seller receives on average the same demand so that none
of them is perceived as more convenient by buyers. In this case,
consumers are distributed uniformly over producers. If $H>0$, instead,
the distribution of demands is not uniform and some producers are seen
as more or less convenient than others. $\Delta$ represents instead
the magnitude of demand fluctuations. Note that because of our
assumptions on the relation between prices and demands, $H$ is a proxy
for the average price whereas $\Delta$ quantifies the typical spread
of prices in the economy. Note also that when $H>0$ an external agent
who watches the economy from the outside trying to identify the best
bargain would manage to find more convenient sellers and make a
profit. When $H=0$, instead, this would not be possible. So
transitions from regimes with $H>0$ to regimes with $H=0$ can be seen
as transitions between inefficient and efficient states of the
economy, where by efficient state we mean one where goods flow from
sellers to buyers in such a way that no information exploitable by an
external agent is generated. States that are optimal from a collective
perspective have both $H=0$ and $\Delta$ small, because on one hand a
uniform demand distribution is desirable and on the other price
fluctuations should be such that agents have as much cost certainty as
possible on a day by day basis. Hence $H$ and $\Delta$ describe
intertwined properties, and it is on their mutual dependence that we
shall focus in what follows.

\begin{figure}[t]
\centering 
\includegraphics*[width=7cm]{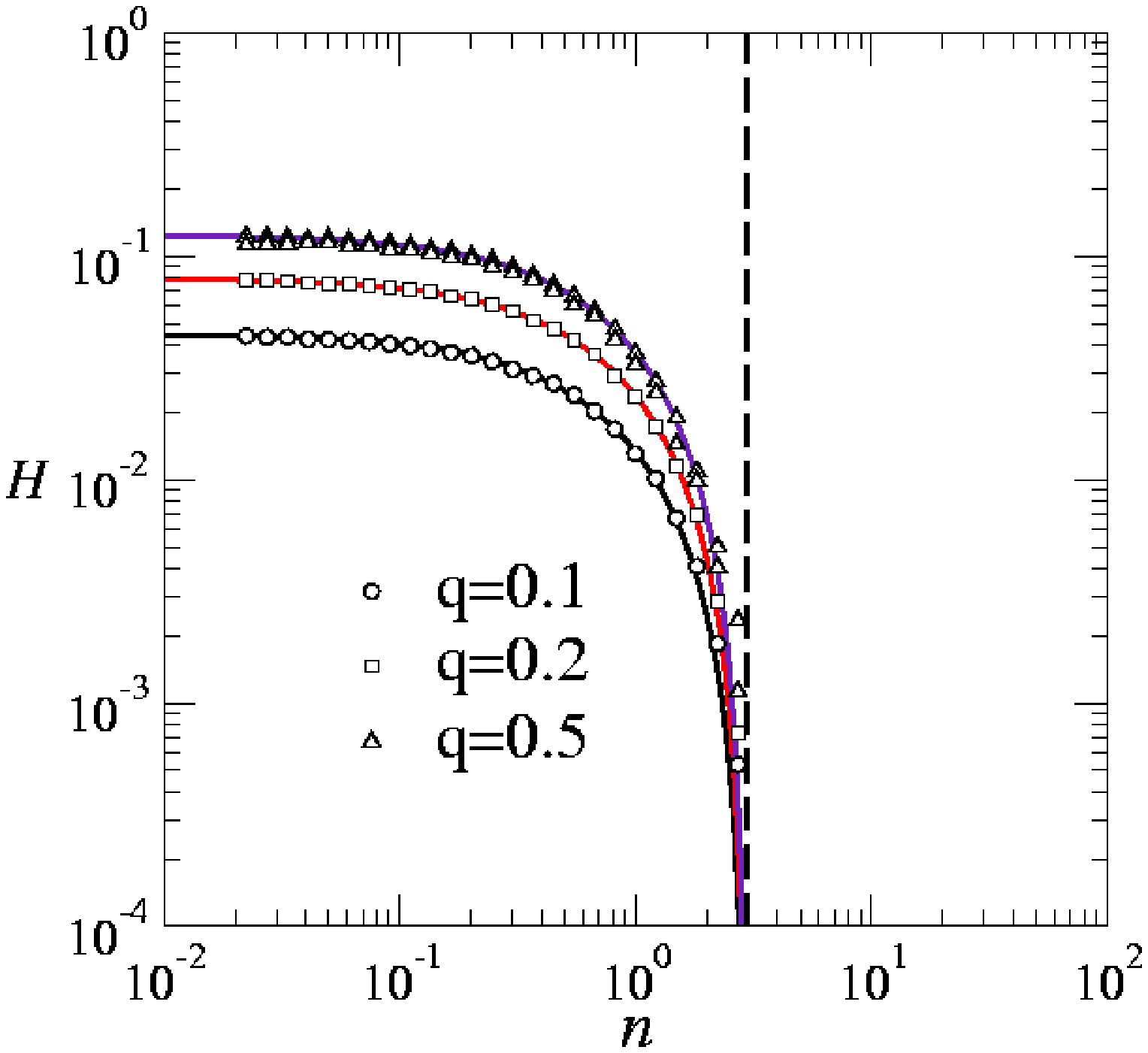}
\centering 
\includegraphics*[width=7cm]{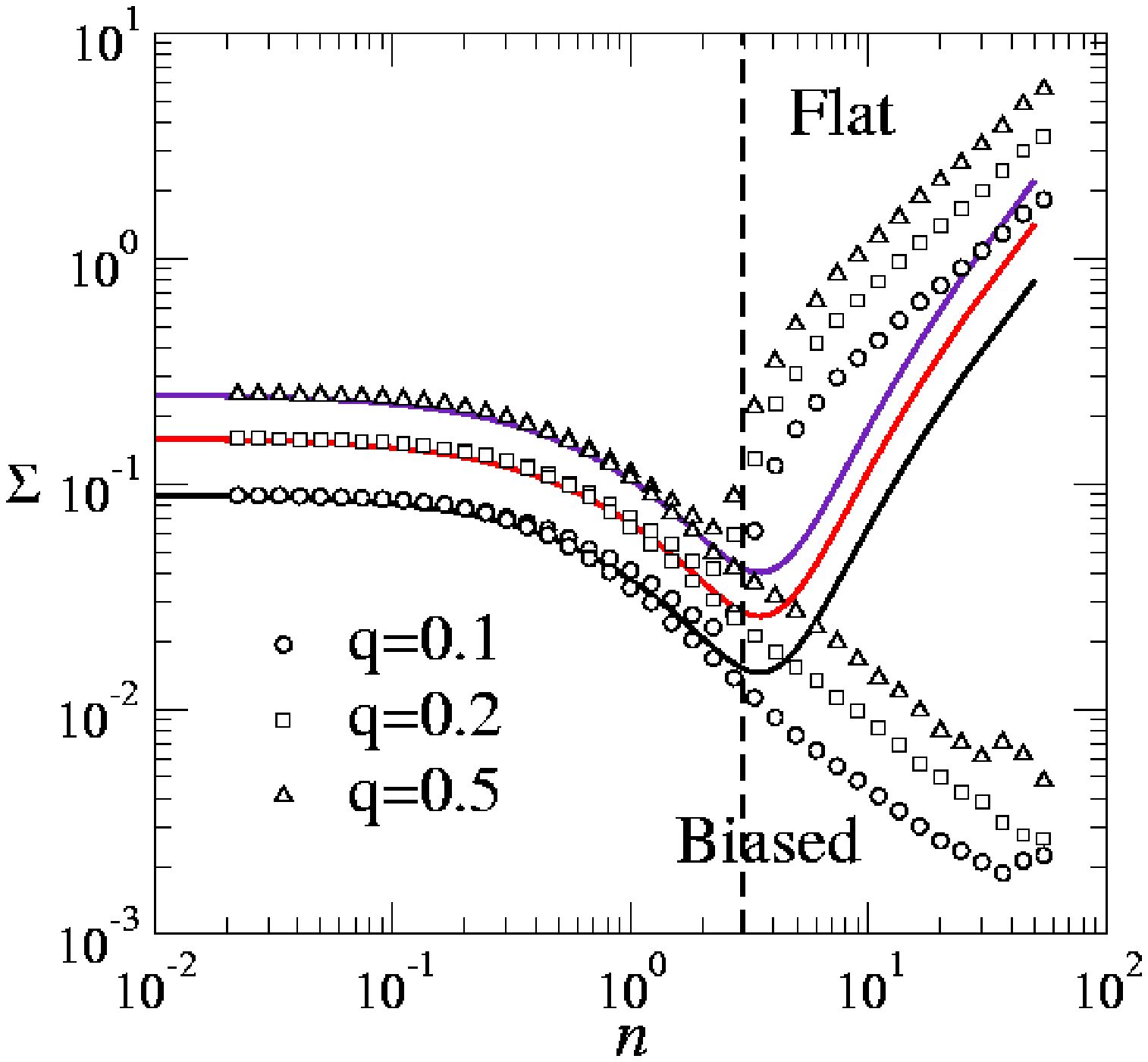}
\caption{Behavior of $H$ and $\Sigma=\Delta+H$ as a function of $n$
for various $q$ and flat ($U_{i1}(0)=U_{i2}(0)$) and biased
($[U_{i1}(0)-U_{i2}(0)]=0.2$) initial conditions (analytical curves
and numerical simulations, from \cite{spie}).}
\label{fish}
\end{figure}

Results are shown in Fig. \ref{fish}.
The behavior of $H$ indicates that as the number of buyers increases
they tend to distribute more and more uniformly over sellers until,
for $n=n_c$, $H$ vanishes and the distribution becomes uniform.  For
$n<n_c$ the economy is inefficient as the uneven distribution of
demands generates exploitable profit opportunities. For $n>n_c$ the
economy is instead efficient. Notice that results are indeed
independent of initial conditions $U_{ig}(0)$ in the inefficient
phase, while for $n<n_c$ ergodicity is broken and the steady state
depends on initial conditions. Furthermore, we see that in the
inefficient phase fluctuations are small whereas when the economy
becomes efficient the dependence on initial conditions may drive the
system to both states with large price fluctuations where ($\Delta\sim
n$), which are rather undesirable, and states with small fluctuations
(where $\Delta\sim 1/n$). This can be interpreted with the following
mechanism. When there are few buyers, many sellers receive small
demands and thus the economy presents many profitable
opportunities. As more and more buyers join the opportunity window
shrinks and players may be forced to switch bundles repeatedly in the
attempt to identify convenient commodities. This leads to the increase
of fluctuations and ultimately to a loss of day-by-day cost certainty.

Like the models described before, this one can also be studied
analytically by resorting to a replica minimization. It is not
difficult to see that the Lyapunov function in this case is precisely
$H$, so buyers collectively act so as to exploit profitable
opportunities as much as possibles. In \cite{spie} a different
solution method, based on dynamical generating functionals, is
employed. We defer a discussion of this technique to Sec. \ref{mamg1}.

\subsection{Route choice behavior and urban traffic}

A most striking example of the influence of different information
structures on the stationary properties of these systems has been
given in the experimental literature on behavioral aspects of
route-choice dynamics in vehicular traffic
\cite{tra1,tra2}. Experiments dealt with groups of people having to
choose at each time step (day) between two alternatives (routes),
having at their disposals a certain externally provided information
about the aggregate daily result, a sort of tunable traffic
bulletin. The payoff for each choice depends on the number of agents
making that choice in such a way that the larger this number the
smaller the payoff. Experiments have shown that while agents were able
to adapt rather well and reach states that were efficient on average,
the overreaction, namely the fluctuations or the difference between
the optimal rate of decision change by agents and the actual rate of
change, displayed a strong dependence on the type of information
supplied, for instance with or without impact correction,
time-dependent, user-dependent etc. In particular, the overall best
states (smallest overreaction) were attained when the information is
user-specific (see however \cite{tra1,tra2} for additional details and
more results).

\begin{figure}[t]
\centering
\includegraphics[width=5cm]{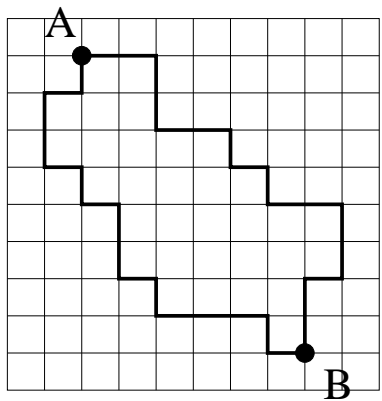}
\caption{Regular grid with two routes for traveling from A to B.}
\label{traf}
\end{figure}

The issue of how the information structure affects macroscopic
properties has been tackled in a traffic-inspired resource allocation
game which can be seen, roughly speaking, as a lattice version of the
previous `fish-market' model \cite{urb}. Let us consider the following
situation. A road network, which for simplicity is taken to be a
square lattice with $L^2$ sites, is given.  On each day, each one of
$N$ drivers has to travel from location A to location B (say,
work/home) following one of $S$ possible routes. The points $A$ and
$B$ are different for different drivers while the routes at their
disposal are taken to be $S$ quenched random self-avoiding walks of
length $\ell$ going from A to B (see Fig. \ref{traf}).
Routes play here the role of the predictors of the El Farol problem
and of the feasible bundles of the fish market: indexing lattice edges
by $\mu$, each route $g$ of every driver $i$ can be written as a
vector $\bsy{q}_{ig}=\{q_{ig}^\mu\}$, where $q_{ig}^\mu=1$ if driver
$i$ passes through edge $\mu$ in route $g$, and $q_{ig}^\mu=0$
otherwise. Drivers are assumed to be inductive and their behavior is
governed by the following rules:
\begin{eqnarray}
\prob\{g_i(t)=g\}=C(t)\exp\l[\Gamma U_{ig}(t)\r]\nonumber\\
Q^\mu(t)=\sum_i q_{i g_i(t)}^\mu\\
U_{ig}(t+1)-U_{ig}(t)=-\frac{1}{P}\sum_\mu q_{ig}^\mu Q^\mu(t)+\frac{1}{2}\l(1-\delta_{g,g_i(t)}\r)\zeta_{ig}(t)\nonumber
\end{eqnarray}
Let us discuss them in some detail. The first one says that agents
choose their preferred route on day $t$, $g_i(t)$, using a
probabilistic model with learning rate $\Gamma>0$. $Q^\mu(t)$ denotes
the traffic load on street $\mu$ on day $t$. The score updating
process is composed of two parts:
\begin{itemize}
\item the first term, $-\frac{1}{P}\sum_\mu q_{ig}^\mu Q^\mu(t)$, says
  that agents prefer less crowded routes;
\item the second term,
  $\frac{1}{2}\l(1-\delta_{g,g_i(t)}\r)\zeta_{ig}(t)$ is non-zero only
  for routes the driver has not taken on any given day and represents
  the information noise, or the inaccuracy with which he knows the
  traffic load on network edges he hasn't visited. $\zeta_{ig}(t)$ is
  a Gaussian noise with mean $\eta$ and correlations \be
  \avg{\zeta_{ig}(t)\zeta_{jh}(t')}=\Delta\delta_{ij}\delta_{gh}\delta_{tt'}
  \ee
\end{itemize}
Different information structures correspond to different values of
$\eta$ and $\Delta$:
\begin{itemize}
\item the case $\eta=\Delta=0$ (no information noise) corresponds to
  the case in which all drivers possess complete knowledge of the
  traffic load on each network edge on every day
\item for $\Delta>0$ the information about unvisited edges is
  user-specific and noisy. In particular
\begin{itemize}
\item for $\eta=0$ the noise is unbiased
\item for $\eta>0$ the driver overestimates the performance of routes
  not taken
\item for $\eta<0$ the driver underestimates the performance of routes
  not taken
\end{itemize}
\end{itemize}
Let us notice, en passant, that smart drivers should be aware of the
fact that the traffic load on a given route would have been larger had
they chosen it and therefore they should underestimate the efficiency
of unused routes. In other words, drivers account for their impact on
the traffic loads when they are able to disentangle their contribution
to it (their `impact') before updating their scores. In this model,
drivers completely account for their contribution to the traffic for
$\eta=-2$. We shall see however that any small $\eta<0$ is sufficient
to alter significantly the collective properties. We shall distinguish
between two types of drivers: `random drivers' with $\Gamma=0$, who
choose their route every day at random with equal probability, and
`optimizers' with $\Gamma=\infty$, who every day choose the route they
expect to be faster.

As usual, one is interested in the collective properties in the steady
state. We have several control parameters, namely $\eta$, $\Delta$,
$\Gamma$ and the vehicle density $c=N/P$. The observables we focus on
are
\begin{eqnarray}
H=\frac{1}{P}\sum_\mu\avg{Q^\mu-\ovl{\avg{Q}}}^2, ~~~~~~~~~~~
\ovl{\avg{Q}}=\frac{1}{P} \sum_\mu\avg{Q^\mu}\\
\sigma^2=\frac{1}{P}\sum_\mu\avg{\l(Q^\mu-\ovl{\avg{Q}}\r)^2}
\end{eqnarray}
where as usual $\avg{\cdots}$ stands for a time average over the
stationary state of the learning dynamics. Just as in the fish-market
model, $H$ describes the distribution of drivers over the street
network in the stationary state. If $H=0$, the distribution is uniform
($\avg{Q^\mu}=\ovl{\avg{Q}}$ for all $\mu$) and it is not possible to
find less crowded streets on the grid. If $H>0$, instead, the
distribution is not uniform and fast pathways do exist. Notice that if
transit times are assumed to be proportional to the street loads
$Q^\mu$, then $\sigma^2$ measures the total traveling time of
drivers. Then, the optimal road usage is achieved when $\sigma^2$ is
minimal. Note that since all routes have the same total length $\ell$
(i.e. $\sum_\mu q_{ig}^\mu=\ell$ for all $i$ and $g$),
$\ovl{\avg{Q}}=c\ell$ is a constant.

\begin{figure}[t]
\centering
\includegraphics*[width=7cm]{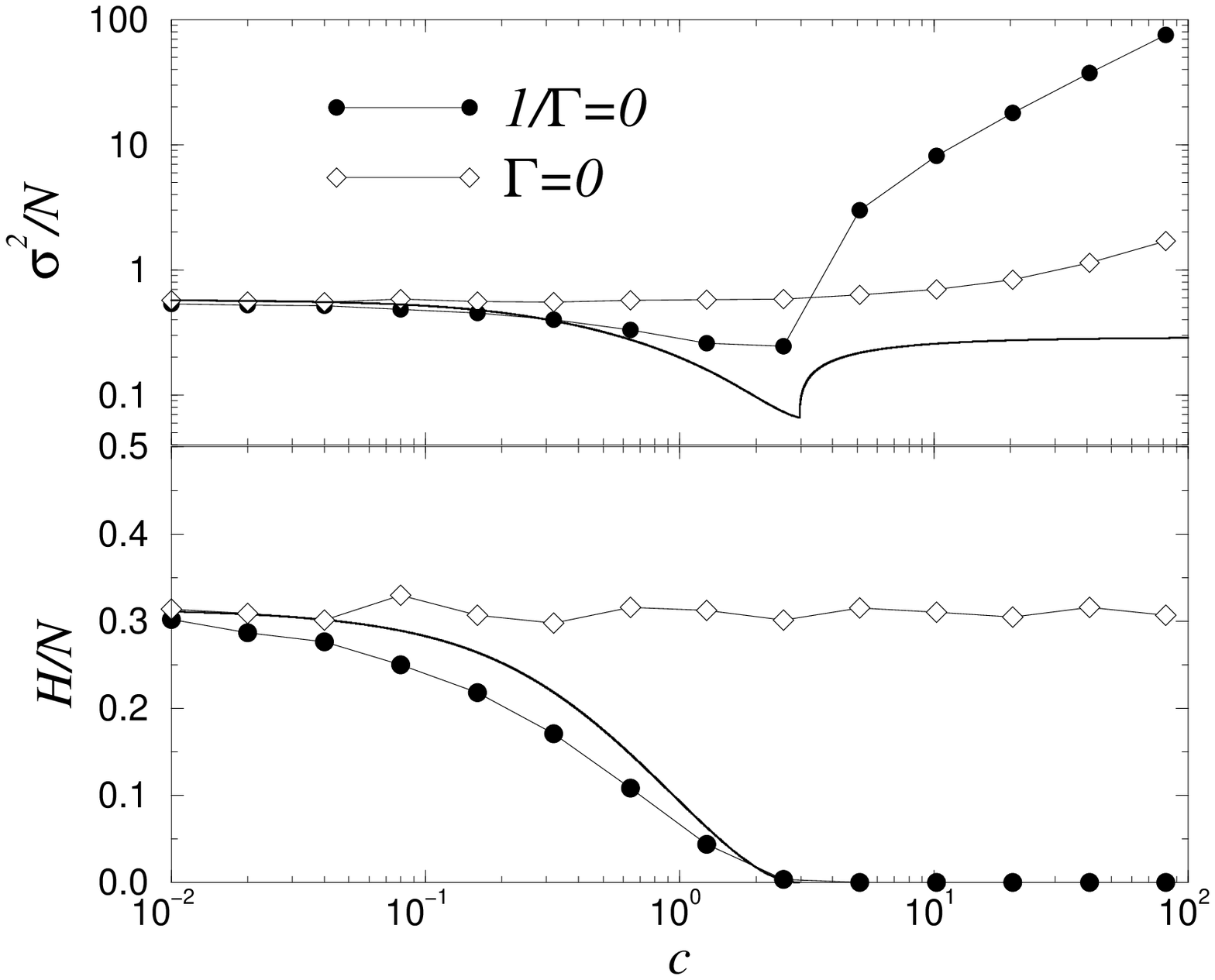}
\includegraphics*[width=7cm]{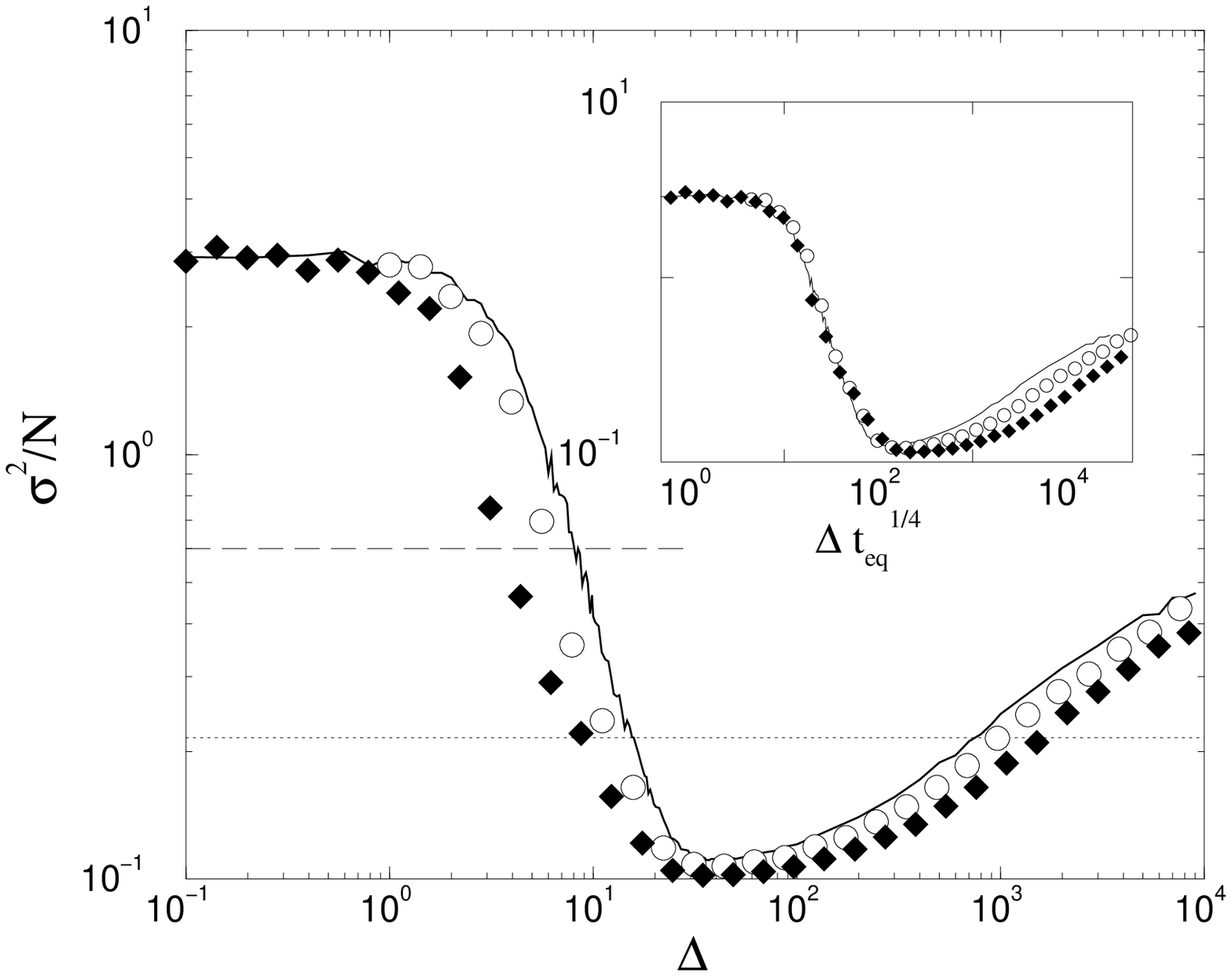}
\caption{Left panel: $\sigma^2/N$ (top) and $H/N$ (bottom) for random
drivers (open symbols) and optimizers (closed symbols). Simulation
parameters: $S=2$, $P=200$, $\ell=50$, $U_{ig}(0)=0$ for all $i$ and
$g$. Averages are taken over at least $50$ disorder samples for each
point. The solid line in the graphs is the analytic estimate of $H$
and $\sigma^2$ (for $\Gamma= 0^+$) for the model with uncorrelated
disorder. Right panel: behavior of $\sigma^2/N$ for a city of $P=200$
streets with $N=1024$ drivers ($c=5.12$, in the inefficient phase) as
a function of the parameter $\Delta$ with $\eta=0$. The horizontal
lines correspond to drivers with $\Gamma=0$ (dashed) and $\Gamma=0^+$
(dotted). Results for optimizers with $\Gamma=\infty$ are shown for
equilibration times $t_{\rm eq}=100$ (full line), $400$ (open circles)
and $1600$ (full diamonds). In the inset, data are plotted versus
$\Delta t_{\rm eq}^{1/4}$ (from \cite{urb}).}
\label{macrotraf}
\end{figure}

Numerical simulations for $\eta=\Delta=0$ reveal the picture displayed
in Fig. \ref{macrotraf}).
We see that random drivers lead to a stationary state where a uniform
distribution of vehicles is never achieved, as $H>0$ for all
$c$. Optimizers, instead, behave in a similar way only for small
vehicle densities.  As $c$ is increased, the traffic load becomes more
and more uniform ($H$ decreases) and fluctuations ($\sigma^2$)
decrease, indicating that inductive drivers manage to behave better
than random ones. At a critical point $c_c\simeq 3$ the distribution
becomes uniform (i.e.  $H=0$) and vehicles fill the available streets
uniformly. Now drivers can't find a convenient way and are forced to
change route very frequently. As a consequence, global fluctuations
increase dramatically. Notice that above the critical point traffic
fluctuations are significantly smaller for random drivers than for
optimizers. Finally, the stationary state depends on the initial
conditions $U_{ig}(0)$ for $c>c_c$: the larger the initial spread, the
smaller the value of $\sigma^2$. The conclusion is that random drivers
lead to an overall more efficient state in conditions of heavy traffic
while optimizers perform better when the car density is low.

Unfortunately, the analytical side of this model is much harder than
the previous examples because the quenched disorder (the feasible
routes) is in this case spatially correlated. It is possible however
to solve analytically a milder version with uncorrelated
disorder. Results (shown in the Fig. \ref{macrotraf}) reproduce the
qualitative behavior described above fairly well, and predict a
critical density of $c_c=2.97\ldots$.).

For $\eta=0$ and $\Delta>0$, the dependence on initial conditions
disappears and is replaced by a non-trivial dynamical behavior (see
again Fig. \ref{macrotraf}).  In the high density phase, where drivers
would behave worse than random with $\Delta=0$, global efficiency can
improve beyond the random threshold if $\Delta>0$. Taking averages
after a fixed equilibration time $t_{\rm eq}$, we find that $\sigma^2$
reaches, for $\Delta\approx 40$, a minimum that is well below the
value of $\sigma^2$ for $\Gamma=0^+$ with the same homogeneous initial
conditions $y_i(0)=0$. For $\Delta\to\infty$ we recover the behavior
of random drivers. However, when we increase $t_{\rm eq}$, the curve
shifts to the left, showing that the system is not in a steady state.
Rescaling $\Delta$ by $t_{\rm eq}^{-1/4}$, the decreasing part of the
plot collapses, while the the rest of the curve flattens. This
suggests that the equilibrium value of $\sigma^2$ drops suddenly as
soon as $\Delta>0$. Loosely speaking: noise-corrupted user-specific
information can avoid crowd effects when the vehicle density is very
high.

\begin{figure}[t]
\centering
\includegraphics*[angle=-90,width=8cm]{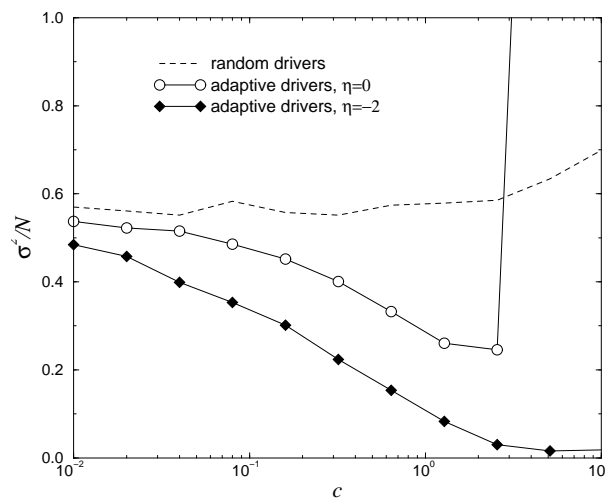}
\caption{ Behavior of $\sigma^2/N$ for a city of $P=200$ streets with
$\eta=-2$. Adaptive (resp. random) drivers have $\Gamma=\infty$
(resp. $\Gamma=0$) (from \cite{urb}).}
\label{trota}
\end{figure}

We finally come to the case $\eta\neq 0$ and $\Delta=0$,
Fig. \ref{trota}.
While one observes no qualitative changes for $\eta>0$, for $\eta<0$
fluctuations are drastically reduced in the supercritical phase. In
particular, for $\eta=-2$ (when, as we said above, drivers completely
account for their contribution to the traffic) the dynamics converges
to a state characterized by no traffic fluctuations ($\sigma^2=H$)
because each driver selects one route and sticks to it.

Hence this setup allows to address the impact of different information
structures, and thus of different types of information broadcasting,
on the collective properties of urban traffic. This is perhaps one of
the most promising research lines with respect to applications opened
by resource allocation games so far.


\section{Optimal properties of large random economies}

\subsection{Introduction}
 
The standard tenet of microeconomics is that economic activity is
aimed at the efficient allocation of scarce resources \cite{Masco}. As
we said before, `allocation' includes exchange, production and
consumption of commodities. The concept of `efficiency' is instead
usually connected to the solutions of constrained maximum and/or
minimum problems, as for instance firms strive to maximize profits at
minimum costs while the goal of consumers is to maximize their utility
subject to their budget constraints. The fundamental concept by which
mathematical economists explain the emergence of efficient states from
the disparate choices of individual agents in economic systems is that
of `equilibrium', that is a state where all agents maximize their
objective functions and the waste of resources -- in the form of
imbalance between demand and supply -- is minimum (actually,
zero). Typical results concern the existence and stability of
equilibria for different types of economies (see below for a precise
definition).  In such settings it is however extremely difficult to
extract meaningful macroscopic laws (comparable with empirical data)
from the mathematical results, in great part because of the
difficulties in handling agents' heterogeneity effectively. In what
follows, we will show that when heterogeneity is taken properly into
account the structure of equilibria of model economies (as well as of
other related optimization problems of microeconomics) proves to be
rich and non-trivial. We shall review the collective properties of a
few exemplary linear optimization problems of microeconomics, whose
setting will be borrowed from the economic literature
\cite{Lancaster}.  We will see that the emerging scenario presents in
all cases two distinct regimes: an expanding phase where technological
innovations lead to an overall economic growth, and a saturated regime
where growth is not achieved by technological innovation but rather by
a diversification of the production.  The key technical role in our
analysis is played by the replica method and the transitions between
expanding and contracting states can be completely characterized by a
few macroscopic order parameters. Remarkably, it will turn out that
the physical order parameters that arise bear an immediate economic
interpretation.

\subsection{Meeting demands at minimum costs}

To begin with, we consider the simple linear model of production to
meet demand satisfying an optimality criterion \cite{gale,spdsa}. This
illustrates the general two-phase phenomenology described above in an
extremely simplified setting. Let there be $N$ processes (or
technologies) labeled by $i$ and $P$ commodities labeled by
$\mu$. Each process allows the transformation of some commodities
(inputs) into others (outputs) and is characterized by an input-output
vector $\boldsymbol{\xi}_i=\{\xi_i^\mu\}$ where negative (positive)
components represent inputs (outputs). Each process can be operated at
any scale $s_i\geq 0$. The scales $s_i$ must be chosen so that the
total amount of commodity $\mu$ that is produced (consumed) matches a
fixed demand (availability): $\sum_i s_i\xi_i^\mu=\kappa^\mu$ for all
$\mu$, where the thresholds $\kappa^\mu$ may be positive (for goods
one wants to be produced) or negative (for goods to be
consumed). Among all feasible states $\{s_i\}$, one may select the one
which minimizes a particular function of the $s_i$. Here we take the
simplest choice of a linear combination $\sum_i s_i p_i$, which can be
thought of as the total operating cost, if $p_i$ is seen as the
operation cost at unit scale.

\begin{figure}[t]\begin{center}
\includegraphics*[width=10cm]{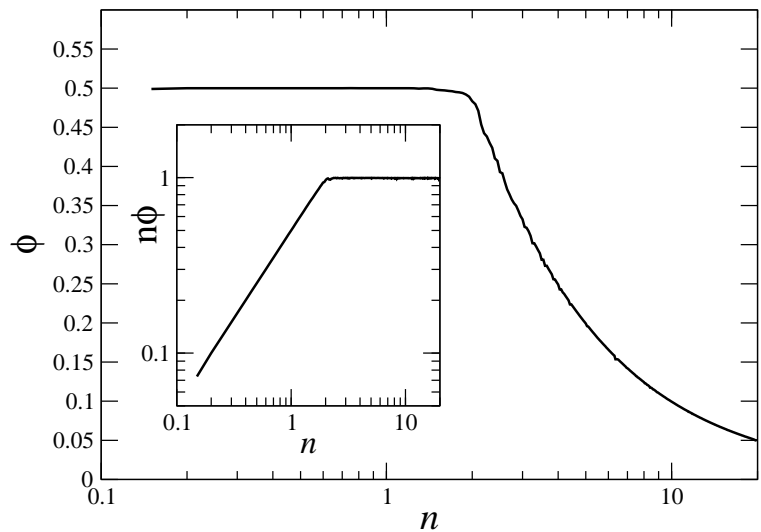}
\caption{\label{funo}Fraction $\phi$ of active processes vs $n$ for
$p_i=1$. Inset: $\phi n$ vs $n$ for the same parameter
values. $\xi_i^\mu$'s are taken to be Gaussian variables with variance
$1/P$ such that $\sum_\mu\xi_i^\mu=-0.001$ for all $i$. $\kappa^\mu$'s
are sampled from the bimodal distribution
$q(\kappa)=\frac{1-m}{2}\delta(\kappa+1)
+\frac{1+m}{2}\delta(\kappa-1)$ with $m=0.1$ (from \cite{spdsa}).}
\end{center} \end{figure}

We ask the following question: how does the operation pattern
(e.g. the fraction of active processes such that $s_i>0$) change when
$N$ increases, i.e. as more technologies become available? Indeed, the
macroscopic structure of the efficient state must be expected to
depend on the ratio $N/P$: for $N\ll P$ a technology will be more
likely to be active ($s_i>0$) than for $N\gg P$, when selection will
be stronger and processes performing the required conversions more
efficiently will be favored. This problem can be tackled by methods of
statistical mechanics in the limit $N\to\infty$ with $n=N/P$ finite
upon assuming that the $\xi_i^\mu$'s are quenched random variables
(similarly to what has been done for other linear optimization
problems such as the knapsack problem \cite{Knaps1,Knaps2,Nishi}). A
further important requirement is that $\sum_\mu\xi_i^\mu<0$ for all
$i$, which ensures that processes cannot be combined to yield a
technology with only outputs. We refer the reader to \cite{spdsa} for
details and focus on the emerging picture (see Fig. \ref{funo}).
 One sees that for small $n$ roughly a half
of the processes are active. This means that as $n$ increases, that is
as more and more technologies become available, the number of active
processes per good increases (see inset) i.e. the arrival of new
technologies favors existing ones. The picture changes radically for
$n\gtrsim 2$, as $\phi$ starts to decrease and $n\phi=1$. Now the
number of active processes equals that of commodities and technologies
undergo a much stronger selection which reduces the probability that a
randomly drawn input-output vector is active. This simple model
describes in a nutshell a transition to a highly competitive state
where all possible productions are saturated by existing technologies
and an increase in activity levels can be achieved only by increasing
$P$. We shall see below that a similar picture extends to the more
complicated case of general equilibrium.

\subsection{Competitive equilibria of linear economies}

An economy can be seen as a complex system of interacting agents
(consumers, firms, banks etc.) with conflicting goals and
complementarities. It is indeed the heterogeneity of the agents which
drives the economic process. Surely, if all agents were identical with
identical endowments, there would be no trade. Modeling an economy as
a system of heterogeneous agents is however a quite complex task
\cite{Kirman}. In this section we review how statistical mechanics may
be helpful in deriving the macroscopic properties of large random
economies. Specifically this approach allows one to derive statistical
laws that provide a picture of how structural properties are affected
by changes of macroscopic parameters. This is the same type of
information than random matrix theory provides about the structure of
heavy nuclei \cite{Wigner1,Wigner2}.

\subsubsection{Definition}

We stick to the standard microeconomic setup (see
e.g. \cite{Lancaster}). An {\it economy} is defined as a system of $N$
firms labeled by $i$, $P$ commodities labeled by $\mu$ and $L$
consumers labeled by $\ell$.  Each firm is endowed with a technology
that allows the transformation of some commodities, called `inputs',
into others, called `outputs'.  Every technology is completely
characterized by its `input-output vector'
$\bsy{\xi}_i=\{\xi_i^\mu\}$, where negative (respectively positive)
components represent quantities of inputs (respectively outputs), and
can be operated at any scale $s_i\geq 0$, meaning that when run at
scale $s_i$ it produces or consumes a quantity $s_i\xi_i^\mu$ of
commodity $\mu$. The price of commodities is given by the `price
vector' $\bsy{p}=\{p^\mu\geq 0\}$. Each consumer is characterized by
his/her initial endowment of commodities
$\bsy{y}_\ell=\{y_\ell^\mu\geq 0\}$ and by his/her utility function
$U_\ell$, associating to every bundle of goods $\bsy{x}=\{x^\mu\geq
0\}$ a real number $U_\ell(\bsy{x})$ representing his/her degree of
satisfaction.

It is assumed that firms choose their activity levels $s_i$ so as to
maximize their profits $\pi_i$ for a fixed price vector $\bsy{p}$:
\be\label{firms} \max_{s_i\geq 0}\pi_i~~~~~~~{\rm with~
}\pi_i=s_i(\bsy{p\cdot\xi}_i) \ee On the other hand, consumers choose
their consumptions $\bsy{x}_\ell$ so as to maximize their utilities
within their budget constraints for a fixed price vector $\bsy{p}$:
\be\label{consumers} \max_{\bsy{x}_\ell\in
  B_\ell}U_\ell(\bsy{x})~~~~~~~{\rm with ~} B_\ell=\{\bsy{x}\geq
0{\rm ~ s.t.~ }\bsy{p\cdot y}_\ell\geq\bsy{p\cdot x}\} \ee Equilibria
are states $(\{s_i^\star\},\{\bsy{x}_\ell^\star\},\bsy{p}^\star)$ for
which (i) the above problems (\ref{firms}) and (\ref{consumers}) are
simultaneously solved for all $i$ and $\ell$ and (ii) the aggregate
demand of each commodity matches the aggregate supply: \be\label{mc}
\sum_\ell \l(\bsy{x}_\ell^\star-\bsy{y}_\ell\r)=\sum_i
s_i^\star\bsy{\xi}_i \ee The `market clearing' condition (\ref{mc})
implies zero waste of resources and ultimately determines the optimal
price vector $\bsy{p}^\star$.

In order to connect the microscopic efficiency to macroscopic laws,
one would like to assess the typical values, relative fluctuations and
distributions of consumptions, operation scales and prices at
equilibrium in a large heterogeneous economy, that is, when agents
have different technologies, endowments etc. This problem can be
tackled in its most general form by applying techniques of spin-glass
physics. However, a rich qualitative description can be obtained
already at a less general level, obtained by introducing the following
assumptions \cite{geneq1,geneq2}:
\begin{enumerate}
\item[a.] {\it Consumers}: there is only one consumer (the `society')
  whose utility function is separable: $U(\bsy{x})=\sum_\mu u(x^\mu)$;
  the functions $u$ are such that $u'>0$ and $u''<0$
\item[b.] {\it Initial endowments}: the initial bundle $\bsy{y}$ is a
  quenched random vector whose components $y^\mu$ are sampled
  independently for each $\mu$ from a distribution $\rho(y)$
\item[c.] {\it Technologies}: the input-output vectors $\bsy{\xi}_i$
  have quenched random components $\xi_i^\mu$ that are identically
  distributed Gaussian random variables with zero mean and variance
  $\Delta_i/P$ satisfying $\sum_\mu\xi_i^\mu=-\epsilon_i$ with
  $\epsilon_i>0$; the quantities $\Delta_i$ are themselves quenched
  random numbers drawn from a distribution $g(\Delta)$ independently
  for each $i$ and $\epsilon_i=\eta\sqrt{\Delta_i}$
\end{enumerate}
Let us discuss them briefly. The assumption $L=1$ simplifies the
thermodynamic limit considerably (in the most general setting, the
latter corresponds to diverging $N$, $P$ and $L$). The separability of
$U$ implies that commodities are a priori equivalent. Hence the
society can increase its utility only by acquiring scarce commodities
(ones with low $y^\mu$) at the expense of abundant commodities (with
high $y^\mu$). Reaching non-trivial optimal states then requires that
(i) some commodities are initially more abundant than others (one can
see that no activity takes place in the case
$\rho(y)=\delta(y-\ovl{y})$) and (ii) the productive sector is able to
provide scarce goods using abundant goods as inputs. We will see that
this last point constitutes a strong selection criterion for
technologies. The convexity assumptions on $u$ follow the economic
literature and are convenient from an analytic viewpoint, as will
become clear later. Finally, the assumptions on technologies
guarantee, as in the previous case, that it is impossible to produce
all commodities without consuming any by simply constructing a
suitable combination of technologies (if this were possible, operation
scales and consumptions would diverge while prices would vanish, a
situation that is often described as the `Land of Cockaigne'). We
shall refer to the case $\eta\to 0$, which turns out to have special
physical properties, as the limit of `marginally efficient
technologies'.

We therefore have a deal of control parameters: $N$, $P$, $\eta$,
$g(\Delta)$, $u(x)$ and $\rho(y)$. In what follows we shall
concentrate mostly on the role of $\eta$ and of the relative number of
technologies $N/P$. In particular, we shall consider the
`thermodynamic limit' $N\to\infty$ with $n=N/P$ finite. 

\subsubsection{Statistical mechanics with a single consumer}

The problem of finding the equilibrium can easily be seen to be
equivalent to calculating \be\label{sopra} \max_{\{s_i\geq 0\}}
U\l(\bsy{y}+\sum_i s_i\bsy{\xi}_i\r) \ee In fact, first, if
\req{sopra} is solved then the society evidently maximizes its
utility. On the other hand, producers also maximize profits since
$\partial_{s_i}U=\sum_\mu
\xi_i^\mu\partial_{x^\mu}u=\lambda\partial_{s_i}\pi_i$, where the last
equality follows from the fact that, by virtue of the budget
constraint, $\partial_{x^\mu}U= \lambda p^\mu$ with $\lambda>0$ a
Lagrange multiplier. Thus prices disappear from the problem in
explicit form. However a remarkable outcome of the statistical
mechanics approach is that average prices and price fluctuations, like
other relevant macroscopic observables, turn out to be directly
connected to or easily derived from the spin-glass order parameters
that emerge from the calculation, as we shall see later on.

The statistical mechanics approach starts with the observation that if
$U$ is a sufficiently regular function one expects a self-averaging
condition to hold, i.e. \be \fl
\lim_{N\to\infty}\frac{1}{N}\max_{\{s_i\geq 0\}} U\l(\bsy{y}+\sum_i
s_i\bsy{\xi}_i\r)= \lim_{N\to\infty}\frac{1}{N}\davg{\max_{\{s_i\geq
0\}} U\l(\bsy{y}+\sum_i s_i\bsy{\xi}_i\r)} \ee where $\davg{\cdots}$
stands for an average over the quenched disorder $\{\bsy{\xi}_i\}$:
\be
\davg{\cdots}=\frac{\langle\cdots\prod_i\delta(\sum_\mu\xi_i^\mu+\epsilon
)\rangle_{\boldsymbol{\xi}}}{\langle\prod_i\delta(\sum_\mu
\xi_i^\mu+\epsilon)\rangle_{\boldsymbol{\xi}}}\ee Now the right-hand
side of the above expression can be evaluated by introducing the
`partition function' \be Z=\int_0^\infty d\boldsymbol{x}~ e^{\beta
U(\boldsymbol{x})} \int_0^\infty
d\boldsymbol{s}~\delta\l(\bsy{x}-\bsy{y}-\sum_i s_i \bsy{\xi}\r) \ee
and defining the `free energy' \be\label{effe}
f(\beta)=\lim_{N\to\infty}\frac{1}{\beta N}\davg{\log Z} .\ee As usual
\be \lim_{N\to\infty}\frac{1}{N}\davg{\max_{\{s_i\geq 0\}}
U\l(\bsy{y}+\sum_i s_i\bsy{\xi}_i\r)}=\lim_{\beta\to\infty} f(\beta)
\ee since in the limit $\beta\to\infty$ configurations that maximize
$U$ give the dominant contribution to the partition function. The
evaluation of $f$ ultimately leads to the identification of a function
$G$ of a vector $\bsy{\omega}$ of macroscopic order parameters such
that \be \lim_{\beta\to\infty}f(\beta)={\rm
extr}_{\bsy{\omega}}G(\bsy{\omega}) \ee where extr means that the
solution is provided by the saddle-point of $G$, that is by the vector
$\bsy{\omega}^\star$ solving $\partial_{\bsy{\omega}}G=0$. The
convexity assumptions made on $u$ ensure that the relevant saddle
point is of replica-symmetric form (as in \req{replicasym}).  Under this
condition, $\bsy{\omega}$ turns out to be a six-component vector
($\bsy{\omega}=\{Q,\gamma,\chi,\widehat{\chi},\kappa,\widehat{\kappa}\}$)
and $G$ takes the form
\begin{eqnarray}\fl \label{G}
G(\boldsymbol{\omega})= \frac{1}{2}Q\widehat{\chi}-\frac{\gamma\chi}{2n}+
\frac{1}{n}\kappa\widehat{\kappa}+\avg{\max_{s\geq 0}\l[
-\frac{1}{2}\Delta\widehat{\chi}s^2+st\sqrt{\Delta(\gamma-\widehat{\kappa}^2)}
-\eta\widehat{\kappa}s\sqrt{\Delta}\r]}_{t,\Delta}\nonumber\\ 
+\frac{1}{n}\avg{\max_{x\geq 0}\l[u(x)-\frac{1}{2\chi}\l(
x-y+t\sqrt{nQ}+\kappa\r)^2\r]}_{t,y}
\end{eqnarray}
where $\avg{\cdots}_x$ denotes an average over the random variable
$x$, $t$ is a unit Gaussian random variable and averages over $\Delta$
and $y$ are performed with distributions $g(\Delta)$ and $\rho(y)$.
Before discussing the economic interpretation of the order parameters
let us notice that $G$ is composed of two ``representative agent''
problems:
\begin{itemize}
\item an `effective profit' maximization by a representative firm,
  whose solution reads 
 \be\label{sstar} s^\star\equiv
  s^\star(t,\Delta)=
\cases{
    \frac{t\sigma-\eta\widehat{\kappa}}{\widehat{\chi}\sqrt{\Delta}}
    &for $t\geq \eta\widehat{\kappa}/\sigma$\\
    0&otherwise}
\ee
where we defined $\sigma=\sqrt{\gamma-\widehat{\kappa}^2}$
\item an `effective utility' maximization by the society with respect
  to the consumption of an effective commodity, whose solution, namely
  \be\label{xstar} x^\star\equiv x^\star(t,y)~~~~~~~{\rm such~ that
  }~~~ \chi u'(x^\star)=x^\star-y+t\sqrt{nQ}+\kappa \ee is always
  positive provided the assumptions on $u$ are satisfied
\end{itemize}
These two `effective' problems -- which have been derived and not
postulated a priori -- are interconnected by the remaining terms.

The saddle-point equations $\partial_{\bsy{\omega}}G=0$ for \req{G}
have the following form:
\begin{eqnarray}\
  Q=\avg{\Delta(s^\star)^2}_{t,\Delta}\\
  \chi=\frac{n}{\sigma}\avg{t s^\star\sqrt{\Delta}}_{t,\Delta}\\
  \kappa=\chi\widehat{\kappa}+n\eta\avg{s^\star\sqrt{\Delta}}_{t,\Delta}\\
  \widehat{\kappa}=\avg{u'(x^\star)}_{t,y}\\
  \sigma=\sqrt{\avg{u'(x^\star)^2}_{t,y}-\avg{u'(x^\star)}_{t,y}^2}\\
  \widehat{\chi}=\frac{\avg{t u'(x^\star)}_{t,y}}{\sqrt{nQ}}
\end{eqnarray}
One sees immediately that $\widehat{\kappa}$ represents the optimal
average (relative) price. In fact, utility maximization under budget
constraint gives $\partial_{x^\mu}U= \lambda p^\mu$, with $\lambda>0$
a Lagrange multiplier that can be set to $1$ without any loss of
generality. It then follows that $\sigma$ yields price fluctuations.
It is remarkable that the macroscopic order parameters introduced with
a purely `physical' method can be seen to possess such clear economic
meanings. It is also remarkable that the following laws can be
derived, with minimal manipulations, from the above set of equations:
\begin{eqnarray}
\avg{x^\star-y}_{t,y}=-n\eta\avg{s^\star\sqrt{\Delta}}_{t,\Delta}\\
\avg{u'(x^\star)\l(x^\star-y\r)}_{t,y}=0
\end{eqnarray}
The former expresses the fact that at the relevant saddle point the
market-clearing condition is satisfied (to compare, just average
\req{mc} for $L=1$ over $\mu$ taking the constraint on technologies
into account). The latter expresses the fact that at the relevant
saddle point the consumer saturates his/her budget when choosing his
consumption, a condition known in economics as Walras' law
\cite{Masco}.

It is possible to obtain a more precise characterization of the
macroscopic properties by calculating the distribution of operation
scales, consumptions and prices at equilibrium. These quantities are
given respectively by
\begin{eqnarray}
  P(s)=\avg{\delta(s-s^\star)}_{t,\Delta}=\int_0^\infty
  g(\Delta)P(s|\Delta)d\Delta\\
  P(x)=\avg{\delta(x-x^\star)}_{t,y}=\int_0^\infty \rho(y)P(x|y)dy\\
  P(p)=\avg{\delta(p-u'(x^\star))}_{t,y}
\end{eqnarray}
where $P(s|\Delta)$ and $P(x|y)$ denote respectively the probability
distributions of operation scales at fixed $\Delta$ and of
consumptions at fixed $y$. These can be calculated easily from
\req{sstar} and \req{xstar}. One finds
\begin{eqnarray}
  P(s|\Delta)=(1-\phi)\delta(s)+\frac{\widehat{\chi}}{\sqrt{2\pi\sigma^2}}
      e^{-\frac{(\widehat{\chi}
      s\sqrt{\Delta}+\eta\widehat{\kappa})^2}{2\sigma^2}}\theta(s)\\
      P(x|y)=\frac{1-\chi u''(x)}{\sqrt{2\pi n Q}}e^{-\frac{(x-y-\chi
      u'(x)+\kappa)^2}{2nQ}}
\end{eqnarray}
where
$\phi=\frac{1}{2}[1-\erf\frac{\eta\widehat{\kappa}}{\sigma\sqrt{2}}]$
is the fraction of active firms (i.e. firms such that
$s_i>0$). Moreover, notice that
$P(s|\Delta)=\sqrt{\Delta}P(s\sqrt{\Delta}|1)$, which implies that \be
P(s)=\frac{2}{s^3}\int_0^\infty k^2 g(k^2/s^2)P(k|1)dk \ee Thus,
power-law distributed operation scales are found for broad classes of
distributions $g(\Delta)$, as $P(s)\propto s^{-3-2\gamma}$ when
$g(\Delta)\simeq \Delta^\gamma$ for $\Delta\ll 1$. Recently, some
empirical evidence has been found that distributions of firm sizes
(defined by the number of employees, profits etc.) have scaling forms
\cite{Takayasu}.

\begin{figure}[t]
\centering
\includegraphics*[width=.45\textwidth]{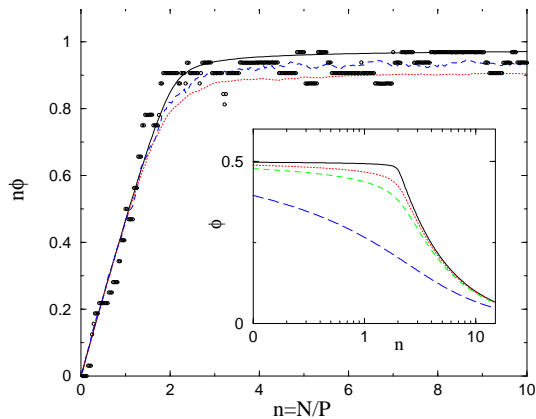}~
\caption{Behavior of $n\phi$ ($\phi=$ fraction of active companies)
  at equilibrium as a function of $n$ for $\eta=0.05$: analytical
  prediction (continuous line), computer experiments with $P=16$
  (dotted line) and for $P=32$ (dashed line) averaged over 100
  disorder samples. Dots represent results of a single realization of
  the technologies. Inset: $\phi$ vs $n$ for $\eta=0.01, 0.05, 0.1,
  0.5$ (top to bottom). From \cite{geneq1}.}
\label{figphi}
\end{figure}

\begin{figure}[t]
\centering
\includegraphics*[width=7cm]{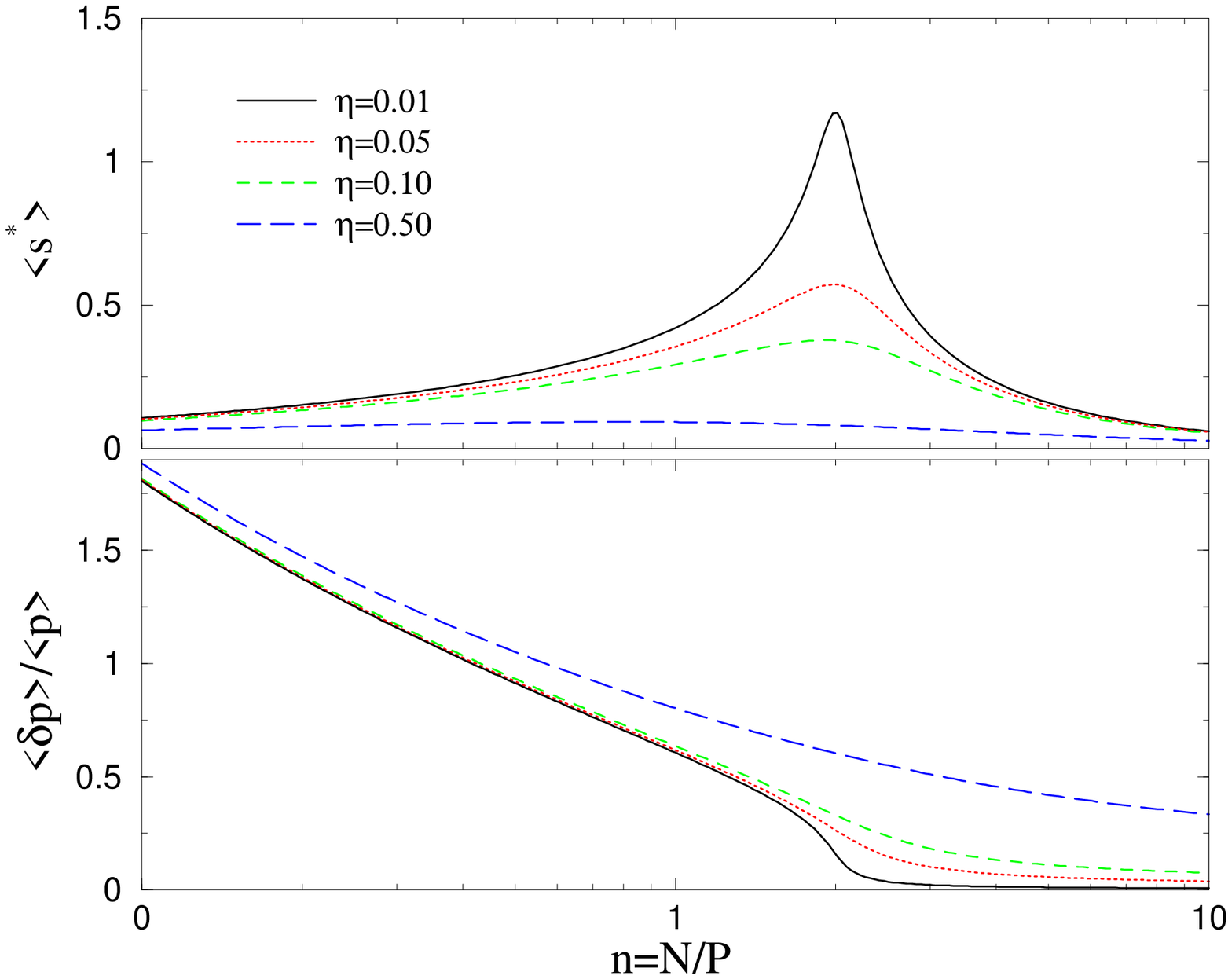}~
\includegraphics*[width=7cm]{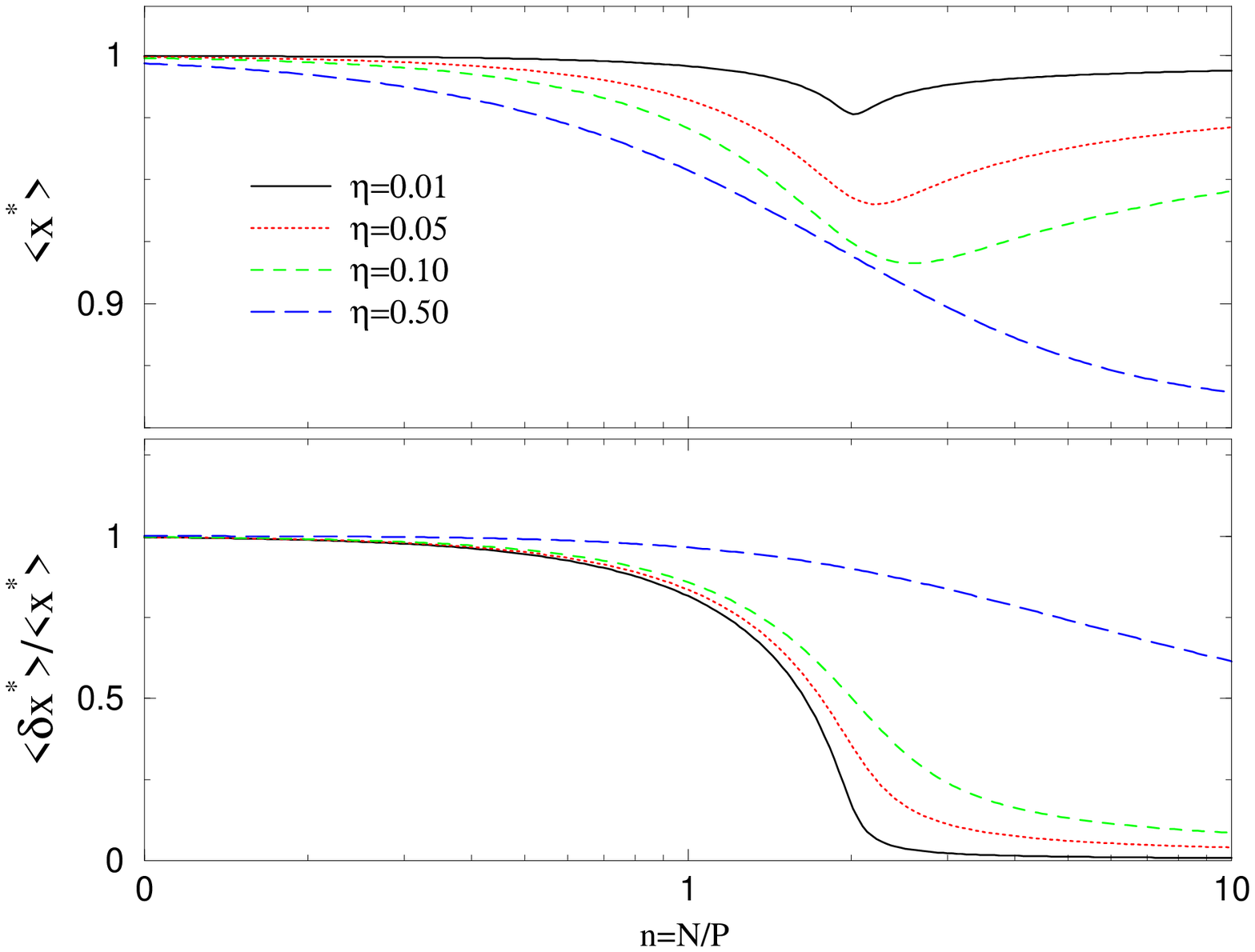}
\caption{Typical macroscopic properties of competitive equilibria for
$g(\Delta)=\delta(\Delta-1)$, $u(x)=\log x$ and $\rho(y)=e^{-y}$. Left
panels: typical operation scale (top) and relative price fluctuations
at equilibrium for different values of $\eta$. Right panels: typical
consumption and relative consumption fluctuations for different values
of $\eta$ (from \cite{geneq1}).}
\label{figgeneq}
\end{figure}

Numerical solution of the saddle-point equations for a generic choice
of the parameters yields the picture illustrated in Figures
\ref{figphi} and \ref{figgeneq}.
The quantity $\phi$ is shown in Fig. \ref{figphi} against numerical
simulations.  One sees that there are two regimes: one where
$\phi\simeq 1/2$ for small $n$, and a second one for large $n$ where
$n\phi\simeq 1$, so that the number of active firms equals that of
commodities, signaling a saturated market.
Fig.\ref{figgeneq}, instead, shows that the average scale of
production increases when $n$ grows as long as $n$ is sufficiently
small. This means that the introduction of new technologies (i.e. from
an increase of $n$) leads to an increased production activity of
existing firms if the number of competitors is low. In parallel,
relative price fluctuations decrease, as does the average level of
consumption, signaling that firms are managing to transform abundant
goods into scarce ones. When $n$ is close to $2$, operation scales
become larger and larger as $\eta$ decreases (i.e. as technologies
become more and more efficient) and ultimately develop a singularity
at $n_c$ in the marginally efficient limit $\eta\to 0$ (see below for
more details about this limit). The fluctuations of relative
consumptions start to drop (the sharper the lower is $\eta$), as the
distribution of consumptions becomes more and more peaked around the
mean value. Identifying abundant (or scarce) goods becomes
increasingly hard. In high $n$ regime, the introduction of new
technologies, by e.g. technological innovation ($N\to N+1$), leads to
a decrease in the average operation scale, i.e.  new profitable
technologies punish existing ones. The economy becomes strongly
selective as firms cannot take advantage of the spread between scarce
and abundant goods any longer. On the other hand, the average
consumption starts growing with $n$, as is expected in a competitive
economy that selects highly efficient technologies. In this phase the
introduction of new commodities (an increase in $P$) leads to an
increase in the scale of operations.

The above results confirm rather clearly that the collective
properties of competitive equilibria display a marked qualitative
change when $n$ increases, as one passes from an expanding to a
saturated regime around $n\simeq 2$. Such a change is a smooth
crossover for any finite $\eta>0$. However, in the limit $\eta\to 0$
in which technologies are `marginally efficient' the crossover becomes
a sharp second-order phase transition characterized by the fact that
$\phi=1/2$ for $n<n_c$ and $\phi<1/2$ otherwise, whereas \be
\avg{s^\star}\simeq |n-n_c|^{-1/2} ~~~~~~~~~~~ |n-n_c|\ll 1\ee (see
\cite{geneq1} for analytical details).  This can be explained
intuitively by a simple geometric argument.  Let us write the initial
endowments as $y^\mu=\ovl{y}+\delta y^\mu$, separating a constant part
($\ovl{y}$) from a fluctuating part ($\delta y^\mu$) such that
$\sum_\mu\delta y^\mu=0$. Now market clearing with $\eta=0$ implies
that $\bsy{\xi}_i\cdot\bsy{y}=\bsy{\xi}_i\cdot\bsy{\delta y}$, so that
all the transformations take place in the space orthogonal to the
constant vector. This means that those technologies with
$\bsy{\xi}_i\cdot\bsy{\delta y}<0$ which reduce the initial spread of
endowments $\boldsymbol{\delta x}_0$ lead to a increase in wealth and
hence will be run at a positive scale. Those with a positive component
along $\boldsymbol{\delta x}_0$ will have $s_i=0$. Given that the
probability to generate randomly a vector in the half-space
$\bsy{\xi}_i\cdot\bsy{\delta y}<0$ is $1/2$, when $N$ is large we
expect $N/2$ active firms. Still the number of possible active firms
is bounded above by $P$, hence when $n=N/P=2$ the space of
technologies becomes complete and $x^\mu=\ovl{y}$ for all $\mu$. There
is no possibility to increase welfare further.

\subsubsection{Case of many consumers}

In the model just described, there are $N$ firms running linear
activities $\xi_i^\mu$, which are vectors in a $P$-dimensional
commodity space, at a scale $s_i\geq 0$.  These firms face a demand
function $Q^\mu(p)$ from consumers, which is the quantity that
consumers will buy at prices $p^\mu$. The profit of firm $i$ is given
by $\pi_i=s_i\sum_{\mu=1}^P p^\mu q_i^\mu$.

Let us consider a more general case. Let us assume there are $L$
consumers, each with an initial endowment $y_\ell^\mu$ of commodity
$\mu$ and each taking a share $\theta_{i\ell}$ in the profit of firm
$i$. We assume that consumers face fixed prices $p^\mu$. So the
initial wealth of consumer $\ell=1,\ldots,L$ is \be
w_\ell=\sum_{\mu=1}^P p^\mu y_\ell^\mu+\sum_{i=1}^N
\theta_{i\ell}\pi_i \ee If consumers are identical, apart from the
initial endowments, and aim at maximizing a utility function
$U(\bsy{x})=\sum_{\mu=1}^P \log x^\mu$ as before, the solution
is relatively straightforward: the problem of consumer $\ell$ is
solved by \be x_\ell^\mu=\frac{w_\ell}{Pp^\mu} \ee (i.e. each consumer
distributes his wealth uniformly over commodities, taking prices into
account). Now the total demand function will be
\begin{equation}\label{demand}
  Q^\mu=\sum_{\ell=1}^L x_\ell^\mu=\frac{W}{P}\frac{1}{p^\mu},~~~~~~~W=\sum_{\ell=1}^L w_\ell
\end{equation}
In a pure exchange economy (without production: $s_i=0~\forall i$) the
above quantity will equal to total initial endowment of each
commodity, i.e. \be Q^\mu=y^\mu\equiv\sum_{\ell=1}^L y^\mu_\ell \ee If
$y_\ell^\mu$ are drawn independently at random with mean $\ovl{y}$ and
variance $D$, then $y^\mu$ will have mean $L\ovl{y}$ and variance $LD$
and the relative fluctuations of the total initial endowments will be
$\delta y/\ovl{y}=\sqrt{D}/(\sqrt{L}\bar y)$, which decreases as $L$
increases. When we allow firms to operate ($s_i>0$), relative
fluctuations in the demand must be expected to be of the same order
\be \frac{Q^\mu-\ovl{Q}}{\ovl{Q}}\sim
\frac{1}{\sqrt{L}},~~~~~~~\ovl{Q}=\frac{1}{P}\sum_{\mu=1}^P Q^\mu.
\ee Therefore, by equation (\ref{demand}), relative price fluctuations
will also be of the order $1/\sqrt{L}$. This simple argument explains
how the different macroscopic quantities re-scale in the presence of
$L$ consumers when $L\to\infty$. We remark that Ref. \cite{geneq1}
shows that the scales of production have a non-trivial behavior in the
limit of extremely uniform initial endowments, which suggest an
essential singularity $\avg{s}\sim \exp(-c/\sqrt{L})$ as $L\to\infty$.

The case of consumers with different utility functions requires a more
involved approach, because the heterogeneity of consumer utility is
likely to imply a non-symmetric demand function (even when prices
$p^\mu$ are all equal). Apart from this, it is reasonable to expect
that the basic insights gained from the above analysis, such as the
presence of a cross-over between two structurally different phases of
the economy, will remain valid.

\subsection{Economic growth: the Von Neumann problem}

Von Neumann's expanding model addresses the issue of computing the
maximum achievable growth rate of a linear production economy
\cite{john}. Economic growth is seen basically as an autocatalytic
chemical process in which technologies play the role of reactions and
commodities of reactants. In spite of its extremely simple setup, the
model has played a key role in the mathematical theory of economic
growth, particularly in view of its connection to dynamical growth via
the so-called turnpike theorems \cite{turnpike}.

The time-dependent model is defined as follows. One considers an
economy with $P$ commodities (labeled $\mu$) and $N$ linear
technologies (labeled $i$), each of which can be operated at a
non-negative scale $S_i\geq 0$ and is characterized by an output
vector $\boldsymbol{a}_i=\{a_i^\mu\}$ and by an input vector
$\boldsymbol{b}_i=\{b_i^\mu\}$, such that $S_i a_i^\mu$ (respectively
$S_i b_i^\mu$) denotes the units of commodity $\mu$ produced
(respectively used) by process $i$ when run at scale $S_i$.  It is
assumed that input/output vectors are fixed in time and that operation
scales are the degrees of freedom to be set, for instance, by
firms. At time $t$, the economy is characterized by an aggregate input
vector $\bsy{I}(t)=\sum_i S_i(t)\bsy{b}_i$ and output vector
$\bsy{O}(t)=\sum_i S_i(t)\bsy{a}_i$. Part of the latter will be used
as the input at period $t+1$ whereas the rest, namely
\begin{equation}\label{cmu}
\bsy{C}(t)\equiv \bsy{O}(t)-\bsy{I}(t+1)
\end{equation}
is consumed. In absence of external sources, in order to ensure
stability it is reasonable to require that inputs at any time do not
exceed the outputs at the previous time, i.e. one must have
$C^\mu(t)\geq 0$ for all $\mu$ at all times. Let us focus on solutions
in which input vectors grow in time at a constant rate, i.e.  of the
form $\bsy{I}(t+1)=\rho \bsy{I}(t)$ with $\rho>0$ a constant (the
growth rate). For these solution, the scales of production have the
form $S_i(t)=s_i \rho^t$, and likewise
$\bsy{C}(t)=\bsy{c}\rho^t$. Therefore the stability condition can be
re-cast in the form
\begin{equation}\label{1}
c^\mu\equiv \sum_i s_i\l(a_i^\mu-\rho b_i^\mu\r)\geq 0~~~~~~~\forall\mu
\end{equation}
The (technological) expansion problem amounts to calculating the
maximum $\rho>0$ such that a configuration $\boldsymbol{s}=\{s_i\geq
0\}$ satisfying the above condition exists (it is easy to show that
such an optimal growth rate exists \cite{gale}).  In such a
configuration the aggregate output of each commodity is at least
$\rho$ times its aggregate input. If the maximum $\rho$, which we
denote by $\rho^\star$, is larger than 1 the economy is `expanding',
whereas it is `contracting' for $\rho^\star<1$. On the other hand, the
actual value of $\rho^\star$ is expected to depend on the input and
output matrices. Intuitively, $\rho^\star$ should increase with the
number $N$ of technologies and decrease when the economy is required
to produce a larger number $P$ of goods.

In \cite{VN} this problem was tacked in the limit $N\to\infty$ with
$n=N/P$ finite under the assumption that $(a_i^\mu,b_i^\mu)$ are
independent and identically distributed quenched random variables for
each $i$ and $\mu$, with the aim of uncovering the emerging collective
properties that are typical of large random realizations of a complex
wiring of input-output relationship. To begin with, let us write
$a_i^\mu=\ovl{a}(1+\alpha_i^\mu)$ and
$b_i^\mu=\ovl{b}(1+\beta_i^\mu)$, where $\ovl{a}$ and $\ovl{b}$ are
positive constants while $\alpha_i^\mu,~\beta_i^\mu$ are zero-average
quenched random variables. Inserting these into (\ref{1}) one easily
sees that, to leading order in $N$, the optimal growth rate
$\rho^\star$ is given by the ratio $\ovl{a}/\ovl{b}$ of the average
output and average input coefficients, hence it is independent of the
specific input-output network. The non trivial aspects of the problem
are related to the corrections to the leading part. We therefore write
the growth rate as
\begin{equation}
\rho=\frac{\ovl{a}}{\ovl{b}}\left(1+\frac{g}{\sqrt{N}}\right)\label{rho}
\end{equation}
so that, assuming $\ovl{a}=\ovl{b}$ for simplicity, (\ref{1}) becomes
\begin{equation}
\frac{c^\mu}{\bar{a}}=\sum_i s_i\left[\alpha_i^\mu-\frac{g}{\sqrt{N}}-
\left(1+\frac{g}{\sqrt{N}}\right)\beta_i^\mu\right]\geq 0~~~~~~~\forall\mu
\label{cimu}
\end{equation}
The problem thus reduces to that of finding the largest value
$g^\star$ of $g$ for which it is possible to find coefficients
$\{s_i\geq 0\}$ satisfying (\ref{cimu}). In the limit $N\to\infty$ one
may resort to a Gardner-type calculus \cite{Gardner}. Defining the
characteristic function \be \chi(\bsy{s})=\prod_\mu \theta \left[
\frac{1}{\sqrt{N}}\sum_i s_i\left[ \alpha_i^\mu-\frac{g}{\sqrt{N}}-
\left(1+\frac{g}{\sqrt{N}}\right) \beta_i^\mu \right]\r] \ee one can
write the typical volume of configuration space occupied by
micro-states satisfying (\ref{1}) for $N\to\infty$ at fixed $g$ is
given by
\begin{equation}\label{Vtyp}
 V_{{\rm typ}}(g)=\lim_{N\to\infty}\frac{1}{N} \davg{\log V(g)}
\end{equation}
where $V(g)$ is the volume of solutions at fixed disorder:
\begin{eqnarray}
V(g)= \int_0^\infty \chi(\bsy{s}) \delta\left(\sum_is_i
-N\right)d\boldsymbol{s}
\end{eqnarray}
(without affecting the optimal growth rate, we introduced a linear
constraint $\sum_i s_i=N$). It is reasonable to expect that, when $g$
increases, $V_{{\rm typ}}(g)$ shrinks, and in particular that $V_{{\rm
typ}}(g)\to 0$ for $g\to g^\star$. Now after carrying out the disorder
average (see \cite{VN} for details), which only depends on \be
k=\davg{\l(\beta_i^\mu-\alpha_i^\mu\r)^2} \ee the key macroscopic
order parameters turns out to be the overlap $q_{\ell\ell'}=(1/N)
\sum_i s_{i\ell}s_{i\ell'}$ between different optimal configurations
$\ell$ and $\ell'$. Because the space of solutions $\{s_i\}$ is a
convex set (by construction), the replica-symmetric approximation, for
which $ q_{\ell\ell'}=q+\chi\delta_{\ell\ell'}$ is in this case
exact. Note that $\chi$, which describes the fluctuations of $s_i$
among feasible solutions, should also vanish as $g\to g^\star$, hence
the conditions $g=g^\star$ and $\chi=0$ are equivalent and the
analysis of optimal states coincides with the study of the $\chi\to 0$
limit of the replica-symmetric solution.

\begin{figure}[t]
\centering
\includegraphics*[width=8cm]{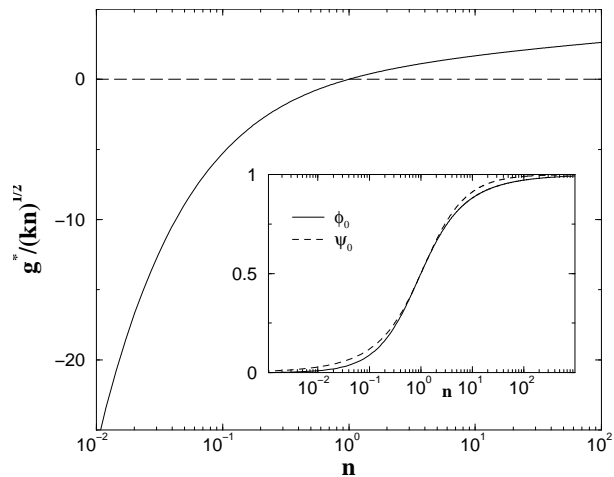}
\caption{Behavior of $g^\star/\sqrt{kn}$ vs $n$.  Inset: $\phi_0$ and
$\psi_0$ (related by \req{forr}) vs $n$ (from \cite{VN}).}
\label{gsta}
\end{figure}

Results for the re-scaled quantity $g^\star/\sqrt{nk}$ are shown in
Fig. \ref{gsta}.
The line separates the region of feasible solutions with $g\leq
g^\star$ from the region of unfeasible solutions.  $g^\star$ crosses
the line $g=0$ (i.e. passes from a regime with growth rate
$\rho<\ovl{a}/\ovl{b}$ to one with growth rate $\rho>\ovl{a}/\ovl{b}$)
at $n_c=1$. In the inset we show the fraction of inactive processes
$\psi_0$ (i.e. such that $s_i=0$) and that of intermediate commodities
$\phi_0$ (i.e. such that $c^\mu=0$) at $g=g^\star$, as a function of
$n$. These are found to be universal functions of $n$ independent of
the details of the disorder distribution, related by \be
\phi_0=n(1-\psi_0) \label{forr} \ee Both $\phi_0$ and $\psi_0$ tend to
one when $n$ increases, meaning that the `expanding phase' at $n>n_c$
is highly selective. The condition \req{forr} has a simple geometrical
interpretation: it implies that the number of active processes equals
that of intermediate commodities at $g^\star$. Noting that for any
$\mu$ such that $c^\mu=0$ we have a linear equation for the scales
$s_i>0$, we see also that (\ref{forr}) simply corresponds to the
requirement that the number of equations should match the number of
variables.

Based on these results one can speculate on how long term growth will
be affected by technological innovation. The latter, defined as the
introduction of new processes, i.e. new feasible ways of combining
inputs to produce desirable outputs \cite{Romer} would just correspond
to an increase in the number $N$ of transformation processes which the
economy has at its disposal. Now the change in the growth rate is
related to the change in $g^\star/\sqrt{n}$, which is given by \be
\delta\rho\simeq
-\frac{\ovl{a}}{\ovl{b}}\frac{g}{n^{3/2}\sqrt{P}}\delta n, \ee
Therefore an increase in $N$ can have a large positive impact on long
term growth when $n$ is small. For technologically mature economies
($n\gg 1$) instead, $g^\star/\sqrt{n}$ increases much more slowly,
hence technological innovation has much smaller effect on long term
growth.


\section{Toy models of financial markets: Minority Games}

\subsection{Introduction}

The Minority Game (MG for short) \cite{CZ} is a strict relative of the
El Farol problem (it corresponds roughly to the case $L=N/2$) that has
been proposed to model speculative trading in financial markets, that
is systems where agents buy and sell asset shares with the only goal
of profiting from price fluctuations. The basic idea is that when most
traders are buying it is profitable to sell and vice-versa, so that it
is always convenient to be in the minority group. Abstracting, one
considers the following situation. We have $N$ agents, each of which
has to formulate at every time step $t$ a binary bid
$b_i(t)\in\{-1,1\}$ (buy/sell). The payoff received at time $t$ by
each agent depends both on his/her action and on the aggregate action
$A(t)=\sum_i b_i(t)$ (the `excess demand') and it is given by
$\pi_i(t)=-b_i(t)A(t)$. Thus, agents in the minority group win. The
minimal measures of efficiency to be employed are the average excess
demand and fluctuations in the steady state: 
\be
\avg{A}=\lim_{T,T_{{\rm eq}}\to\infty}\frac{1}{T-T_{{\rm
      eq}}}\sum_{t=T_{{\rm eq}}}^T A(t)~~~~~~~{\rm
  and}~~~~~~~\sigma^2=\avg{A^2} 
\ee 
where $T_{{\rm eq}}$ is an
equilibration time. An efficient state is one where $\avg{A}=0$ and
$\sigma^2$ is small. Notice that the number of people which could have
been accommodated in the minority is $|A|/2$, hence $\sigma$ is a
measure of the waste of resource. What remains to be specified is how
agents make their decisions. Agents who buy or sell at random with
equal probability at every time step lead to a state where $\avg{A}=0$
and $\sigma^2=N$.  Of course, it is the way in which agents take their
decisions (which needs to be specified) and their interactions that
gives rise to the complex collective behavior.

The MG is a useful toy model that allows to elucidate the collective
behavior of systems of heterogeneous interacting agents by addressing
directly the interplay between microscopic behavior and macroscopic
properties (fluctuations, predictability, efficiency, etc.). From a
purely theoretical viewpoint, the detailed study of the emergence of
cooperation in competitive systems makes the Minority Game a benchmark
model of interacting agents. It has however also turned out to be able
to reproduce, to some extent, the rich statistical phenomenology of
financial markets, that are well known (and at least since
\cite{Fama}) to be characterized by clear statistical regularities,
often referred to as ``stylized facts''\footnote{An ever increasing
number of such facts are documented in the literature. The best known
of these are the following: (a) asset returns are approximately
uncorrelated beyond a time scale or the order of tens of minutes; (b)
the unconditional distribution of returns displays a power-law tail
with an exponent ranging from $2$ to $4$ for different stocks and
markets; (c) the distribution of returns over a time scale $\tau$
becomes more and more Gaussian as $\tau$ increases; (d) volatility is
positively autocorrelated over time scales as long as several days,
implying that periods of high volatility cluster in time (`volatility
clustering'). See \cite{Stylized} for details.}.

There are at present a few comprehensive books that cover many aspects
of the MG, from both the theoretical viewpoint and the financial
market viewpoint \cite{mgbook,coolen}. Here we shall consider some
basic and extended aspects of the model that are only marginally
treated elsewhere. In this section we shall concentrate mainly on the
original model, first presenting a more thorough derivation of the
minority rule, then a simple version of the MG and finally discussing
the standard model. The next section is instead devoted to some
extensions that have a particularly interesting physical content.

\subsection{From agents' expectations to the minority (and majority) rule}

The connection between MGs and financial markets can be established
na\"\i vely by observing that markets are instruments for allocating
goods. This, combined with the no arbitrage hypothesis according to
which no purchase or sale by itself may result in a risk-less profit,
suggests that markets should in principle be zero-sum
games. Transaction costs make it a game that is unfavorable on
average, i.e. a Minority Game. It would however be important to
understand whether the minority mechanism can be derived from a
particular microscopic scheme. This is indeed possible \cite{Mats}.

Let us imagine a market in which $N$ agents submit their orders
$a_i(t)$ for a certain asset simultaneously at every time step
$t=1,2,\ldots$. Let $a_i(t)>0$ mean that agent $i$ contributes
$a_i(t)$\euro~ to the demand for the asset while $a_i(t)<0$ means that
$i$ sells $-a_i(t)/p(t-1)$ units of asset, which is the current
equivalent (i.e. at price $p(t-1)$) of $|a_i(t)|$\euro.  With
$a_i(t)=\pm 1$ and $A(t)=\sum_i a_i(t)$, the demand is given by
$D(t)=\frac{N+A(t)}{2}$, whereas the supply is
$S(t)=\frac{N-A(t)}{2p(t-1)}$. Finally, assume that the price is fixed
by the market clearing condition, $p(t)=D(t)/S(t)$, i.e.  \be
p(t)=p(t-1)\frac{N+A(t)}{N-A(t)}.
\label{MCCmg}
\ee Taking the logarithm of both sides and expanding to the leading
order one gets \be \log p(t)-\log p(t-1)\simeq \frac{A(t)}{\lambda}
\ee with $\lambda=N$. The quantity on the left-hand side is normally
called the `return' of the asset. $A(t)$ is instead the excess demand,
namely the difference between demand and supply. This equation
expresses the dynamics of prices in terms of an aggregate quantity
$A(t)$ that all agents contribute to form \cite{Farmer1}. $A(t)$ may
thus be considered a proxy for the return.

Now take agent $i$ and assume he must decide whether to buy or sell at
time $t$. To do this, he should compare the expected profit (or
utility) of the two actions, which depends on what the price will be
at time $t+1$. For instance the utility he would face at time $t+1$ if
he buys $1$\euro~ of asset at time $t$ (i.e. $a_i(t)=1$) is given by
\be 
u_i(t)=\frac{p(t+1)}{p(t)}-1
\label{uimg}
\ee 
($u_i(t)>0$ if $p(t+1)>p(t)$). At this stage the price $p(t+1)$ is
unknown to him (and presumably to everybody else).  Therefore if our
agent $i$ wants to use Eq. \req{uimg} to make his choice at time $t$,
he has to replace $p(t+1)$ by the expectation he has at time $t$ of
what the price will be at time $t+1$, denoted by
$\mathbb{E}_t^{(i)}[p(t+1)]$. Let us assume that that \cite{Mats} \be
\mathbb{E}_t^{(i)}[p(t+1)]=(1-\psi_i) p(t)+\psi_i p(t-1)
\label{Expp}
\ee The parameter $\psi_i$ allows to distinguish two types of traders,
depending on whether $\psi_i$ is positive or negative. Agents with
$\psi_i>0$ believe that market prices fluctuate around a fixed value
(the `fundamental'), so that the future price is an average of past
prices. For this reason these agents are called `fundamentalists'. 
They may also be called {\em contrarians} since they believe
that the future price increment $\Delta p(t+1)=p(t+1)-p(t)$ is
negatively correlated with the last one 
\be 
E^{(i)}_t[\Delta p(t+1)]
=-\psi_i\Delta p(t).  
\ee  
On the
other hand, if $\psi_i<0$ the agent believes that the future price
increment will occur in the direction of the trend defined by the last
two prices, so that future price increments $\Delta p(t+1)$ are
positively correlated with the past ones, as if the price were
following a monotonic trend. This type of agents are called `trend
followers'.

The expected utility for buying at time $t$ will be
$\mathbb{E}_t^{(i)}[u_i(t)|a_i(t)=+1]=-\psi_i[p(t)-p(t-1)]/p(t)$
which, using \req{MCCmg}, becomes \be
\mathbb{E}_t^{(i)}[u_i(t)|a_i(t)=+1]=-2\psi_i A(t)/[N+A(t)] \ee A
similar calculation can be carried out for the expected utility for
selling at time $t$. The net result is that the expected utility for
action $a_i(t)$ at time $t$ can be written as \be
\mathbb{E}_t^{(i)}[u_i(t)]=-2\psi_i a_i(t)\frac{A(t)}{N+a_i(t)A(t)}.
\label{Etiui}
\ee Notice that agents who took the majority action
$a_i(t)=\sign[A(t)]$ expect to receive a payoff
$-2\psi_i|A(t)|/[N+|A(t)|]$ whereas agents in the minority group
expect to get $2\psi_i|A(t)|/[N-|A(t)|]$. It is clear that the
expected payoff of fundamentalists (resp. trend-followers) is positive
when they are in the minority (resp. majority) group. Therefore
Minority Games are simple schemes for describing the behavior of
contrarians whereas Majority Games are appropriate for
trend-followers.

In real markets, both groups are present and the resulting price
dynamics stems from a competition between the two groups
\cite{Luxm}. Which group dominates and shapes the price dynamics
depends on the evolution of traders' expectations, which in turn
depends on the behavior of price itself. Common sense suggests that
when everybody is going to buy the price will rise and it will be
convenient to buy. Accordingly, speculative markets in certain regimes
(e.g. bubbles) should look more like Majority Games rather than
Minority Games (and vice-versa in other regimes). If all traders base
themselves on the same price history, expectations should converge and
traders would end up playing either a Majority or a Minority Game. But
of course agents revise and calibrate their expectations according to
the real price history so fundamentalists and trend-followers coexist
symbiotically in real markets. The problems with arguments in support
of either the Minority or the Majority Game essentially arise from the
fact that the objective assessment of the validity of a trading
strategy is a complex inter-temporal problem that cannot be based on
the result of a single transaction: whether buying today is profitable
or not depends on what the price will be when one sells. Hence the
payoff of a single transaction is hardly a meaningful concept unless
one considers round-trip (buy/sell or sell/buy) transactions. From
this point of view the MG is a rather crude approximation. Yet, we
shall see below that it provides a remarkably rich and realistic
picture of financial markets as complex adaptive systems. Models of
interacting fundamentalists and trend-followers will be addressed in
the following section.

\subsection{The simplest Minority Game}\label{zimpol}

Before considering the model in its full complexity, it is instructive
to to take a glimpse at a minimal version with inductive agents in
which the collective behavior can be easily understood with simple
mathematics \cite{Mats}. Let us suppose that traders employ a
probabilistic rule of the form \be\label{logit}
\prob\{b_i(t)=b\}=C(t)\exp\l[b\Delta_i(t)\r]~~~~~~~b\in\{-1,1\} \ee
where $C(t)$ is a normalization factor and $\Delta_i(t)$ accounts for
the agent's expectations about what will be the winning action (if
$\Delta_i(t)>0$ then he/she will choose $b_i(t)=1$ with higher
probability). The `score function' $\Delta_i$ is updated according to
\be\label{simpledyn} \Delta_i(t+1)-\Delta_i(t)=-\Gamma A(t)/N \ee with
$\Gamma>0$ a constant, so that if $A(t)<0$ agents increase $\Delta_i$
and the probability of choosing action $1$. Let us finally assume that
the initial conditions $\Delta_i(0)$ are drawn from a distribution
$p_0(\Delta)$ with standard deviation $s$. How does the collective
behavior depend on the parameters $\Gamma$ and $s$?

\begin{figure}[t]
\centering
\includegraphics*[width=8cm]{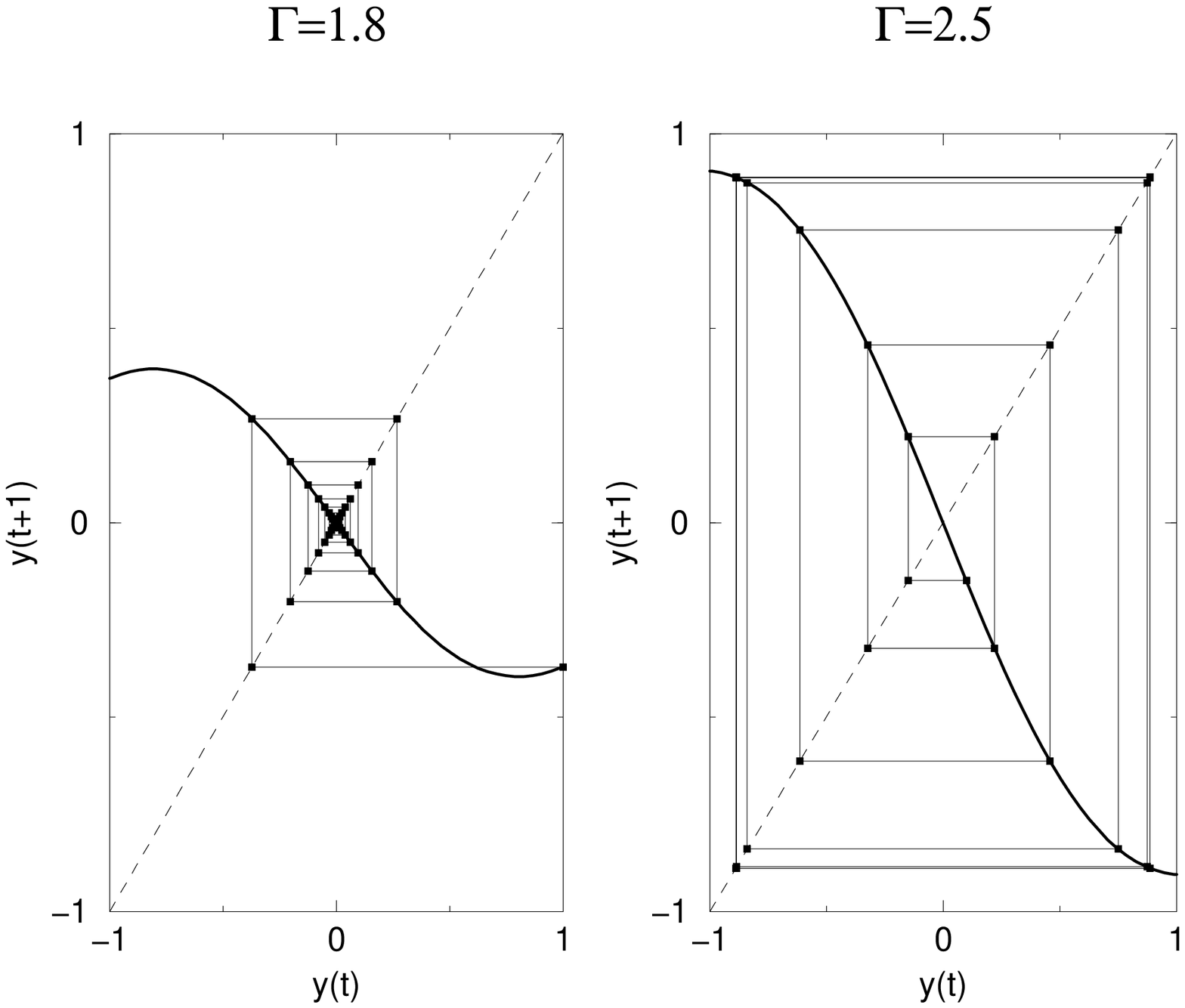}
\includegraphics*[width=7cm]{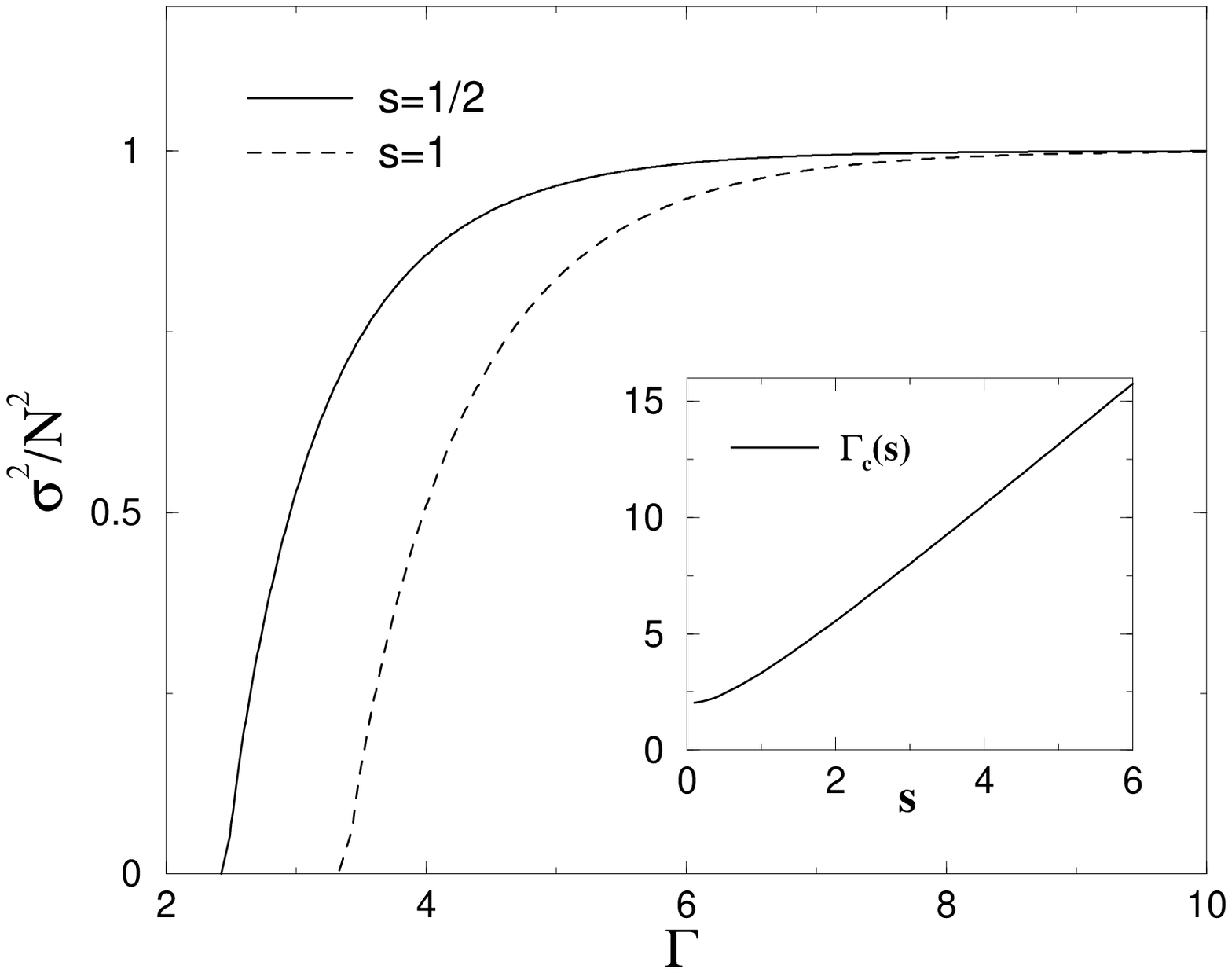}
\caption{Left panels: the map $y(t)$ for $\Gamma=1.8<\Gamma_c$ and
  $\Gamma=2.5>\Gamma_c$ for $s=0$.  Right panel: global efficiency
  $\sigma^2/N^2$ as a function of $\Gamma$ for two different sets of
  initial conditions: $\Delta_i(0)$ is drawn from a Gaussian
  distribution with variance $s^2$. The full line corresponds to
  $s=1/2$ whereas the dashed line is the result for $s=1$. The inset
  reports the critical learning rate $\Gamma_c$ as a function of the
  spread $s$ of initial conditions.}
\label{figsimplemg}
\end{figure}

Notice that $y(t)=\Delta_i(t)-\Delta_i(0)$ does not depend on $i$, for
all times.  For $N\gg 1$, the law of large numbers allows us to
approximate $A(t)$ by its average with probability distribution
(\ref{logit}). This yields an approximate dynamical equation for
$y(t)$: \be\label{map} y(t+1)\simeq y(t)-\Gamma\avg{\tanh
[y(t)+\Delta(0)]}_0 \ee where the average $\avg{\ldots}_0$ is on the
distribution $p_0$ of initial conditions.  Eq. \req{map} admits a
fixed point $y(t)=y^\star$, with $y^\star$ the solution of $\avg{\tanh
[y^\star+\Delta(0)]}_0\equiv\avg{A}=0$.  Let us assume that this
solution is stable. This describes a stationary state where the
relative scores $\Delta_i(t)$ are displaced by a quantity $y^\star$
from the initial conditions.  This gives \be \sigma^2=\sum_{i=1}^N
\l(1-\avg{a_i}^2\r)=N\l[1-\avg{\tanh[y^\star+\Delta(0)]^2}_0\r] \ee
Notice that $\sigma^2\propto N$ and it {\em decreases} with the spread
of the distribution of initial conditions. A linear stability analysis
of Eq. \req{map} shows that these solutions are stable only when \be
\Gamma<\Gamma_c=\frac{2}{1-\avg{\tanh[y^\star+\Delta(0)]^2}_0}=
\frac{2N}{\sigma^2}.  \ee When $\Gamma>\Gamma_c$ one finds periodic
solutions of the form $y(t)=y^\star+z^\star(-1)^t$ where $y^\star$ and
$z^\star$ satisfy certain prescribed conditions. The parameter
$z^\star$ plays the role of an order parameter of the transition at
$\Gamma_c$ ($z^\star=0$ for $\Gamma<\Gamma_c$). Again we have
$\avg{A}=0$, but now \be \sigma^2\simeq N^2
\frac{\avg{\tanh[y^\star+z^\star+\Delta(0)]}_0^2+
\avg{\tanh[y^\star-z^\star+\Delta(0)]}_0^2}{2} \ee i.e. fluctuations
are proportional to $N^2$. Hence this is a much less efficient state.
The orbits of the dynamics of $y(t)$ for $\Gamma<\Gamma_c$ and
$\Gamma>\Gamma_c$ are shown in Fig. \ref{figsimplemg} together with
the behavior of $\sigma^2/N^2$.
We conclude that the more heterogeneous the initial condition is, the
more efficient is the final state and the more the fixed point
$y^\star$ is stable. The transition from a state where
$\sigma^2\propto N$ to a state with $\sigma^2\propto N^2$ will turn
out to be a generic feature of MGs.

\subsection{The Minority Game}

In the simple case discussed above, agents base their choice only on
their past experience. The standard Minority Game describes a more
general situation in which traders use both their past experience and
some (endogenous or exogenous) information pattern. The model is
defined as follows \cite{ChM}. There are $N$ agents labeled $i$. At
each time step $t$ agents receive one of $P$ possible information
patterns $\mu(t)$ (whose precise nature will be discussed below) based
on which each trader must formulate a binary bid
$b_i(t)\in\{-1,1\}$. To this aim, each of them is endowed with $S$
strategies $\bsy{a}_{ig}=\{a_{ig}^\mu\}$ ($g=1,\ldots,S$) that map
informations $\mu\in\{1,\ldots,P\}$ into actions
$a_{ig}^\mu\in\{-1,1\}$. Each component $a_{ig}^\mu$ of every strategy
is selected randomly and independently from $\{-1,1\}$ with equal
probability for every $i$, $g$ and $\mu$ at time $t=0$ and is kept
fixed throughout the game.  Agents keep tracks of the performance of
their strategies by means of valuations functions or scores $U_{ig}$
that are initialized at some value $U_{ig}(0)$ and whose dynamics
reads \be\label{ldmg} U_{ig}(t+1)-U_{ig}(t)=-a_{ig}^{\mu(t)}A(t)/N \ee
where $A(t)=\sum_i b_i(t)$ is the excess demand at time $t$. At each
round, every agent picks the strategy $g_i(t)=\argmax_g U_{ig}(t)$
carrying the highest valuation and formulates the corresponding bid:
$b_i(t)=a_{g_i(t)}^{\mu(t)}$. In this way, agents adopt at each time
the strategy they expect to deliver the highest profit (the score of
strategies forecasting the correct minority action increase in time).

The nature of the information patterns $\mu(t)$ is still to be
specified. In principle, the natural choice corresponds to taking the
string of the past $m$ minority actions (hence $P=2^m$) as the
information fed to agents at every time step, with the idea to
describe a closed system where agents process and react to an
information they produce themselves collectively. We refer to this
choice as the case of endogenous information. On the other hand, one
may think of replacing for the sake of simplicity the above
information (which has a non-trivial dynamics itself) with an integer
drawn at random at each time step from $\{1,\ldots,P\}$ with uniform
probability. This corresponds to the case of random exogenous
information \cite{Cavagna}. Again, this replacement induces a major
simplification in the structure of the model by turning a complex
non-Markovian system with feedback into a Markovian one. In addition
and at odds with the El Farol problem, it was shown that collective
properties are roughly unaffected when real information is substituted
with random information. These results suggest that, to some extent,
the feedback is irrelevant as far as collective properties are
concerned. We shall hence focus on the case of exogenous information
for the following sections. A more careful discussion of the subtle
case of endogenous information will be deferred to Sec. \ref{endo}. In
summary, the Minority Game is completely defined by the following
rules:
\begin{eqnarray}
g_i(t)=\argmax_g U_{ig}(t)\nonumber\\
A(t)=\sum_i a_{i g_i(t)}^{\mu(t)}\\
U_{ig}(t+1)-U_{ig}(t)=-a_{ig}^{\mu(t)}A(t)/N\nonumber
\end{eqnarray}

\begin{figure}[t]
\centering
\includegraphics*[width=8cm]{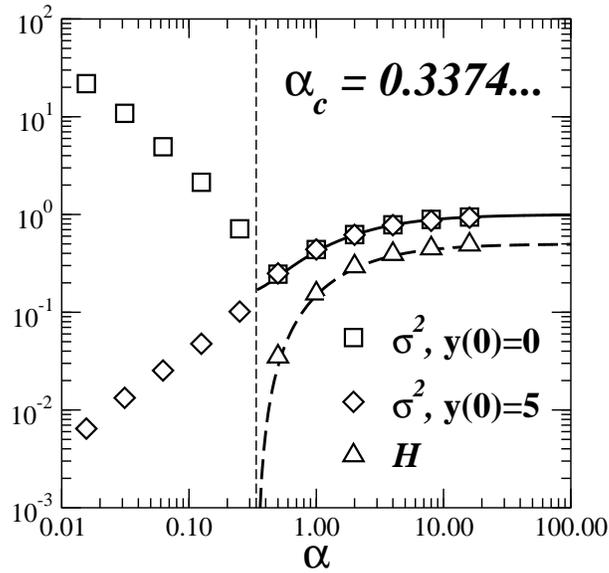}
\caption{Behavior of $\sigma^2/N$ and $H/N$ versus $\alpha$
(analytical and numerical) for different initial conditions
$y(0)=U_{i1}(0)-U_{i2}(0)$.}
\label{figsigmg}
\end{figure}

Let us now discuss the macroscopic properties of the model. Early
works focused on the cooperative properties of the system in the
stationary state. The central quantity of interest is the numerical
difference between buyers and sellers at each time step, $A(t)$. It is
easy to anticipate that none of the two actions $-1$ and $1$ will
systematically be the minority one, i.e. that $A(t)$ will fluctuate
around zero. Were it not so, agents could easily improve their scores
by adopting that strategy which visits most often that side. The size
of fluctuations of $A(t)$, instead, displays a remarkable non-trivial
behavior. The variance $\sigma^2=\avg{A^2}$ of $A(t)$ in the
stationary state measures the efficiency with which resources are
distributed, since the smaller $\sigma^2$, the larger a typical
minority group is. In other words $\sigma^2$ is a reciprocal measure
of the {\em global efficiency} of the system. Early numerical studies
have shown that the relevant control parameter of the model is the
relative number of information patterns $\alpha=P/N$. The behavior
$\sigma^2$ is illustrated in Fig. \ref{figsigmg}.
With $\alpha$ fixed, one typically observes that $A(t)\simeq\sqrt{N}$
or equivalently that $\sigma^2=O(N)$.  When $\alpha\gg 1$ the
information space is too wide to allow for a coordination and agents
essentially behave randomly as $\sigma^2/N\simeq 1$, the value
corresponding to random traders. As $\alpha$ decreases, that is as
more and more agents join the game or as the possible number of
information patterns decreases, $\sigma^2/N$ decreases suggesting that
agents manage to exploit the information in order to coordinate to a
state with better-than-random fluctuations. It turns out that these
steady states are ergodic, that is they are reached independently of
the initial conditions $U_{ig}(0)$. Lowering $\alpha$ further,
ergodicity is lost and the steady state depends on $U_{ig}(0)$. For
the so-called flat initial conditions, $U_{ig}(0)=0$ for all $i$ and
$g$, which describe agents with no a priori bias toward one of their
strategies, one is driven into highly inefficient steady states where
$\sigma^2$ diverges as $\alpha$ decreases approximately as
$\sigma^2\simeq 1/\alpha$. Notice that this implies $\sigma^2\simeq
N^2$. This behavior for has been attributed to the occurrence of
``crowd effects''. Remarkably this ergodicity breaking transition is
related to a phase transition with symmetry breaking that was first
discovered by Savit and coworkers \cite{Savit} for the case of
endogenous information. Reporting the frequency with which the
minority action was $1$ conditional on the value of $\mu$, they
observed that for $\alpha\ll 1$ the minority was falling on either
side with equal probability irrespective of $\mu$. But when $\alpha\gg
1$ the minority happened to be more likely on one side, depending on
the value of $\mu$. These observations have been sharpened in a study
that allowed to locate the phase transition at the point
$\alpha_c\simeq 0.34$ for $S=2$ where $\sigma^2$ attains its minimum
(see next section for details). The transition separates a symmetric
($\alpha<\alpha_c$) from an asymmetric phase ($\alpha>\alpha_c$). The
symmetry which is broken is that of the average of $A(t)$ conditional
on the history $\mu$, $\avg{A|\mu}$.  The idea is that if
$\avg{A|\mu}\neq 0$ for a certain $\mu$ then the knowledge of $\mu$
alone suffices for a non-trivial statistical prediction of the sign of
$A(t)$. In the asymmetric phase, $\avg{A|\mu}\neq 0$ for at least one
$\mu$. Thus the sign of $A(t)$ is predictable, to some extent, on the
basis of $\mu$ alone. A measure of the degree of predictability is
given by the function \be H=\frac{1}{P}\sum_{\mu=1}^P\avg{A|\mu}^2.
\label{Hmu}
\ee In the symmetric phase $\avg{A|\mu}=0$ for all $\mu$ and hence
$H=0$. $H$ is a decreasing function of the number $N$ of agents (at
fixed $P$): newcomers exploit the predictability of $A(t)$ and hence
reduce it. The behavior of $H$ is also reported in
Fig. \ref{figsigmg}. Notice that it acts like a `physical' order
parameter.

\subsection{Statistical mechanics of the MG: static approach}\label{mgth}

We shall discuss in this review two lines along which the statistical
mechanics of the Minority Game with random external information can be
studied.  The first one is a static theory whose crucial steps are (a)
finding a (random) Lyapunov function of the dynamics that allows one
to identify the steady states of the learning process with its minima;
(b) calculating the latter via the replica method. The second one
consists in constructing a dynamical mean-field theory using the
learning dynamics as a starting point. The two approaches are
essentially complementary: the statics gives more information about
the predictability and allows to interpret the collective properties
in terms of a minimized quantity; the dynamics focuses on ergodicity
and is a more appropriate setting to discuss fluctuations. Below we
will outline the static approach to the standard MG for the case
$S=2$, deferring a discussion of the dynamical method to
Sec. \ref{mamg1}. Other possibilities, like the `crowd-anticrowd'
theory \cite{cac} will not be discussed here (an account can be found
in \cite{NFJBOOK}).

It is helpful for a start to introduce the auxiliary variables
\cite{ChM} \be \bsy{\xi}_i=\frac{\bsy{a}_{i1}-\bsy{a}_{i2}}{2},~~~~~
\bsy{\omega}_i=\frac{\bsy{a}_{i1}+\bsy{a}_{i2}}{2},~~~~~
y_i(t)=\frac{U_{i1}(t)-U_{i2}(t)}{2} \ee in terms of which \req{ldmg}
can be re-cast as \be\label{dyna1}
y_i(t+1)-y_i(t)=-\frac{1}{N}\xi_i^{\mu(t)}A(t) \ee The advantage lies
in the fact that the dependence of $\bsy{a}_{ig_i(t)}$ on the strategy
valuation can be made explicit by noticing that $g_i(t)=1$ if
$y_i(t)>0$ and $g_i(t)=2$ if $y_i(t)<0$ (we shall therefore refer to
$y_i$ as the `preference' of agent $i$). As a consequence, the
relevant microscopic dynamical variable is the Ising spin
$s_i(t)=\sign[y_i(t)]$. On has in particular
\begin{eqnarray}
  \bsy{a}_{ig_i(t)}=\bsy{\omega}_i+s_i(t)\bsy{\xi}_i\\
  A(t)=\sum_i\l[\omega_i^{\mu(t)}+s_i(t)\xi_i^{\mu(t)}\r]\equiv\Omega^{\mu(t)}+\sum_i
  s_i(t)\xi_i^{\mu(t)}
\end{eqnarray}
The dynamics \req{dyna1} is non-linear in a way that doesn't allow to
write it in the form of a gradient descent. However, as in the El
Farol problem, one may regularize the dynamics via a learning rate
$\Gamma>0$ such that \cite{thermal} \be\label{rules}
\prob\{g_i(t)=g\}=C(t) ~e^{\Gamma U_{ig}(t)}~~~~~~~C(t)={\rm
normalization} \ee It is then possible to construct the
continuous-time limit of \req{dyna1} in view of the fact that the
dynamics possesses a `natural' characteristic time scale given by
$P$. Proceeding as shown for the El Farol case, one arrives at the
following continuous-time Langevin process \cite{CTL}:
\begin{eqnarray}
  \dot y_i(\tau)=-\ovl{\xi_i\Omega}-\sum_j\ovl{\xi_i\xi_j}
  \tanh[y_j(\tau)]+z_i(\tau)\label{langmg}\\
  \avg{z_i(\tau)z_j(\tau')}\simeq\frac{\Gamma\sigma^2}{\alpha
  N}\ovl{\xi_i\xi_j}\delta(\tau-\tau')\label{noizmg}
\end{eqnarray}
where $\tau=\Gamma t/P$ is a re-scaled time and $\sigma^2$ is the
volatility\footnote{Eq. (\ref{noizmg}) is based on a time-independent
volatility approximation which happens to be very well satisfied away
from the critical line. We refer the interested reader to \cite{CTL}
for further details.} and the over-line denotes an average over
$\mu$. One sees that in the limit $\Gamma\to 0$, in which the dynamics
becomes deterministic, the system performs a gradient descent with a
well-defined Hamiltonian. Indeed, in order to extract the steady state
from the above process, one may take its time average:
\be\label{dyna1avg}
\dot{\avg{y_i}}=-\ovl{\xi_i\Omega}-\sum_j\ovl{\xi_i\xi_j}m_j,~~~~~~~
m_i=\avg{\tanh(y_i)}\in[-1,1] \ee It is now clear that the stationary
values of the variables $m_i$ can be obtained from the minimization of
\be H= \frac{1}{P}\sum_\mu\l[\Omega^\mu+\sum_i\xi_i^\mu m_i\r]^2 \ee
which coincides with the predictability in the steady state. Hence
agents coordinate so as to make the market as unpredictable as
possible. This conclusion remains correct even for $\Gamma>0$: indeed
$m_i$ are still given by the minima of $H$, though the dynamics is no
more deterministic (see \cite{CTL}). Actually, within the
approximation of Eq. (\ref{noizmg}), it can be shown (see
Sec. \ref{rates}) that for $\alpha>\alpha_c$ the steady state is
independent of $\Gamma$.

As usual, minimization of $H$ is achieved through the replica trick as
\be \lim_{N\to\infty}\frac{1}{N}\davg{\min_{\bsy{m}}\frac{H}{N}}
=\lim_{\beta\to\infty}\lim_{r\to 0}\lim_{N\to\infty}\frac{1}{\beta r
N} \log\davg{\l[\Tr_{\bsy{m}}e^{-\beta H}\r]^r} \ee The calculation is
detailed at length in the literature (see e.g. \cite{physa}). The
resulting phase structure is as follows:
\begin{itemize}
\item for $\alpha$ larger than a critical value
  $\alpha_c=0.3374\ldots$ there is a unique ($\Gamma$-independent)
  minimum with $H>0$
\item for $\alpha<\alpha_c$, there is a
  continuous of minima where ${H}$ vanishes. The minimum
  selected by the dynamics depends on initial conditions (and on
  $\Gamma$)
\end{itemize}

Hence the system at $\alpha_c$ undergoes a phase transition from a
predictable to an unpredictable phase. Such a static transition
corresponds to a dynamical instability in the dynamics of preferences
for $\Gamma=0$.  To see this, let us first mention that in numerical
simulations one observes that $y_i$ either grows linearly with time or
stays finite.  Based on this, one can conclude that solutions of
\req{dyna1avg} are of the form $\avg{y_i}=v_i t$, with \be
v_i=-\ovl{\xi_i\Omega}-\sum_j\ovl{\xi_i\xi_j}m_j \ee and that there
are two possibilities:
\begin{itemize}
\item either $v_i\neq 0$ and $y_i(t)$ diverges as $t\to\infty$, in
  which case $m_i=sign(v_i)$ the agent ends up using just one of his
  strategies (we call these agents `frozen')
\item or $v_i=0$ and $\avg{y_i}$ stays finite, in which case
$-1<m_i<1$ and the agent keeps flipping between his strategies (we
call these agents `fickle')
\end{itemize}
Let us consider the dynamics of preferences for fickle agents. Setting
$y_i(\tau)=\avg{y_i}+\epsilon_i(\tau)$ where $\epsilon_i(\tau)$
describes small fluctuations about the average, one can expand
\req{langmg} to first order in $\epsilon_i(t)$: \be
\dot \epsilon_i(\tau)=-\sum_{j{\rm
~fickle}}\ovl{\xi_i\xi_j}(1-m_j^2)\epsilon_j(t) \equiv-\sum_{j{\rm
~fickle}} T_{ij}\epsilon_j(t) \ee where
$T_{ij}=\ovl{\xi_i\xi_j}(1-m_j^2)$. As long as the matrix
$\bsy{T}=(T_{ij})$ is positive definite, the above dynamical system
will be linearly stable. Now $\bsy{T}=\bsy{U V}$ with
$U_{ij}=\ovl{\xi_i\xi_j}$ and $V_{ij}=(1-m_i^2)\delta_{ij}$. But for
fickle agents ($|m_i|<1$) all eigenvalues of $\bsy{V}$ are positive
definite, so that $\det(\bsy{T})$ vanishes together with
$\det(\bsy{U})$. The spectrum of the random matrix $\bsy{U}$ can be
evaluated using random matrix theory. For our purposes it suffices to
calculate the minimum eigenvalue, which turns out to be
$\lambda_0=\frac{1}{2}\l(1-\sqrt{\frac{1-\phi}{\alpha}}\r)^2$. The
instability sets in when $\lambda_0=0$, that is when 
\be\label{conda}
1-\phi=\alpha 
\ee 
This equation and the distinction between fickle and frozen agents
only depend on $m_i$, which are determined for $\alpha\ge\alpha_c$ by
the unique minimum of $H$, independently of $\Gamma$. Hence
Eq. (\ref{conda}) and the location $\alpha_c$ of the phase transition,
are independent of $\Gamma$.

\subsection{The role of learning rates and decision noise}\label{rates}

It is interesting to consider briefly the impact that the introduction
of a finite learning rate $\Gamma$ has on the properties of the model.
Let us begin by noting that $\Gamma$, which at the level of agents
plays a role similar to an `inverse temperature', at the collective
level acts instead as an effective `temperature', since it tunes the
fluctuating random component in agent's dynamics (see \req{noizmg}).
The larger $\Gamma$ or, equivalently, the smaller the minimum score
difference agents can appreciate (this quantity is roughly of order
$1/\Gamma$), the more the response fluctuates and the longer it takes
to average fluctuations out and reach a steady state.

We have anticipated above that $\Gamma$ affects the steady state only
in the sub-critical phase. Its effect is particularly strong on the
volatility, which can be written as \be
\sigma^2=H+\sum_i\ovl{\xi_i^2}(1-m_i^2)+\sum_{i\neq j}\ovl{\xi_i
\xi_j} \avg{(\tanh y_i-m_i)(\tanh y_j-m_j)} 
\label{ssss}
\ee The dependence on $\Gamma$ is only present in the last term on the
right-hand side, which measures fluctuations of $\tanh y_i$ around its
mean. The average is over the distribution of $y_i$ (which in turn
depends on $\sigma^2$ via the noise). The latter can be computed from
the Fokker-Planck equation associated to \req{langmg}, which itself
depends on $\sigma^2$ (see \ref{noizmg}). Hence $\sigma^2$ is
determined by the solution of a self-consistent problem \cite{CTL}.
For $\alpha>\alpha_c$, fluctuations of $y_i$ are independent and hence
the third term of  (\ref{ssss}) is identically zero. As a result,
$\sigma^2$ is independent of $\Gamma$, as confirmed to a remarkable
degree of accuracy by numerical simulations \cite{CTL}. When
$\alpha<\alpha_c$ a correlation arises from the fact that the dynamics
is constrained to the subspace of $\bsy{y}$ which is spanned by the
$P$ vectors $\bsy{\xi}^\mu$, and which contains the initial condition
$\bsy{y}(0)$. The dependence on initial conditions and the dependence
on $\Gamma$ both arise as a consequence of this fact. Again, numerical
simulations fully confirm this picture \cite{CTL}.

It is worth remarking that, the smoothed choice rule \req{rules} can
also be written as \be s_i(t)={\rm sign}\l[y_i(t)+\zeta_i(t)/\Gamma\r]
\ee where $\zeta_i(t)$ are independent identically distributed random
variables with probability density
$p(\zeta)=\frac{1}{2}\l[1-(\tanh\zeta)^2\r]$.  Indeed, for $\Gamma=0$
the noisy part of the argument of the sign dominates and the agent
selects his strategy at random with equal probability at each time
step, while for $\Gamma\to\infty$ one recovers the original
deterministic rule $s_i(t)={\rm sign}[y_i(t)]$.

On the basis of this observation, Coolen {\em et al.} \cite{multnoise}
introduce a different type of decision noise, called `multiplicative
noise', defined as \be s_i(t)={\rm
sign}\l[y_i(t)\l(1+\zeta_i(t)/\Gamma\r)\r] \ee which corresponds to
\be \prob\{s_i(t)=\pm 1\}=C(t) ~e^{\pm\Gamma {\rm
sign}[y_i(t)]}~~~~~~~C(t)={\rm normalization} \ee It is evident that
in this case frozen agents are affected as well. Indeed, the critical
point $\alpha_c$ turns out to depend rather strongly on $\Gamma$:
when $\Gamma$ gets smaller the informationally efficient phase shrinks
as the critical point shifts to smaller values of $\alpha$.

\subsection{The role of market impact}

Ever since J. Nash's pioneering work in game theory, that of Nash
equilibrium (NE) has been a reference concept in socio-economic
systems of interacting agents. A NE is in some sense an optimal state
of strategic situations, one in which no agent has incentives to
deviate from his behavior unilaterally. It is easy to see that, a
priori, the Minority Game possesses a huge number of such states when
$N\gg 1$. In fact, there is one symmetric NE in mixed strategies,
where agents draw their bid $b_i$ at random at every time step with
${\rm Prob}\{b_i=+1\}=1/2$ for all $i$. This state has $\sigma^2=N$
and $H=0$. If $N$ is even, there are also ${N\choose N/2}$ pure
strategy NE where half of the players take $b_i=+1$ and the other half
takes $b_i=-1$. Moreover, states where $N-2k$ agents play mixed
strategies and the remaining $2k$ play pure strategies $b_i=+1$ and
$b_i=-1$, are also NE.  Thus the game possesses an exponentially large
number of Nash equilibria. One can then ask whether the steady state
of the model is one of them. The answer is a resounding no. In this
section we will study this issue and discuss the important question of
why it is so. Why are inductive agents playing sub-optimally? We shall
see that at the heart of the matter lies the consideration which
agents have of their {\em market impact}, i.e. of their impact on the
aggregate quantity $A(t)$. In fact, the inability to coordinate on a
NE follows from the na\"\i ve idea that in a system of $N$ agents
every single agent `weights' $1/N$ and is thus negligible in the
statistical limit $N\to\infty$. Once this assumption is dropped and
agents account for their own impact, the resulting steady state
improves dramatically and eventually a NE may be reached.

To begin with, it is instructive to study the role of market impact in
the simplest MG with $P=1$ discussed in Sec. \ref{zimpol}, in which
agents must choose at each time step between the two actions
$a_{i}\in\{-1,1\}$. Let us consider the following modification of the
learning dynamics \req{simpledyn}: \be
\Delta_i(t+1)-\Delta_i(t)=-\frac{\Gamma}{N}[A(t)-\eta a_i(t)].
\label{learneta}
\ee The term proportional to $\eta$ in \req{learneta} describes the
fact that agent $i$ accounts for his own contribution to $A(t)$. One
indeed sees that \req{learneta} reduces to \req{simpledyn} for
$\eta=0$, whereas for $\eta=1$ agent $i$ considers only the aggregate
action of other agents, $A(t)-a_i(t)=\sum_{j\neq i}a_j(t)$, and does
not react to his own action $a_i(t)$. Values of $\eta$ between $0$ and
$1$ tune the extent to which agents account for their ``market
impact''.

It is easy to see that the dynamics for $\eta=1$ behaves in the long
run in a radically different way than for $\eta=0$. Let us take the
average of \req{learneta} in the steady state and define
$m_i=\avg{a_i}$. We note that \be
\avg{\Delta_i(t+1)}-\avg{\Delta_i(t)}=-\frac{\Gamma}{N}\left[ \sum_j
m_j-\eta m_i\right]= -\frac{\Gamma}{N}\frac{\partial
H_\eta}{\partial\eta}
\label{DYNN}
\ee where \be H_\eta = \frac{1}{2}\left(\sum_i
  m_i\right)^2-\frac{\eta}{2}\sum_i m_i^2.  \ee This implies that the
  stationary values of the $m_i$'s are given by the minima of
  $H_\eta$.  Notice that $H_1$ is a harmonic function of the
  $m_i$'s. Hence it attains its minima on the boundary of the
  hypercube $[-1,1]^N$. So for $\eta=1$ all agents always take the
  same actions $a_i(t)=m_i=+1$ or $a_i(t)=m_i=-1$ and the waste of
  resources is as small as possible, as $\sigma^2=0$ or $1$ if $N$ is
  even or odd, which is a tremendous improvement with respect to the
  case $\eta=0$ (where $\sigma^2\sim N$ or $N^2$). These states are
  indeed Nash equilibria of the associated $N$ persons minority
  game. This argument can be extended with some work to all $\eta>0$,
  and one can show that the stationary states of the learning process
  for any $\eta>0$ are Nash equilibria. Hence as soon as agents start
  to account for their market impact ($\eta>0$) the collective
  behavior of the system changes abruptly and inefficiencies are
  drastically reduced. Furthermore, the asymptotic state is not unique
  ($H_\eta$ possesses more than one minimum!) and the one in which the
  system settles is selected by the initial conditions. The set of
  equilibria is discrete and the system jumps discontinuously from an
  equilibrium to another, as the initial conditions $\Delta_i(0)$
  vary. This also contrasts with the $\eta=0$ case, where the
  equilibrium shifts continuously as a function of the initial
  conditions.

Let us now consider the full MG with market impact correction with
public information \cite{physa} (the above picture is representative
of the situation in the MG in the limit $\alpha\to 0$), whose learning
dynamics reads \be\label{eta}
U_{ig}(t+1)-U_{ig}(t)=-\frac{a_{ig}^{\mu(t)}}{N}\l[A(t)-\eta\l(a_{i
g_i(t)}^{\mu(t)}-a_{ig}^{\mu(t)}\r)\r] \ee As before, $\eta$ allows to
interpolate between the {\em naive} `price-taking' behavior of the
standard MG in which agents are unaware of their market impact
($\eta=0$) and a more sophisticated behavior where agents account for
it. Note indeed that with $\eta=1$ the reinforcement
$U_{ig}(t+1)-U_{ig}(t)$ is proportional to the actual payoff that
agent $i$ would have got had he actually played strategy $g$ at time
$t$. Hence in a way the above learning process assumes that agents are
able to disentangle their contribution from the aggregate $A(t)$. This
may not be realistic in practical situations. For example imagine
that, as in the original version of the MG, agents only observe the
sign of $A(t)$ and not its value. This information is not enough to
infer the sign of $A(t)-a_{i g_i(t)}^{\mu(t)} +a_{ig}^{\mu(t)}$ and
hence the payoff they would have received if they had played strategy
$g$ instead of $g_i(t)$. However, agents can approximately account for
the market impact by rewarding the strategy they have played by a
reinforcement factor $\eta$, i.e.  \be
U_{ig_i}(t+1)-U_{ig}(t)=-a_{ig}^{\mu(t)}\frac{A(t)}{N}+\frac{\eta}{N}\delta_{g
g_i(t)}\nonumber \ee In fact, the collective behavior of the learning
dynamics above is identical to that obtained with \req{eta}. This is
because what matters in the long run is the time average of the
processes, which is the same because
$\ovl{\avg{a_{ig}a_{ig'}}}\simeq\delta_{g,g'}$. 

At first sight, the term proportional to $\eta$ looks negligible with
respect to $A(t)$ because it is of order one whereas
$A(t)=O(\sqrt{N})$. However while $A(t)$ fluctuates around zero,
$\delta_{s,s_i(t)}$ has always the same sign. When the term
proportional to $A(t)$ is averaged over the $P=\alpha N$ states $\mu$
it also becomes of order one. Hence the effect of the two terms is
comparable in the long run. (A similar phenomenon occurs in spin
glasses where the naive mean filed theory has to be corrected by the
Onsager reaction term to eliminate self-interaction effects.). For
generic $\eta$ ($0\leq\eta\leq 1$) the steady state is described by
the minima of \be H_\eta=H-\eta\sum_i\ovl{\xi_i^2}(1-m_i^2)
\label{Heta}
\ee where $H=\ovl{\avg{A|\mu}^2}$ is the predictability. Note that
$H_1=\sigma^2$, so players who fully account for their impact
effectively minimize fluctuations.

\begin{figure}[t]
\centering 
\includegraphics[width=8cm]{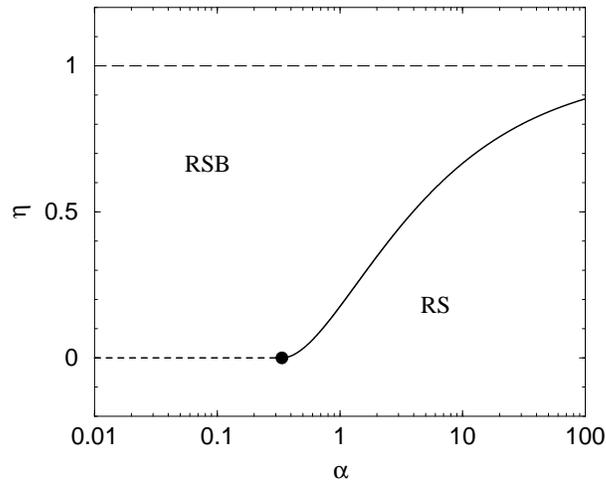}
\caption{Phase diagram of the Minority Game in the $(\alpha,\eta)$
  plane. The RS region corresponds to the replica-symmetric phase and
  the RSB region to the replica symmetry broken phase (from
  \cite{andemar}). The mark corresponds to the critical point
  $\alpha_c\simeq 0.3374$. Above it, the RSB $\to$ RS transition is
  second order; below it, it is discontinuous.}
\label{phasediag}
\end{figure}

Unfortunately, the study of the ground state properties of $H_\eta$
requires techniques which are more sophisticated than those used for
the MG. Indeed for $\eta>0$ the simple replica-symmetric solution that
we have discussed so far becomes unstable against perturbations that
break replica permutation symmetry (this is related to the fact that
$H_\eta$ has more than one minimum) and one needs to study more
complicated solution types \cite{andemar}.  The ensuing phase
structure is shown in Fig. \ref{phasediag}
The critical line (analog to the de Almeida-Thouless line of
spin-glass theory) can be calculated straightforwardly using the
dynamical stability argument mentioned at the end of Sec.
\ref{mgth}. It suffices to replace $U_{ij}=\ovl{\xi_i\xi_j}$ with
$U_{ij}=\ovl{\xi_i\xi_j}-\eta\delta_{ij}\ovl{\xi_i^2}$. The resulting
condition reads $1-\phi=\alpha(1-\sqrt{\eta})^2$ and coincides with
the critical line for replica-symmetry breaking.

\begin{figure}[t]
\centering 
\includegraphics[width=8cm]{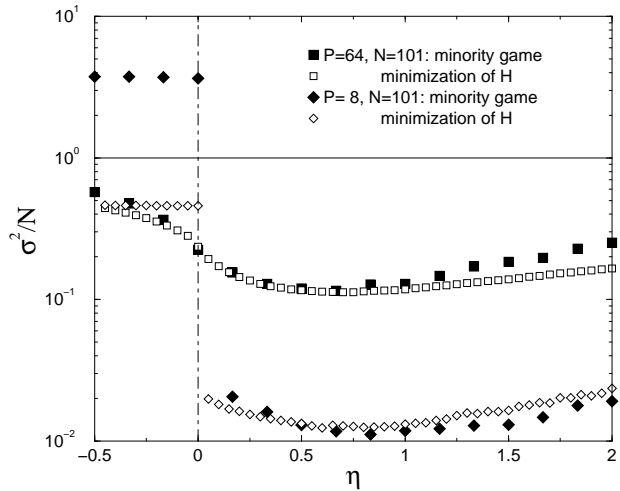}
\caption{$\sigma^2/N$ as a function of $\eta$ for $S=2$ and
$\alpha\simeq 0.079<\alpha_c\simeq 0.3374$ and $\alpha\simeq
0.63>\alpha_c$.  Results both of numerical simulations of the minority
game and of the numerical minimization of $H_\eta$ are shown. In both
cases the replica symmetry breaks at $\eta=0$ (from \cite{physa}).}
\label{figsigvseta}
\end{figure}

The MG behavior ($\eta=0$) is separated from the Nash equilibrium
behavior ($\eta=1$) by a phase transition which is continuous for
$\alpha>\alpha_c$. Remarkably for $\alpha<\alpha_c$ the transition
occurs at $\eta=0$ and it becomes discontinuous. As shown in Fig.
\ref{figsigvseta}, nothing dramatic happens when crossing the
transition for $\alpha>\alpha_c$. For $\alpha<\alpha_c$ instead
$\sigma^2/N$ features a discontinuous jump across the transition line
at $\eta=0$.
The origin of the discontinuity lies in the dynamic degeneracy of the
system for $\alpha<\alpha_c$ and $\eta=0$. Even an infinitesimal
change in $\eta$ can dramatically alter the nature of the minima of
$H_\eta$: for negative $\eta$ there is only one minimum which becomes
shallower and shallower as $\eta\to 0^-$. At $\eta=0$ the minimum is
always unique but it is no more point-like. Rather it is a connected
set.  An infinitesimal positive value of $\eta$ is enough to lift this
degeneracy. The set of minima becomes suddenly disconnected.  At fixed
$\alpha<\alpha_c$, varying $\eta$ across the transition $H_\eta$
changes continuously -- with a discontinuity in its first derivative
-- whereas the remaining fluctuation terms in $\sigma^2/N$ change
discontinuously with a jump. The potential implications of this result
are quite striking: rewarding the strategy played more than those
which have not been played by a small amount is always advantageous.
In particular, {\em an infinitesimal reward is sufficient to reduce
  fluctuations by a finite amount, for $\alpha<\alpha_c$}.

\begin{figure}[t]
\centering
\includegraphics[width=8cm,angle=-90]{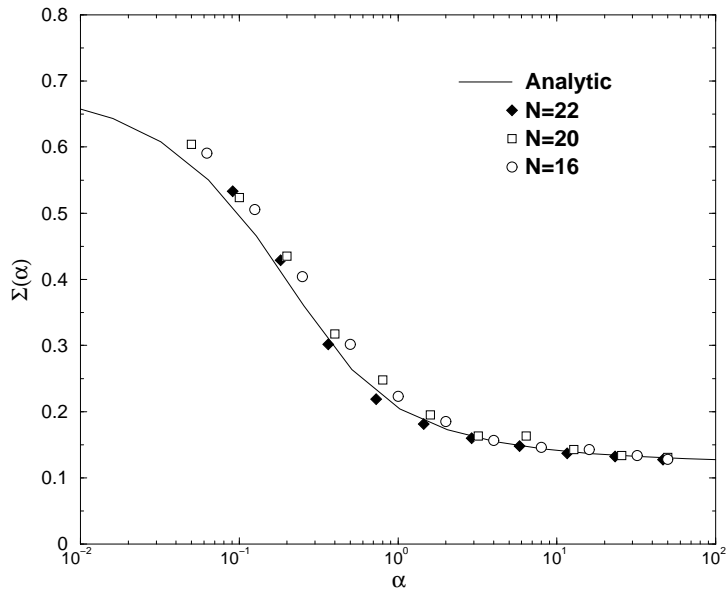}
\caption{Logarithm of the average number of NE divided by $N$ as a
function of $\alpha$ (from \cite{andemar}).}
\label{nasheq}
\end{figure}

Let us finally come to the case $\eta=1$, corresponding to NE, in
which, as we said, steady states coincide with the states of minimum
$\sigma^2$. One understands that these minima occur when agents play
only one of their available strategies\footnote{There may also be
other NE, which correspond to saddle points of $\sigma^2$ and are
hence stationary points of the multi-population replicator
dynamics. Agents do not play evolutionarily stable strategies in these
NE and as we shall see the dynamics of learning never converges to
these states. Hence we do not consider these NE further.}, since
$\sigma^2$ attains minima in the corners of the configuration space
$[-1,1]^N$. The statistical properties of the minima of $\sigma^2$ can
again be analyzed with tools of statistical mechanics. As is clear
from Fig. \ref{phasediag}, for $\eta=1$ one is always in the phase
with broken replica symmetry because $\sigma^2$ attains its minima on
a disconnected set of points. For $S=2$ strategies per agent it has
been shown analytically via the so-called annealed approximation that
the number of NE (i.e.  of minima of $\sigma^2$) is exponentially
large in $N$ (see Fig.  \ref{nasheq}).
It is clear that the global efficiency of NE is better than in the
standard MG, since fluctuations are smaller. Furthermore, increasing
the number $S$ of strategies the efficiency of NE increases
(i.e. $\sigma^2$ decreases) as shown in \cite{physa}. This contrasts
with what happens in the MG, where the efficiency generally decreases
when $S$ increases.  Therefore, not only agents in the MG play
sub-optimally, but the more resources they have the larger is the
deviation of their behavior from an optimum.

We are still left with the question: why do agents in the MG play
sub-optimally? In order to answer, let us consider the case of an {\em
  external} agent with $S$ strategies, an agent who does not take part
in the game but just observes its outcome from the outside. From this
position, each of his strategies delivers an average {\em virtual}
gain $\pi_g^{{\rm vir}}=-\ovl{a_g\avg{A}}$ ($g=1,\ldots,S$). Given
that the strategies $a^\mu_g$ are drawn randomly, the $\pi_g^{{\rm
    vir}}$'s are independent random variables. Moreover, since
$\pi_g^{{\rm vir}}$ is the sum of $P\gg 1$ independent variables
$a_g^\mu\avg{A^\mu}/P$, their distribution is Gaussian with zero mean
and variance \be {\rm Var}\l(\pi_g^{{\rm
    vir}}\r)=\frac{1}{P^2}\sum_{\mu=1}^P{\rm Var}(a^\mu_g)
\avg{A|\mu}^2 =\frac{H}{P}.  \ee Clearly, the strategy $g^\star$
bearing the highest expected profit $\pi_{g^\star}^{{\rm vir}}$ is
superior to all others.  It would be most reasonable for this agent to
just stick to this strategy.

However, the same agent {\em inside} the game will typically use not
only strategy $g^\star$ since every strategy, when used, delivers a
{\em real} gain which is reduced with respect to the virtual one by
the ``market impact''. Imagine the ``experiment'' of injecting the new
agent in a MG.  Then, neglecting the reaction of other agents to the
new-comer, one would have that $\avg{A|\mu}\to \avg{A|\mu}+a^\mu_g$.
Then the real gain of the newcomer is: \be \pi_g^{{\rm real}}\simeq
-\ovl{a_g\avg{A}}-\avg{a_g\,a_g}=\pi_g^{{\rm vir}}-1.  \ee The agent
will then update the score of the strategy he uses (say $g$) with the
real gain $\pi_g^{{\rm real}}$ and those of the strategies he does not
use (say $g'$) with the virtual one, so that $U_{g'}=\pi_{g'}^{{\rm
    real}}+1-\ovl{a_g a_{g'}}\simeq \pi_{g'}^{{\rm real}}+1$.
Therefore agents in the MG over-estimate the performance of the
strategies they do not play.  Then if strategy $g$ is played with a
frequency $f_g$, the virtual score increases {\em on average} by \be
v_g=U_g(t+1)-U_g(t)=\pi_g^{{\rm real}}-f_g+1
\label{vs}
\ee The fact that a good strategy $g$ is used frequently reduces its
perceived success\footnote{More precisely the frequency $f_g$ with
  which the agent plays strategy $g$ will be such that the rate of
  increase of the scores is the same $v_g=v^\star$ for all strategies
  with $f_g>0$. Strategies which are not played ($f_g=0$) have
  $\pi_g^{{\rm real}}+1<v^\star$. Considering the reaction of other
  agents does not modify these conclusions.} and leads agents to mix
their best strategy with less performing ones. This is a consequence
of the fact that agents neglect their impact on the market. It is now
clear why, given that the market impact reduces the perceived
performance $v_g$ of strategies by an amount which equals the
frequency $f_g$ with which strategies are played, agents can improve
their performance if they {\em reward} the strategy which they have
played by some extra points (the $\eta$ factor). This contributes a
term $\eta f_g$ to the rate of growth of strategy $g$ so \req{vs}
becomes $v_g=\pi_g^{{\rm real}}-(1-\eta)f_g+1$. Any $\eta>0$ reduces
the market impact and improves agent's performance. In particular for
$\eta=1$ agents properly account for the market impact and indeed in
this case the growth rate $v_s$ of their strategies do not depend on
the way they play.

\subsection{Exogenous vs endogenous information}\label{endo}

In the El Farol problem and in the MG the state $\mu(t)$ is determined
by the outcome of past games, as in \req{dyninfo}. In other words
$\mu(t)$ is an {\em endogenous} information which encodes information
on the game itself: agents record which has been the winning action in
the last $m=\log_2 P$ games and store this information in the binary
representation of the integer $\mu$. How do the results which we
derived for {\em exogenous} information, i.e. when $\mu$ is just
randomly drawn at each time, change if we go back to endogenous
information?

This issue has been the subject of much debate and considerable
analytical and numerical work was required to settle it. We will limit
ourselves here to a sketch of the line of reasoning and of the
results. As we said, it was at first believed, based on computer
simulation, that the MGs with exogenous and endogenous information
yield the same macroscopic pictures. However the situation turned out
to be more subtle. In fact, \req{dyninfo} implies that the dynamics of
$\mu(t)$ depends on the collective behavior of the game outcome
$A(t)$. The key quantity to understand the dynamics of information
patterns is the stationary state distribution of the process $\mu(t)$
which is induced by the dynamics of $A(t)$. As in the El Farol model,
this process is a diffusion on a De Bruijn graph, where the transition
probabilities depend on the statistics of $A(t)$ conditional on a
particular site $\mu$ of the graph. When the dynamics of $A(t)$ has a
strong stochastic component, which occurs when many agents play in a
probabilistic fashion (i.e. when $|m_i|=|\avg{s_i}|< 1$), all possible
transition $\mu\to \mu'$ occur with a positive, finite
probability. Hence the stationary state distribution has a support on
all the states $\mu\in\{1,\ldots, P\}$. At odd with the case of
exogenous information, some state may be visited more often than some
other state, but all states are visited. This leads ultimately to the
same qualitative scenario as in the completely random case and
explains the early numerical finding on the irrelevance of the origin
of the information in the MG\footnote{Rather than the origin of
information, Ref. \cite{Cavagna} speaks of irrelevance of memory. The
term ``memory'' is used in an improper way. Actually the memory of
agents is stored into their scores $U_{ig}(t)$.}. Roughly speaking,
one can say that this scenario holds whenever \be
\frac{1}{N}\sum_{i=1}^N m_i^2<1, \ee which in sufficient to ensure
that agents behave in a probabilistic way. 

\begin{figure}[t]
\begin{center}
\includegraphics*[width=8cm]{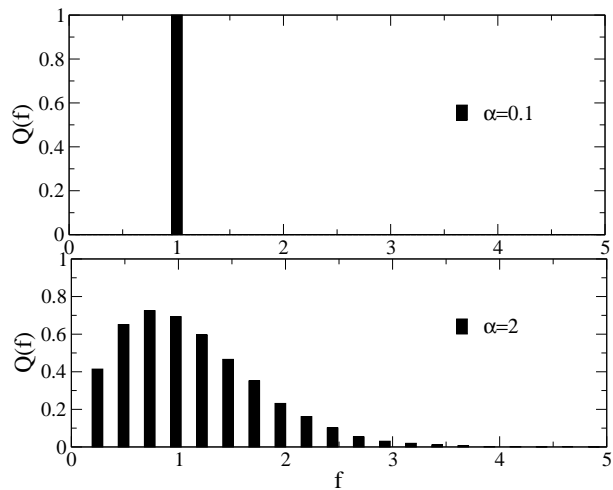}
\caption{\label{p_rho}Relative distribution of frequencies $Q(f)$ for
at $\alpha=0.1$ (top) and $\alpha=2$ (bottom). Simulations performed
with $\alpha N^2=30000$, with averages over 100 disorder samples per
point.}
\end{center}
\end{figure}

To be more precise, one can analyze the steady-state distribution of
history frequencies $\rho(\mu)$ relative to the uniform case, which is
given by
\begin{equation}
Q(f)=\frac{1}{P}\sum_\mu\delta[f-P\rho(\mu)]
\end{equation}
(if $\rho(\mu)=1/P$ for all $\mu$, $Q(f)$ is a delta-distribution at
$f=1$) as was done e.g. in \cite{relev}. This quantity is reported in
Fig. \ref{p_rho}.
One sees that in the supercritical regime the distribution is indeed
not uniform. This explains why, from a quantitative viewpoint it turns
out that macroscopic observables actually depend on the type of
information in the asymmetric regime $\alpha>\alpha_c$ where the
deviations of the history frequency distribution from uniformity are
more significant. The arguments just described, though approximate,
are able to account for these deviations rather well. Recently, the
dynamics of the MG with endogenous information was solved exactly by
the generating functional method \cite{coolong}, confirming the
general picture outlined above.

Clearly, the situation changes drastically when one considers the MG
corrected for the market impact with $\eta=1$. We know that all agents
ultimately freeze in this case, so that the learning dynamics
converges to a state with $Q=1$, i.e. with no stochastic fluctuations;
therefore in the long run $A(t)$ becomes a function of $\mu(t)$ alone.
This means that the dynamics of $\mu(t)$ becomes deterministic: it
locks into periodic orbits of the order of $\sqrt{P}$ values of $\mu$.
As a consequence, only a tiny fraction of information patterns are
generated by the dynamics of $A(t)$ and these few on the periodic
orbit are visited uniformly (one after the other). This dynamic
reduction of the size information space from $P$ to a number of order
$\sqrt{P}$ implies a similar reduction of the effective value of
$\alpha$ to something close to $0$. Given that $\sigma^2/N$ decreases
with $\alpha$, we conclude that the performance of the system with
endogenous information improves with respect to the case of exogenous
information. For intermediate values of $\eta$ and endogenous
information the system interpolates between the two extreme behaviors
of the standard MG ($\eta=0$) -- where the origin of information is to
some extent irrelevant -- and of the sophisticated agents ($\eta=1$)
case -- where a dynamic selection of a small subset of states $\mu$
occurs.

\section{Extensions and generalizations}

We shall discuss now a few variations on the MG theme, mostly inspired
by problems related to financial markets, in particular by the origin
of the peculiar intermittent and non-Gaussian (`fat tailed')
fluctuation patterns they generate.  In the reference model of price
dynamics, which is the simplest one accounting for no-arbitrage
hypothesis and market's efficiency, the logarithm of prices performs a
random walk and hence returns are gaussian. On the other hand, several
complex agent-based models are able to reproduce a realistic
phenomenology to a high degree but with little analytic control. In
the context of MGs we shall see that heavy tails in the distribution
of returns and clustering in time emerge close to the phase
transition, which suggests that markets operate close to
criticality. Realistic behaviour persists also when agents have a
finite score memory, but it disappears as soon as agents account for
their market impact. We shall also briefly discuss MGs with many
assets, in which agents have to choose among several assets with
different information content. Then we shall move on to Majority Games
and review the properties of mixed models in which fundamentalists and
trend-followers interact. A discussion of a model with asymmetric
(private) information closes the section.

\subsection{Grand-canonical Minority Game and stylized facts}\label{gcmgth}

The following model introduces volume fluctuations in the MG, as the
number of agents involved in the game varies from one time step to the
next. In the grand-canonical MG \cite{gcmg}, each agent $i$ has at his
disposal only one quenched random trading strategy
$\bsy{a}_i=\{a_i^\mu\}$ and has to choose whether to join the market
($\phi_i(t)=1$) or not ($\phi_i(t)=0$) at every time step. In order to
make this decision the agent compares the expected profit from joining
the market to a fixed standard. The model is completely defined by the
following scheme:
\begin{eqnarray}
\phi_i(t)=\theta[U_i(t)]\nonumber\\
A(t)=\sum_i\phi_i(t)a_i^{\mu(t)}\\
U_i(t+1)-U_i(t)=-a_i^{\mu(t)}A(t)-\epsilon_i\nonumber
\end{eqnarray}
The quantity $\epsilon_i$ represents the benchmark: $\epsilon_i<0$
means that agents have an incentive to take part in the market
because, for instance, they are urged to sell or exchange assets;
$\epsilon_i>0$ implies that agents receive a fixed positive payoff by
staying away from the market, like a fixed interest from a bank.
Alternatively, $\epsilon_i$ can be seen as the a priori incentive of
agent $i$ to enter the market: if $\epsilon_i<0$
(resp. $\epsilon_i>0$) the agent has a small incentive to enter
(resp. stay out). One can consider two different types of agents: {\it
producers}, who always enter the market and are characterized by
$\epsilon_i=-\infty$; and {\it speculators}, who instead aim at taking
profit of fluctuations and are characterized by a finite $\epsilon_i$.
We set
\begin{eqnarray}
\epsilon_i=\epsilon~~~~~~~~~~~{\rm for~} 1\leq i\leq N_s\nonumber\\
\epsilon_i=-\infty~~~~~~~{\rm for~} N_s+1\leq i\leq N_s+N_p\equiv N\nonumber
\end{eqnarray}
where $N_s$ and $N_p$ stand for the number of speculators and
producers, respectively. Speculators act on the market only if they
expect to receive a payoff higher than the benchmark; producers act no
matter what.

\begin{figure}[t]
\centering
\includegraphics[width=.45\textwidth]{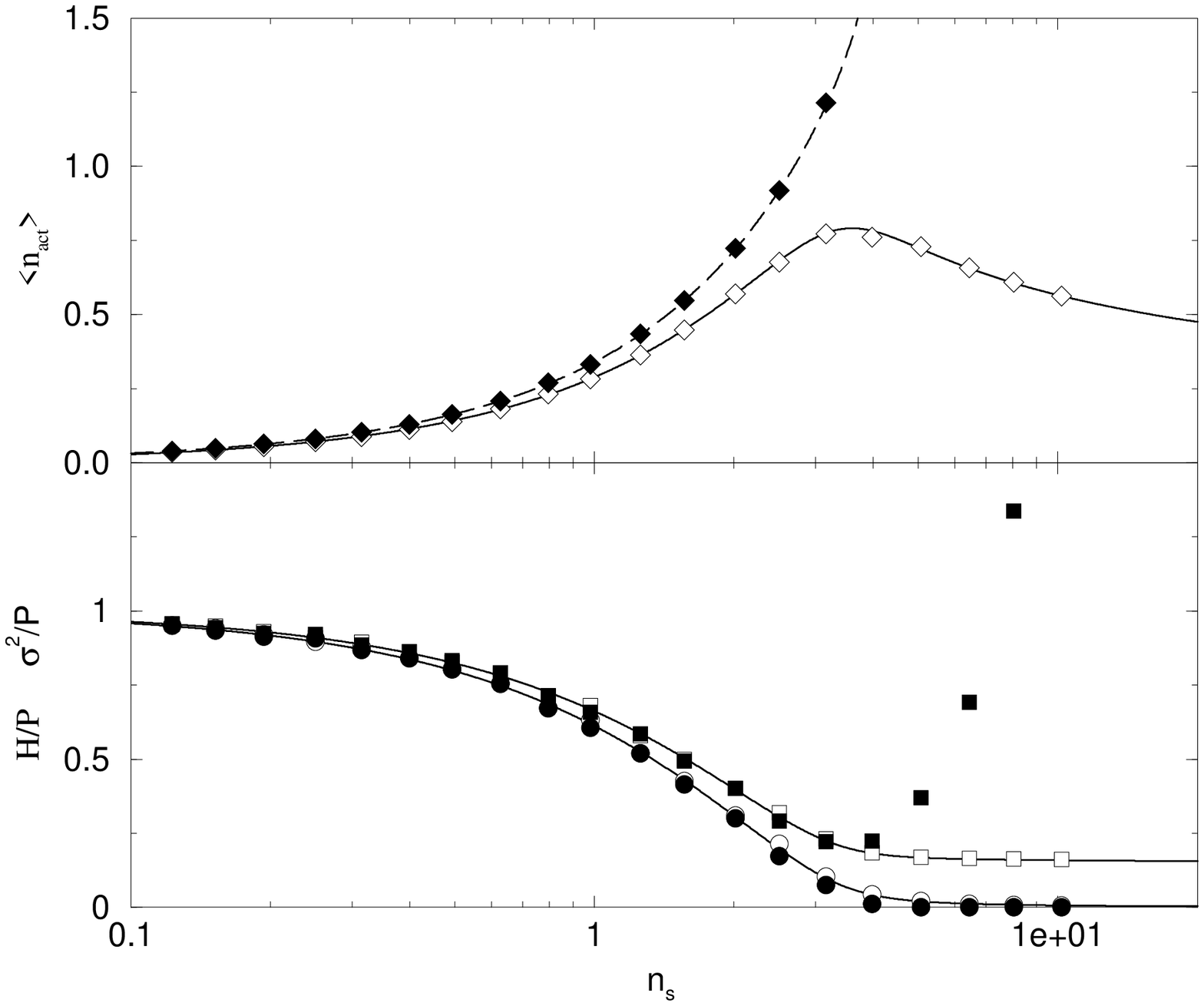}
\includegraphics[width=.45\textwidth]{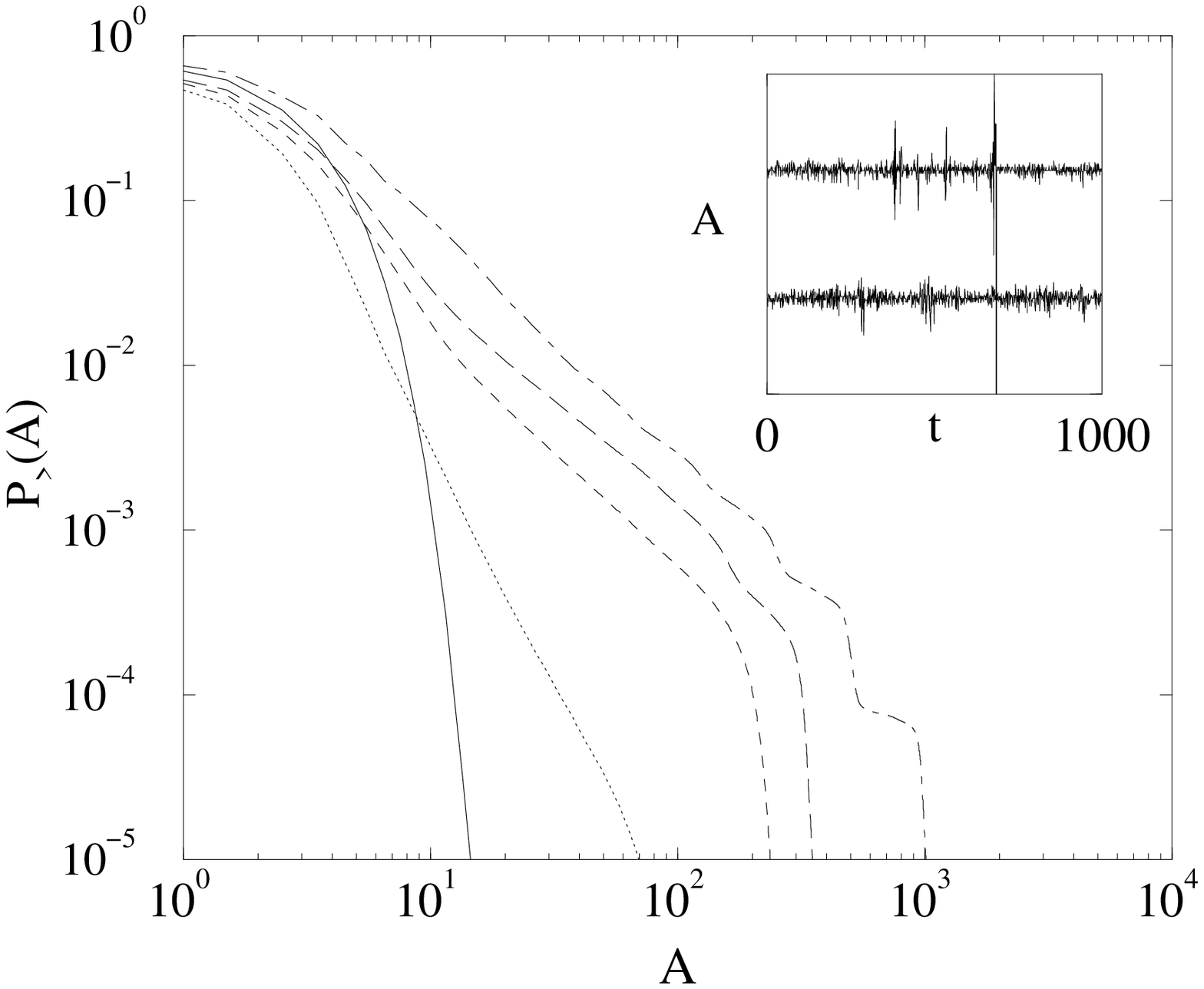}
\caption{Left panel: relative number of active agents (top),
volatility and predictability per pattern (bottom) as a function of
$n_s$ for $\epsilon=0.1$ (open markers) and $\epsilon=-0.01$ (full
markers). Right panel: cumuluative probability distribution
$P_>(A)=\prob\{|A(t)|>x\}$ versus $x$ in the steady state. Inset: time
series $A(t)$ versus $t$ for $n_s=20$ (top) and $n_s=200$
(bottom). From \cite{gcmg}.}
\label{figgcmg}
\end{figure}

The relevant control parameters are the relative number of speculators
and producers, respectively: $n_s=N_s/P$ and $n_p=N_p/P$. As usual,
one is interested in the behavior of the volatility $\sigma^2$ and of
the predictability $H$. Besides, it is interesting to analyze also the
relative number of active speculators, defined as \be
n_{act}=\frac{1}{P}\sum_i \avg{\phi_i} \ee Results are shown in 
Fig. \ref{figgcmg}.
On sees that with a fixed number $n_p$ of producers, the market
becomes more and more unpredictable, i.e. $H$ decreases, as the number
$n_s$ of speculators increases, independently of the value of
$\epsilon$. At the same time also the volatility $\sigma^2$ decreases
as agents play in an increasingly coordinated way. In a market with
few speculators ($n_s<1$ in Fig.), most of the fluctuations in $A(t)$
are due to the random choice of $\mu(t)$ (i.e.  $\sigma^2\simeq H$)
and the number $n_{\rm act}$ of active speculators grows approximately
linearly with $n_s$. When $n_s$ increases further, the market reaches
a point where it is barely predictable. Now the collective behavior
becomes $\epsilon$-dependent:
\begin{itemize}
\item for $\epsilon<0$ the relative number of active speculators
continues growing with $n_s$ even if the market is unpredictable
$H\simeq 0$. The volatility $\sigma^2$ has a minimum and then it
increases with $n_s$
\item for $\epsilon>0$, instead, the relative number of active traders
  decreases and finally converges to a constant. This means that the
  market becomes highly selective: only a negligible fraction of
  speculators trade ($\phi_i(t)=1$) whereas the majority is inactive
  ($\phi_i(t)=0$). The volatility $\sigma^2$ also remains roughly
  constant in this limit
\end{itemize}
In other words, $\epsilon=0$ for $n_s\ge n_s^\star(n_p)$
($n_s^\star(1)=4.15\ldots$) is the locus of a first order phase
transition across which $N_{\rm act}$ and $\sigma^2$ exhibit a
discontinuity.

So far for collective properties; what about stylized facts? Numerical
simulations reproduce anomalous fluctuations similar to those of real
financial markets close to the phase transition line. As shown in Fig.
\ref{figgcmg}, the distribution of $A(t)$ is roughly Gaussian for
small enough $n_s$ (it {\it must} tend to a Gaussian when $n_s\to 0$),
and has fatter and fatter tails as $n_s$ increases. The same behavior
is seen for decreasing $\epsilon$: fat tails emerge in the vicinity of
the critical point. In particular the distribution of $A(t)$ shows a
power law behavior $P(|A|>x)\sim x^{-\beta}$ with an exponent which
can be estimated to be $\beta\simeq 2.8, 1.4$ for $n_s=20, 200$
respectively and $\epsilon=0.01$. With $n_s=100$ the exponent takes
values $\beta\simeq 1.4, 2.3, 3.1$ for $\epsilon= 0.01, 0.1, 0.5$.
Note that empirical values of $\beta$ typically range from 2 to 4.
Finally: volatility clustering is observed in conjunction with the
power-law tails (see inset).

\begin{figure}[t]
\centering
\includegraphics[width=8cm]{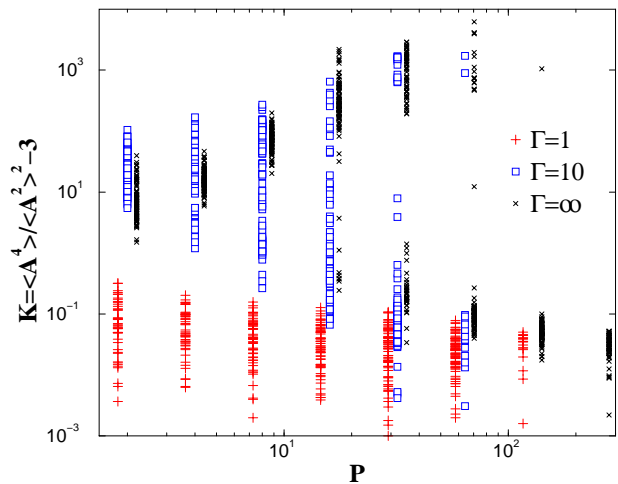}
\caption{Kurtosis of $A(t)$ in simulations with $\epsilon=0.01$,
$n_s=70$, $n_p=1$ and several different system sizes $P$ for
$\Gamma=1,10$ and $\infty$.}
\label{figkurt}
\end{figure}

Let us analyze more closely the emergence of power-law tails in the
distribution of $A(t)$ and of volatility clustering. In Fig.
\ref{figkurt} the kurtosis excess (if $x$ is a generic random variable
with zero mean, $K$ is defined as $K=\frac{\avg{x^4}}{\avg{x^2}^2}-3$;
loosely speaking, it is a convenient proxy for the distance of a
certain distribution from a Gaussian, for which $K=0$) $K$ of the
distribution is shown as a function of the system size and of the
learning rate $\Gamma$ for a `regularized' model with choice rule \be
\prob\{\phi_i(t)=1\}=1/[1+e^{-\Gamma U_i(t)}] \ee
One sees that as the system size increases (or if one introduces a
small enough learning rate $\Gamma$,see below) the distribution tends
to a Gaussian as $K$ decreases with $P$. Moreover we see that for a
rage of parameters the appearance of fat tails is sample-dependent, as
both samples with and without fat tails may occur.

This behaviour is reminiscent of well-known finite-size effects in the
theory of critical phenomena: in the $d$-dimensional Ising model, for
example, at temperature $T=T_c+\gamma$ critical fluctuations (e.g. in
the magnetization) occur as long as the system size $N$ is smaller
than the correlation volume $\sim\gamma^{-d\nu}$. But for
$N\gg\gamma^{-d\nu}$ the system shows the normal fluctuations of a
paramagnet. Some light on the finite-size effects in our case can be
shed by studying the continuous-time limit of the score updating
dynamics.  Regularizing the choice rule to \be
{\rm prob}\{\phi_i(t)=1\}=1/[1+e^{-\Gamma U_i(t)}] \ee with learning rate
$\Gamma$, and applying the machinery described in Sec. \ref{mgth}, one
can transform the discrete-time learning dynamics into the
continuous-time Langevin process
\begin{eqnarray}
\dot U_i(t)=-\ovl{a_i\avg{A}_y}-\epsilon+\eta_i(t)\label{gcmgctl}\\
\avg{\eta_i(t)\eta_j(t')}=\frac{\sigma^2}{N}\ovl{a_i a_j}\delta(t-t')
\end{eqnarray}
Notice that the noise strength is proportional to the time dependent
volatility $\sigma^2=\avg{A^2}$. The noise term is a source of
correlated fluctuations because $\ovl{a_ia_j\avg{A^2}}/N\simeq
1/\sqrt{N}$ is small but non zero, for $i\neq j$ if $N$ is finite.
This noise competes with the deterministic part of \req{gcmgctl}: if
the former outweighs the latter, then one expects that the dynamics
will sustain collective correlated fluctuations in the $U_i(t)$ which
otherwise would be washed away. In order to obtain an approximate
analytic condition for the onset of volatility clustering one may then
compare the noise correlation term, which is of order
$\ovl{a_ia_j\avg{A^2}_{y}}/N\sim\sigma^2/P^{3/2}$ for $i\neq j$, with
the square of the deterministic term of \req{gcmgctl}, which is given
by $\l[\ovl{a_i\avg{A}_{y}}+\epsilon\r]^2\simeq
\l[\sqrt{H/P}+\epsilon\r]^2$. Rearranging terms, one finds that
volatility clustering can be expected to set in when \be
\frac{H}{\sigma^2}+2\epsilon\sqrt\frac{H}{P}\frac{P}{\sigma^2}+
\epsilon^2\,\frac{P}{\sigma^2} \simeq\frac{B}{\sqrt{P}}
\label{condvolclus}
\ee where $B$ is a constant. This prediction finds remarkable
confirmations in numerical experiments \cite{gcmg}. Recalling the
analogy with magnetic systems made at the beginning of this section,
one understands that \req{condvolclus} and $H/P\sim \epsilon^2$ imply
that the same occurs in the GCMG with $d\nu=4$. In other words, {\it
the critical window shrinks as $N^{-1/4}$ when $N\to\infty$}.
However, because of the long range nature of the interaction,
anomalous fluctuations either concern the whole system or do not
affect it at all. In the critical region the Gaussian phase coexists
probabilistically with a phase characterized by anomalous
fluctuations. This, like the discontinuous nature of the transition at
$\epsilon=0$, is typical of first order phase transitions.

\subsection{Market ecology}

One of the first modification of the MG has investigated the effects
of introducing an explicit asymmetry in the two possible actions
\cite{jasymm}. This is the case of the El Farol bar problem: the
actions `go' or `don't go' to the bar are not symmetric because (i) if
one takes the wrong action there is still a difference between going
to a crowded bar and not going to an uncrowded bar and (ii) the
comfort level corresponds to a share of 60\% of agents attending. If
each agent takes the opposite choice one ends up in an inefficient
attendance of 40\%. The outcomes of the MG are instead symmetric: If
every agent switches to the opposite choice, all the payoffs remain
unchanged. Quite generally this leads to study games where the payoffs
to agent $i$ at time $t$ is given by \be \pi_i(t)=-a_{i
g_i(t)}^{\mu(t)}\l[A_0^{\mu(t)}+\sum_j a_{j g_j(t)}^{\mu(t)}\r]
\label{asymmMG}
\ee where $\bsy{A}_0=\{A_0^\mu\}$ is some fixed vector. In particular,
\cite{jasymm} investigated the case where $A_0^\mu=L$ independently of
$\mu$, as in the El Farol bar, and where information is endogenous.
Interestingly, because of the fact that due to \req{dyninfo} some
values of $\mu$ occur more often than others, the conclusion that the
collective behavior is independent of whether the information $\mu$ is
endogenously generated or is exogenous (i.e. random), which was
roughly correct for the standard MG, is not true in this case.

There is however a second motivation for considering a model based on
\req{asymmMG} which was explored in \cite{CMZha,ccmz}. Considering the
MG as a model of a financial market, it can be argued that there are
different types of market participants with different goals. Some
trade to gain money from transactions with no particular interest in
the asset they buy and sell. Only price fluctuations matter for this
kind of traders, which one usually calls `speculators'.  Another type
of market participants are those who use the market for exchanging
goods. This is indeed the reason why markets exist. This type of
agents is interested in the asset itself: they will buy it or sell it
irrespective of the history of recent fluctuations: this type of
agents can be called {\em producers}. While speculators have a range
of behavioral rules which process the available information in search
of {\em arbitrage opportunities}, producers use a trading rule which
is constant in time. Producers are part of the financial world and
their behavior is correlated with the state of the world $\mu$ which
is thought to capture all relevant economic information: in other
words, they only have one strategy at their disposal. This type of
traders play a role similar to that of {\em hedgers}\footnote{A hedge
is an action (e.g. buy/sell) done with the aim of reducing the risk of
another action.}: they inject information into the market. Their
trading activity is completely predictable given the state of the
world $\mu$ and the term $A_0^\mu$ represents their aggregate
contribution to the market.

It is easy to understand that in a market composed of producers only
the distribution of price changes would be nearly Gaussian: in fact,
$A_0^\mu$ can be regarded as the sum of $N_p$ random terms, where
$N_p$ is the number of producers. The process associated to producers
can be considered as the {\em fundamentals}, i.e. the price process
which reflects the economic performance of the asset. Roughly
speaking, one may expect that speculative trading will {\em color}
this process and transform its statistical properties. Actually the
discussion may be extended to a further type of agents, the so-called
{\em noise traders}. These persons totally disregard the state of the
world $\mu$ or have no information at all on it. They rather follow
rules of behavior which are statistically uncorrelated with $\mu$
(such as the moon phases) and with the behavior of other agents. The
presence of these agents does not introduce any new qualitative
features. The question is: how do all these ``species'' of traders
interact?

\begin{figure}[t]
\centering
\includegraphics*[width=8cm]{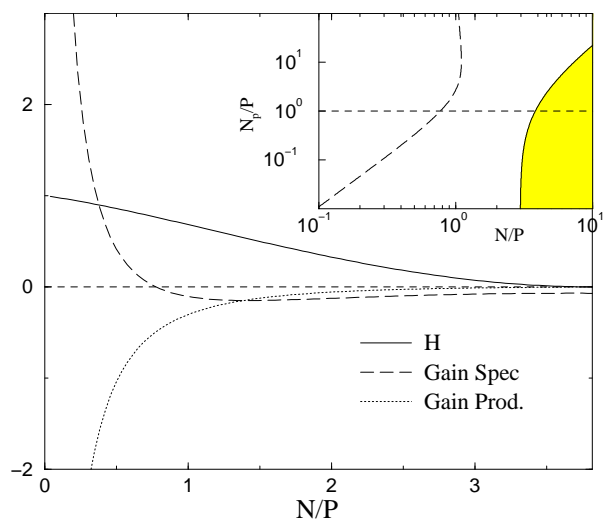}
\caption{Average gains of producers and speculators as a funcion of
the (reduced) number $N/P$ of adaptive agents (speculators). The plot
refers to a system with $N_p=P$ passive agents (producers). The gain
of speculators is positive only when they are few and it decreases
when new speculatos join the market. Producers losses are reduced by
speculators. The predicatbility $H$ is also plotted. Inset: Phase
diagram in the space of the reduced numbers of speculators and
producers. The shaded region to the right of the solid line is the
symmetric phase where $H=0$. The gain of speculators vanishes on
dashed line and it is positive in the region to the left.}
\label{nsnpw2k}
\end{figure}

An intuitive argument runs more or less as follows. First, note that
in a market composed of producers price changes would depend only on
$\mu$. Such a highly predictable market is very favorable for
speculators who may derive considerable gains. However when more and
more speculators join the market, its predictability decreases and the
profit of speculators gets more and more meager. This effect is
illustrated in Fig. \ref{nsnpw2k}, which also shows that producers
instead benefit from the presence of speculators because their losses
are reduced.
When the number of speculators increases beyond a critical value,
which depends on the relative number $N_p/P$ of producers, the market
enters the symmetric phase where $H=0$ and the outcome $A(t)$ becomes
unpredictable from $\mu$. This shows that the relation between these
two species is more similar to symbiosis than to competition:
producers feed speculators by injecting information in the market and
benefit, in their turn, of the liquidity provided by speculators.

\subsection{Multi-asset Minority Games}\label{mamg1}

\subsubsection{Definitions and results}

Minority Games with many assets have been introduced in order to
investigate how speculative trading affects the different assets in a
market \cite{Rodgers, Chau}. A tractable version of these models has
been considered in \cite{new}, with the aim of studying how agents
modify the composition of their portfolios depending on the
`complexities' or information contents of the different assets.

The model consists essentially of two coupled MGs with one strategy
each. Let us consider the case of a market with two assets
$\gamma\in\{-1,1\}$ and $N$ agents. At each time step $\ell$, agents
receive two information patterns $\mu_\gamma\in\{1,\ldots,P_\gamma\}$,
chosen at random and independently with uniform probability. As
always, $P_\gamma$ is taken to scale linearly with $N$, and their
ratio is denoted by $\alpha_\gamma=P_\gamma/N$. Every agent $i$
disposes of one trading strategy per asset,
$\bsy{a}_{i\gamma}=\{a_{i\gamma}^{\mu_\gamma}\}$, that prescribe an
action $a_{i\gamma}^{\mu_\gamma}\in\{-1,1\}$ (buy/sell) for each
possible information pattern of asset $\gamma$. Each component
$a_{i\gamma}^{\mu_\gamma}$ is selected randomly and independently with
uniform probability and is kept fixed throughout the game. Traders
keep tracks of their performance in the different markets through a
score function $U_{i\gamma}(\ell)$. The behavior of agents is
summarized by the following rules: 
\begin{eqnarray}\label{learn1}
s_i(t)=\sign[y_i(t)]\nonumber\\
A_\gamma(t)=\sum_{j=1}^N
a^{\mu_\gamma(t)}_{j\gamma}\delta_{s_j(t),\gamma}\\
U_{i\gamma}(t+1)-U_{i\gamma}(t)=-a^{\mu_\gamma(t)}_{i\gamma}
A_\gamma(t)/\sqrt{N}\nonumber
\end{eqnarray}
where $A_\gamma(t)$ represents the `excess demand' or the total bid of
asset $\gamma$, while $y_i(t)=\sum_\gamma\gamma
U_{i\gamma}(t)$. The Ising variable $s_i$ indicates the asset in
which player $i$ invests at time $t$, which is simply the one with the
largest cumulated score. As usual, it is the minus sign on the
right-hand side of (\ref{learn1}) that enforces the minority-wins rule
in both markets.  It is possible to characterize the asymptotic
behaviour of the multi-agent system (\ref{learn1}) with a few
macroscopic observables. In the present case, besides traditional
observables such as the predictability $H$ and the volatility
$\sigma^2$, defined respectively as
\begin{eqnarray}
  H=\sum_{\gamma\in\{-1,1\}}\frac{1}{NP_\gamma}\sum_{\mu_\gamma=1}^{P_\gamma}
  \avg{A_\gamma|\mu_\gamma}^2=H_++H_-\label{acca}\\
  \sigma^2=\frac{1}{N}\sum_\gamma\avg{A_\gamma^2}=
  \sigma^2_++\sigma^2_-\label{s2}
\end{eqnarray}
it is important to analyze the relative propensity of traders to
invest in a given market, namely \be
m=\frac{1}{N}\sum_{i=1}^N\avg{s_i}
\label{parapa}\ee A positive (resp. negative) $m$ indicates that
agents invest preferentially in asset $+1$ (resp. $-1$). 

\begin{figure}
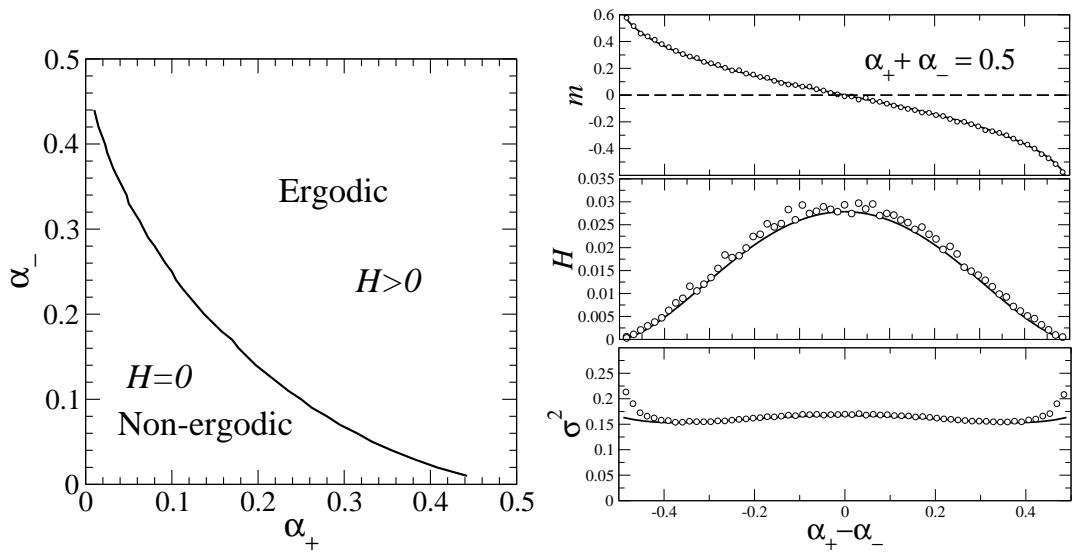

\centering
\includegraphics*[width = 7cm]{phased_can.eps}
\includegraphics*[width = 7cm]{FIGURA_MA.eps}
\caption{Left panel: analytical phase diagram of the canonical
two-asset Minority Game in the $(\alpha_+,\alpha_-)$ plane. Right
panel: behavior of $m$ (top), $H$ (middle) and $\sigma^2$ (bottom)
versus $\alpha_+-\alpha_-$ for $\alpha_++\alpha_-=0.5$. Markers
correspond to simulations with $N=256$ agents, averaged over 200
disorder samples per point. Lines are analytical results (from
\cite{new}).}
\label{Phase_diagram.fig}
\end{figure}

 The phase structure of the model is displayed in Fig.
\ref{Phase_diagram.fig}.
The $(\alpha_+,\alpha_-)$ plane is divided in two regions separated by
a critical line. In the ergodic regime, the system produces
exploitable information, i.e. $H>0$, and the dynamics is ergodic, that
is the steady state turns out to be independent of the initialization
$U_{i\gamma}(0)$ of (\ref{learn1}). Below the critical line, instead,
different initial conditions lead to steady states with different
macroscopic properties (e.g. different volatility), but traders manage
to wash out the information and the system is unpredictable
($H=0$). This scenario essentially reproduces the standard MG phase
transition picture.

The behaviour of the macroscopic observables $m$, $H$ and $\sigma^2$
along the cut $\alpha_++\alpha_-=1/2$ (in the ergodic phase) is also
reported in Fig. \ref{Phase_diagram.fig}.  One sees that agents play
preferentially in the market with smaller information complexity,
which is particularly inconvenient as it coincides with the one with
less exploitable information. This is a somewhat paradoxical result
since a na\"\i ve argument would suggest that agents are attracted by
information rich markets. It actually turns out that this simple
argument is incorrect and the observed behavior is due to the fact
that agents are constrained to trade in one of the two markets.
Rather than seeking the most profitable asset, agents simply escape
the asset where their loss is largest. The conclusion is indeed
reversed when traders may stay out of the market and have negative
incentives to trade (that is, when they have an incentive not to
trade). In this case, which corresponds to a grand-canonical
multi-asset MG, the information-rich asset is chosen preferentially
\cite{new}, though the phase structure becomes more complex than usual
as new phases (with broken ergodicity {\it and} global predictability)
arise. Note however that in this framework no correlations among the
assets emerge, i.e.  $\avg{A_\gamma A_{-\gamma}}=0$. Indeed
\begin{equation}\label{Acor}
  \avg{A_+ A_-}=\sum_{i,j}\avg{a_{i+}^{\mu_+}
a_{j-}^{\mu_-}\frac{1+s_i}{2}\frac{1-s_j}{2}}
\end{equation}
Now, the dynamical variables $U_{i\gamma}(t)$ evolve on timescales
much longer (of order $P_\gamma$) than those over which the
$\mu_\gamma$ evolve. Hence we can safely assume that the distribution
of $s_i$ is independent of $\mu_\gamma$ and factorize the average
$\langle a_{i,+}^{\mu_+}a_{j,-}^{\mu_-}\rangle =\langle
a_{i,+}^{\mu_+}\rangle\langle a_{j,-}^{\mu_-}\rangle$ over the
independent information arrival processes $\mu_{\pm}(t)$. Given that
$\avg{a_{i,\pm}^{\mu_{\pm}}}\simeq 0$ the conclusion $\avg{A_+
A_-}\simeq 0$ follows immediately. The reason for this is that
traders' behavior is aimed at detecting excess returns in the market
with no consideration about the correlation among assets. This
conclusion is against the empirical evidence, as in real financial
markets correlation between stocks are overwhelmingly positive (if it
wasn't so, making money in a financial market would be much
easier!). The microscopic origin of this phenomenon is a rather
difficult issue, which will surely receive much attention in the near
future.

Below we describe the dynamical solution of this model, as an example
of the application of the path-integral formalism to this type of
problems.

\subsubsection{Dynamics (path-integral approach)}

The dynamical approach to the stationary macroscopic properties of
Minority Games is based on the use of dynamical generating functionals
\`a la Martin-Siggia-Rose \cite{MSR} to turn the original multi-agent
process into a single stochastic equation for the behavior of a single
`effective agent', similarly to what is done to study the dynamics of
spin systems with quenched disorder after \cite{dedo}.  This
procedure, which was first applied to Minority Games in
\cite{HeimCool}, allows ultimately to derive closed equations for
correlation functions, response functions, and all other relevant
time-dependent macroscopic parameters. Typically, the resulting
equations are too complicated to be solved at all times. However, with
suitable Ans\"atze one may restrict the analysis to specific solvable
regimes (in this case, we shall focus on ergodic steady
states). Dynamical phase transitions can then be identified from the
breakdown of the assumed behavior. The method is very general, it
doesn't rely on the existence of a Hamiltonian nor on the validity of
detailed balance, but requires an analytical {\it tour de force} for
solving the most general MGs. Luckily, some reasonable starting
simplification help to make it less cumbersome. One is Markovianness,
which in MGs corresponds to models with random external
information. Another is changing the updating rule from the usual
`on-line' learning, in which agents modify their preferences at each
time step, to a `batch' learning, in which agents update their
preferences only after they have seen all possible information
patterns Strictly speaking, the batch process is not equivalent to the
on-line process but in many cases, including that which we consider
here, the two are qualitatively identical. Both simplifications will
be made in this section, where we expound the dynamical solution of
the canonical multi-asset MG. The method is described in detail for
other models and more general cases in \cite{coolen}.

So we consider two coupled GCMGs, interpreted as a system with two
assets characterized by different sizes of information sets and, on
the agents' side, by different strategies and valuation
functions. From \req{learn1}, one sees that the preferences evolve
according to
\begin{equation}
y_i(t+1)-y_i(t)=-\sum_{\gamma\in\{-1,1\}}\gamma
a_{i\gamma}^{\mu_\gamma(t)} A_\gamma(t)/\sqrt{N}
\end{equation}
The `batch' approximation is obtained by averaging the right-hand side
over the $\mu_\sigma$'s. This leads, after a time re-scaling (for
simplicity, we denote the re-scaled time again by $t$), to
\begin{equation}\label{batchcan}
y_i(t+1)-y_i(t)=-\sum_{\gamma\in\{-1,1\}} 
n_\gamma \sum_{j=1}^N J_{ij}^\gamma \phi_{j\gamma}(t)
\end{equation}
where $n_\gamma=1/\alpha_\gamma$ and
$J_{ij}^\gamma=(1/N)\sum_{\mu_\gamma} a_{i\gamma}^{\mu_\gamma}
a_{j\gamma}^{\mu_\gamma}$ are quenched random couplings of Hebbian
type. We also introduced the variable
\begin{equation}
\phi_{i\gamma}(t)=\gamma\delta_{s_i(t),\gamma}=
\frac{1}{2}\l[\gamma+s_i(t)\r]
\end{equation}
All moments like $m_i(t)=\avg{s_i(t)}$ and
$c_{ij}(t,t')=\avg{s_i(t)s_j(t')}$ -- the brackets standing for an
average over all possible time evolutions of the system -- and in turn
macroscopic quantities like the magnetization $m=\davg{\sum_i
m_i(t)/N}$ or the autocorrelation function $C(t,t')=\davg{\sum_i
c_{ii}(t,t')/N}$ can be derived formally from the generating
functional
\begin{equation}
Z[\bsy{\psi}]=\avg{e^{\ii\sum_t\bsy{\psi}(t)\cdot\bsy{s}(t)}}
\end{equation}
by taking suitable derivatives with respect to the auxiliary
generating fields $\bsy{\psi}=\{\psi_i\}$; for instance
\begin{equation}
C(t,t')=-\frac{\ii}{N}\sum_i\lim_{\bsy{\psi}\to \bsy{0}}
\frac{\partial^2\davg{Z[\bsy{\psi}]}}{\partial\psi_i(t)\partial\psi_i(t')}
\end{equation}
 The $\avg{\cdots}$ average is performed by imposing that the $s_i$
satisfy \req{batchcan} at each time step:
\begin{equation}\label{zed}\fl
\davg{Z[\boldsymbol{\psi}]}=\int p[\boldsymbol{y}(0)]~ e^{\ii\sum_t
\boldsymbol{\psi}(t)\cdot \boldsymbol{s}(t)}\davg{\prod_t
W[\boldsymbol{y}(t)\to\boldsymbol{y}(t+1)]} d\boldsymbol{y}(t)
\end{equation}
with transition matrix fixed by (\ref{batchcan}):
\begin{equation}\fl
W[\boldsymbol{y}(t)
\to\boldsymbol{y}(t+1)]=\prod_i\delta\l[y_i(t+1)-y_i(t)-h_i(t)+
\sum_{\gamma\in\{-1,1\}} n_\gamma \sum_{j=1}^N J_{ij}^\gamma
\phi_{j\gamma}(t)\r]\nonumber
\end{equation}
The fields $h_i(t)$ will be used to generate response functions. At
this point the following steps need to be taken:
\begin{enumerate}
\item[a.] Introduce the order parameters
\begin{eqnarray}
Q(t,t')=\frac{1}{N}\sum_{i=1}^N s_i(t) s_i(t')\nonumber\\
L(t,t')=\frac{1}{N}\sum_{i=1}^N \widehat{y}_i(t) \widehat{y}_i(t')\nonumber\\
K(t,t')=\frac{1}{N}\sum_{i=1}^N s_i(t)\widehat{y}_i(t')\\
a(t)=\frac{1}{N}\sum_{i=1}^N s_i(t)\nonumber\\
k(t)=\frac{1}{N}\sum_{i=1}^N \widehat{y}_i(t)\nonumber
\end{eqnarray}
in (\ref{zed}) via such identities as 
\begin{equation}
1=\int dQ(t,t')\delta\l[NQ(t,t')-\sum_{i=1}^N s_i(t)
s_i(t')\r];
\end{equation}
\item[b.] Use the integral representation for the
$\delta$-distributions;
\item[c.] Average over the quenched disorder after isolating the relevant
terms with the help of the variables
\begin{eqnarray}
x_\gamma^{\mu_\gamma}(t)=
\frac{1}{\sqrt{P_\gamma}}\sum_i \phi_{i\gamma}(t)
a_{i\gamma}^{\mu_\gamma}\\
w_\gamma^{\mu_\gamma}(t)=
\frac{1}{\sqrt{P_\gamma}}\sum_i
\widehat{y}_i(t)a_{i\gamma}^{\mu_\gamma}
\end{eqnarray}
\end{enumerate}
These steps require standard manipulations at most. After a
factorization over $i$ and $\mu_\gamma$, one arrives at 
\begin{equation}\label{zetaz}
\davg{Z[\boldsymbol{\psi}]}=\int
D\boldsymbol{\Theta}D\boldsymbol{\widehat{\Theta}}
~e^{N\l[\Psi(\boldsymbol{\Theta},\boldsymbol{\widehat{\Theta}})+
\Omega(\bsy{\widehat{\Theta}})+\Phi(\bsy{\Theta})\r]}
\end{equation}
where $\Theta(t,t')=\{Q(t,t'),L(t,t'),K(t,t'),a(t),k(t)\}$ is the
vector of order parameters,
$\widehat{\Theta}(t,t')=\{\widehat{Q}(t,t'),
\widehat{L}(t,t'),\widehat{K}(t,t'), \widehat{a}(t),\widehat{k}(t)\}$
is the conjugate vector of Lagrange multipliers, while the functions
$\Psi$, $\Phi$ and $\Omega$ are given by
\begin{eqnarray}\fl
\Psi=\ii \sum_t\l[a(t)\widehat{a}(t)+\ell(t)\widehat{\ell}(t)+\r]
\nonumber\\
+\ii \sum_{t,t'}\l[Q(t,t')\widehat{Q}(t,t')+
L(t,t')\widehat{L}(t,t')+K(t,t')\widehat{K}(t,t')\r]\\
\fl \Omega=\frac{1}{N}\sum_i\log\int\prod_t d\widehat{y}(t)dy(t)
p[y(0)]~e^{-\ii\sum_t\l[\widehat{a}(t)s(t)+\widehat{\ell}(t)
\widehat{y}(t)\r]}\nonumber\\\fl
\times e^{\ii\sum_i\psi_i(t)s(t)+\ii\sum_t\widehat{y}(t)\l[
y(t+1)-y(t)-h_i(t)\r]-\ii\sum_{t,t'}\l[\widehat{Q}(t,t')s(t)s(t')
+\widehat{L}(t,t')\widehat{y}(t)\widehat{y}(t')+
\widehat{K}(t,t')s(t)\widehat{y}(t')\r]}\\
\fl \Phi=\sum_\gamma\Big\{-\frac{\alpha_\gamma}{2}\log\|n_\gamma
\boldsymbol{D}_\gamma\| \nonumber 
\\\fl+\alpha_\gamma\log\int d\boldsymbol{\widehat{w}}
e^{-\frac{n_\gamma}{2}\sum_{t,t'}L(t,t')\widehat{w}_\gamma(t)
\widehat{w}_\gamma(t')-\frac{1}{2}
\sum_{t,t'}\l[\boldsymbol{A}_\gamma^T
(n_\gamma\boldsymbol{D}_\gamma)^{-1}\boldsymbol{A}_\gamma\r](t,t')
\widehat{w}_\gamma(t)\widehat{w}_\gamma(t')}\Big\}
\end{eqnarray}
where
\begin{eqnarray}
D_\gamma(t,t')=\frac{1}{4}\l[1+\gamma a(t)+\gamma a(t')+Q(t,t')\r]\\
A_\gamma(t,t')=\delta_{tt'}-\frac{\ii n_\gamma}{2} \l[\gamma
k(t')+K(t,t')\r]
\end{eqnarray}

In the limit $N\to\infty$ the integral (\ref{zetaz}) is dominated by
the saddle-point where the order parameters take the values
\begin{eqnarray}
C(t,t')=\avg{s(t)s(t')}_\star~~~~~~~
L(t,t')=\avg{\widehat{y}(t)\widehat{y}(t')}_\star\nonumber\\
K(t,t')=\avg{s(t)\widehat{y}(t')}_\star~~~~~~~
a(t)=\avg{s(t)}_\star\nonumber\\
k(t)=\avg{\widehat{y}(t)}_\star~~~~~~~~~~~~~~~~~
\widehat{C}(t,t')=\ii\frac{\partial\Phi}{\partial C(t,t')}\\~
\widehat{L}(t,t')=\ii\frac{\partial\Phi}{\partial L(t,t')}~~~~~~~~~~
\widehat{K}(t,t')=\ii\frac{\partial\Phi}{\partial K(t,t')}\nonumber\\
\widehat{a}(t)=\ii\frac{\partial\Phi}{\partial a(t)}~~~~~~~~~~~~~~~~~~~
\widehat{k}(t)=\ii\frac{\partial\Phi}{\partial k(t)}\nonumber
\end{eqnarray}
where 
\begin{equation}
\avg{\cdots}_\star=\frac{1}{N}\sum_i\frac{\int\cdots
M(\{y(t)\},\{\widehat{y}(t)\})\prod_t dy(t)d\widehat{y}(t)}{\int
M(\{y(t)\},\{\widehat{y}(t)\})\prod_t dy(t)d\widehat{y}(t)}
\end{equation}
denotes an average performed with the measure
\begin{eqnarray}\fl
M(\{y(t)\},\{\widehat{y}(t)\})=p[y(0)]~e^{\ii\sum_t\widehat{y}(t)\l[
y(t+1)-y(t)-h_i(t)\r]-\ii\sum_t\l[\widehat{a}(t)s(t)+\widehat{\ell}(t)
\widehat{y}(t)\r]}\nonumber\\\times
e^{-\ii\sum_{t,t'}\l[\widehat{C}(t,t')s(t)s(t')
+\widehat{L}(t,t')\widehat{y}(t)\widehat{y}(t')+
\widehat{K}(t,t')s(t)\widehat{y}(t')\r]}
\end{eqnarray}

Now comparing the above averages with the derivatives of $\davg{Z}$
with respect to $\bsy{\psi}$ and $\bsy{h}$ one easily sees that, in
the limit $N\to\infty$, $Q(t,t')$ may be identified with the
autocorrelation function $C(t,t')$, $a(t)$ turns out to coincide with
the magnetization $m(t)$, whereas $K(t,t')$ may be related to the
response function
\begin{equation}
G(t,t')=\lim_{N\to\infty}\frac{1}{N}\sum_i
\frac{\partial\davg{\avg{s_i(t)}}}{\partial h_i(t')}
\end{equation}
through $K(t,t')=\ii G(t,t')$. Working out the remaining equations,
and in particular the expression of $\Phi$, one finds in addition that
\begin{eqnarray}
\bsy{L}=\bsy{k}
=\boldsymbol{\widehat{C}}=\boldsymbol{\widehat{a}}=\boldsymbol{0}
\nonumber\\
\boldsymbol{\widehat{K}}^T=-\frac{1}{2}\sum_\gamma
\boldsymbol{A}_\gamma^{-1}~~~~~~~~~~\boldsymbol{\widehat{k}}=
-\frac{1}{2}\sum_\gamma\gamma\boldsymbol{A}_\gamma^{-1}\\
\boldsymbol{\widehat{L}}=-\frac{\ii}{2}\sum_\gamma\l[
\boldsymbol{A}_\gamma^{-1}(n_\gamma\boldsymbol{D}_\gamma)
\boldsymbol{A}_\gamma^{-1}\r]\nonumber
\end{eqnarray}
Therefore $M$ can be seen as describing the single-agent process with
noise $z(t)$ given by
\begin{eqnarray}
y(t+1)-y(t)=-\sum_{\gamma,t'}\l[\boldsymbol{1}+
\frac{n_\gamma}{2}\boldsymbol{G}\r]^{-1}(t,t')\phi_\gamma(t')+z(t)\\
\avg{z(t)z(t')}=\sum_\gamma\l[\l(\boldsymbol{1}+
\frac{n_\gamma}{2}\boldsymbol{G}\r)^{-1}\l(n_\gamma\boldsymbol{D}_\gamma\r)
\l(\boldsymbol{1}+ \frac{n_\gamma}{2}\boldsymbol{G}\r)^{-1}\r](t,t')
\label{nv}
\end{eqnarray}
which is completely equivalent to the original multi-agent system in
the limit $N\to\infty$.

Let us now focus on the asymptotic properties of the stationary state,
considering the simplest possibility.  Making for the asymptotic
behavior of $\bsy{C}$ and $\bsy{G}$ the assumptions of
{\it time-translation invariance},
\begin{eqnarray}\label{tti}
\lim_{t\to\infty}C(t+\tau,t)=C(\tau)\\
\lim_{t\to\infty}G(t+\tau,t)=G(\tau)
\end{eqnarray}
{\it finite susceptibility},
\begin{equation}\label{fir}
\lim_{t\to\infty}\sum_{t'\leq t}G(t,t')<\infty
\end{equation}
and {\it weak long-term memory},
\begin{equation}\label{wltm}
\lim_{t\to\infty}G(t,t')=0~~~\forall t'~{\rm finite}
\end{equation}
ergodic stationary states of the dynamics can be fully characterized
in terms of a few parameters. These are, in particular, the persistent
autocorrelation
\begin{equation}\label{cmax}
c=\lim_{\tau\to\infty}\frac{1}{\tau}\sum_{t<\tau}C(t)
\end{equation}
the magnetization
\begin{equation}
m=\lim_{t\to\infty}\frac{1}{t}\sum_{t'}m(t')
\end{equation}
and the susceptibility (or integrated response)
\begin{equation}\label{chimax}
\chi=\lim_{\tau\to\infty}\sum_{t\leq\tau}G(t)
\end{equation}
In this regime, the quantities
\begin{equation}
\widetilde{y}=\lim_{t\to\infty}\frac{y(t)}{t}~~~~~~~
s=\lim_{t\to\infty}\frac{1}{t}\sum_{t'}s(t')~~~~~~~
z=\lim_{t\to\infty}\frac{1}{t}\sum_{t'}z(t')
\end{equation}
are easily seen to be related by
\begin{equation}
\widetilde{y}=-\sum_\gamma\kappa_\gamma\frac{s+\gamma}{2}+z
\end{equation}
where
\begin{eqnarray}
\kappa_\gamma=\frac{2}{2+n_\gamma\chi}\\
\avg{z^2}=\sum_\gamma \frac{\alpha_\gamma\l(1+2\gamma
m+c\r)}{(2\alpha_\gamma+\chi)^2}
\end{eqnarray}

We have the following scenarios:
\begin{enumerate}
\item if $\widetilde{y}>0$, then $s=1$ (the agent is frozen on asset
$1$): this occurs if $z>\kappa_+$
\item if $\widetilde{y}<0$, then $s=-1$ (the agent is frozen on asset
$-1$): this occurs if $z<-\kappa_-$
\item if $\widetilde{y}=0$, then $s=s^\star\equiv
\frac{2z-\sum_\gamma\gamma\kappa_\gamma}{\sum_\gamma\kappa_\gamma}$
(the agent is fickle): this occurs if $-\kappa_-<z<\kappa_+$
\end{enumerate}
Separating the contribuctions of different cases we end up with the
following equations for $m$, $c$ and $\chi$:
\begin{eqnarray}\fl 
m=\avg{\theta(z-\kappa_+)}_z+\avg{s^\star\theta(z+\kappa_-)
\theta(\kappa_+-z)}_z-\avg{\theta(-\kappa_--z)}_z\nonumber\\\fl
c=\avg{\theta(z-\kappa_+)}_z+\avg{(s^\star)^2\theta(z+\kappa_-)
\theta(\kappa_+-z)}_z+\avg{\theta(-\kappa_--z)}_z\\ \fl \sum_\gamma
\frac{\alpha_\gamma \chi}{2\alpha_\gamma+\chi}=\avg{\theta(z+\kappa_-)
\theta(\kappa_+-z)}_z\nonumber
\end{eqnarray}
where $\avg{\cdots}_z$ is an average over the static Gaussian noise
$z$. The Gaussian integrals can be easily computed and these equations
can be solved numerically for $c$, $m$ and $\chi$. Notice that
$n_+>n_-$ (or $\alpha_+<\alpha_-$) implies $\kappa_+<\kappa_-$ so that
the probability that an agents `freezes' on asset $\gamma$ is larger
for $\gamma=+1$, i.e. for the asset with less information. This
conclusion is immediately clear from the above equations. A little
more work is required to see that $H$ is given (apart from factors
$\alpha_\gamma$) by the persistent part of the noise variance
(\ref{nv}):
\begin{equation}
H=\sum_\gamma\frac{\alpha_\gamma^2\l(1+2\gamma
m+c\r)}{(2\alpha_\gamma+\chi)^2}
\end{equation}
These expressions finally yield the analytical curves shown in
Fig. \ref{Phase_diagram.fig}.

\subsection{The Majority Game}

The simplest way to get a glimpse on the macroscopic properties of the
Majority Game is to consider the simplified information-free context
of Sec. \ref{zimpol}, where the model is described by the rules
\begin{eqnarray}
\prob\{b_i(t)=b\}=C \exp\l[b\Delta_i(t)\r]\\
\Delta_i(t+1)-\Delta_i(t)=\Gamma A(t)/N
\end{eqnarray}
by which agents reward the action taken by the majority and increase
the probability of choosing $b_i(t+1)=\sign\l[A(t)\r]$. An analysis
similar to that outlined in the case of the Minority Game easily leads
to the conclusion that the dynamics of $y(t)=\Delta_i(t)-\Delta_i(0)$
(which is $i$-independent) admits the solution $y(t)=y_0+vt$ where
$v=\pm\Gamma$. In this state, agents behave coherently ($b_i(t)=b$ for
all $i$). Consequently, $\avg{A}$ is either $N$ or $-N$ and
$\sigma^2=O(N^2)$ independently of $\Gamma$.

The above conclusion that Majority Games generate huge fluctuations is
rather intuitive. However the full Majority Game turns out to be a
surprisingly rich model \cite{Kozlo}. It is defined by the following
setup:
\begin{eqnarray}
g_i(t)=\argmax U_{ig}(t)\nonumber\\
A(t)=\sum_i a_{i g_i(t)}^{\mu(t)}\\
U_{ig}(t+1)-U_{ig}(t)= a_{ig}^{\mu(t)} \l[A(t)-\eta\l(
a_{i g_i(t)}^\mu-a_{ig}^\mu\r)\r]\nonumber
\end{eqnarray}
where $\mu(t)\in\{1,\ldots,P\}$ stands for the information pattern
presented to agents at time $t$ (taken to be external and random) and
$\eta$ tunes the agents' ability to learn to respond to the action of
all other agents by disentangling their own contribution to the game's
outcome.

Using the notation introduced in Sec. \ref{mgth}, it is easy to see
that \be v_i\equiv\avg{y_i(t+1)-y_i(t)}=\ovl{\xi_i\Omega}+
\sum_j\ovl{\xi_i\xi_j}m_j-\eta\ovl{\xi_i^2}m_i \ee where
$m_i=\avg{\sign(y_i)}$. Hence the dynamics minimizes the function \be
H_\eta=-\frac{1}{2}\sum_{i,j}\ovl{\xi_i\xi_j}m_i
m_j-\sum_i\ovl{\xi_i\Omega}m_i+\frac{\eta}{2}\sum_i\ovl{\xi_i^2}m_i^2
\ee Adding the constant $-\ovl{\Omega^2}/2$ to complete a square with
the first to terms above, one sees that $H_\eta$ is a downward concave
function of the $m_i$'s, which implies that minima occur on the
corners of the definition domain $[-1,1]^N$. Thus the solution with
$v_i=0$ corresponding to fickle agents is ruled out in this case and
the only remaining solutions are those with $v_i\neq 0$ (and
$y_i(t)/t$ finite as $t\to\infty$), corresponding to frozen agents.
For these, \be m_i=\sign(v_i)=
\sign\l(\ovl{\xi_i\Omega}+\sum_j\ovl{\xi_i\xi_j}m_j-\eta\ovl{\xi_i^2}m_i\r)
\ee Notice that since the relevant steady states have $m_i=\pm 1$ the
last term in $H_\eta$ plays the role of a mere constant. Hence impact
factors do not alter the steady state properties of the Majority Game.
(Also due to agents' freezing, the `batch' and `on-line' version yield
the same stationary properties as fluctuations play no role in this
case.) Furthermore, it is clear that any configuration $\{m_i\}$ which
is a solution of these equations for some value of $\eta\in [0,1]$
will also be a solution for all $\eta'<\eta$. Hence the set ${\cal
S}_\eta$ of stationary states is such that ${\cal S}_\eta\subset {\cal
S}_{\eta'}$ for $\eta'<\eta$ and, in particular, ${\cal S}_1\subset
{\cal S}_\eta$ for all $\eta<1$. It is also easy to see that the state
with minimal value of $H_\eta$ lies in ${\cal S}_1$ for all $\eta\in
[0,1]$. This shows that Nash equilibria are stationary states of the
majority game for all values of $\eta$, but the converse is not true
(except for $\eta=1$ of course).

It is possible to draw a complete picture of the model's behavior by
studying the minima of $H_\eta$ explicitly via the replica method. The
calculation has been carried out in \cite{Kozlo} under the assumption
that the two strategies of the same agent can be to some degree
correlated, which is allowed if one takes the disorder distribution
\be \fl P(a_1,a_2)=\frac{w}{2}\l(\delta_{a_1,1}\delta_{a_2,1}+
\delta_{a_1,-1}\delta_{a_2,-1}\r)+
\frac{1-w}{2}\l(\delta_{a_1,1}\delta_{a_2,-1}+
\delta_{a_1,-1}\delta_{a_2,1}\r) \ee Notice that
$w=\prob\{a_{i1}^\mu=a_{i2}^\mu\}$. It turns out that, depending on
the parameters, the system can be in one of two phases: a `retrieval'
phase characterised by attractors with a macroscopic overlap
$A^1=O(N)$ with a given pattern (say, $\mu=1$) and a spin glass phase
with no retrieval ($A^\mu=O(\sqrt{N})$). The occurrence of `retrieval'
may be thought of as the emergence of crowd effects such as fashions
and trends, when a large fraction of agents behave similarly in some
respect, or to economic concentration, when, for example, one
particular place is arbitrarily selected for large scale investments.
Interestingly, one finds that the development of these crowd effects
requires: (i) that the number of agents is large compared to the
number of resources ($\alpha$ small), (ii) a sufficient
differentiation between strategies of agents ($w<2/3$) and (iii) a
large enough initial bias (i.e. an initial macroscopic overlap)
towards a particular resource, fashion or place.  Finally crowd
effects can be sustained under more general conditions (i.e. in the
spin glass phase) if agents do not behave strategically, i.e. if they
neglect their impact on the aggregate ($\eta$ small). This phenomenon
can be attributed to the self-reinforcing term
$(1-\eta)\overline{\xi_i^2}s_i$ in the dynamics which causes a
dramatic increase in the number of stationary states as $\eta$
decreases (which can be seen quantitatively by analyzing the entropy).

\subsection{Models with interacting trend-followers and contrarians}

It is rather easy to understand that the two main groups of traders,
that is fundamentalists and trend-followers, contribute opposite
forces to the price dynamics. Fundamentalists believe that the market
is close to a stationary state and buy (sell) when they repute the
stock to be underpriced (overpriced), thus inducing anti-correlation
in market returns and holding the price close to its `fundamental'
value. Trend-followers, instead, extrapolate trends from recent price
increments and buy or sell assuming that the next increment will occur
in the direction of the trend, thus creating positive return
correlations and large price drifts (`bubbles').  Chartist behavior,
which can also be driven by imitation, is known to cause market
instability. Fundamentalists act instead as a restoring force that
dumps market inefficiencies and excess volatility. The next question
we address concerns the macroscopic properties of models in which
contrarians and trend-followers interact.

\begin{figure}[t]
\centering
\includegraphics*[width=8cm]{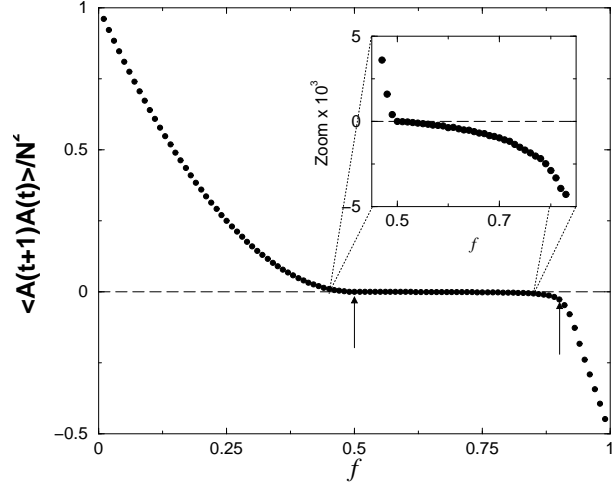}
\caption{(from Ref. \cite{Mats}) 
Autocorrelation of returns as a function of the
fraction $f$ of fundamentalists in the market. Autocorrelation is
taken in the stationary state of a system of $N=10^4$ agents with 
$\Gamma=2.5$. Arrows mark the transitions between the three regimes
described in the text, which occur at $f=0.5$ and at
$f=0.9$. The inset shows a detail of the central part of the
graph.}
\label{matteus}
\end{figure}

As usual, we start from the simple model with no information. Let us
assume that a fraction $f$ of agents are trend followers whereas the
remaining $(1-f)N$ are fundamentalists. The dynamics is governed by
the following scheme:
\begin{eqnarray}
\prob\{b_i(t)=b\}=C \exp\l[b\Delta_i(t)\r]\\
\Delta_i(t+1)-\Delta_i(t)=\epsilon_i\Gamma A(t)/N
\end{eqnarray}
where $\epsilon_i=1$ for trend-followers (say for
$i\in\{1,\ldots,fN\}$) and $\epsilon_i=-1$ for fundamentalists (say
$i\in\{fN+1,\ldots,N\}$). Assuming that $\Delta_i(0)=0$ for
simplicity, we can approximate $A(t)/N$ with its average and see that
the dynamics of $y(t)=\Delta_i(t)-\Delta_i(0)\equiv\Delta_i(t)$ is
given by \be y(t+1)-y(t)=(2f-1)\Gamma\tanh[y(t)]\label{parapa1} \ee
Linear stability analysis of \req{parapa1} leads to the following
scenario. For $f<1/2$ we have two regimes:
\begin{itemize}
\item for $\Gamma<\frac{1}{1-2f}$ the fixed point $y^\star=0$ is
  stable. One has $\avg{A}=0$ and $\sigma^2=O(N)$ as in the
  information-free Minority Game with subcritical $\Gamma$
\item for $\Gamma>\frac{1}{1-2f}$ the fixed point $y^\star=0$ is
  unstable. One has $\avg{A}=0$ and $\sigma^2=O(N^2)$ as in the
  information-free Minority Game with supercritical $\Gamma$
\end{itemize}
For $f>1/2$ instead the fixed point $y^\star=0$ is unstable and the
solution $y(t)=y_0+v t$ with $v=\pm(2f-1)\Gamma$ appears. Here, both
trend-followers and contrarians behave coherently: $b_i(t)=b$ for all
$i\in\{1,\ldots,fN\}$ and $b_i(t)=-b$ for all $i\in\{fN+1,\ldots,N\}$.
As a result, $\avg{A}$ is either $(2f-1)N$ or $(2f-1)N$ and
$\sigma^2=O(N^2)$ as in the information-free Majority Game.  The
conclusion we draw is that the expectations of the majority group (be
it fundamentalists or trend-followers) are fulfilled in the steady
state. This is confirmed by studying the autocorrelation of returns as
a function of $f$ in the steady state, see Fig. \ref{matteus}.

This conclusion extends to the full model, whose properties have been
analyzed in \cite{dgm}. The mixed Majority-Minority Game is defined by
\begin{eqnarray}
g_i(t)=\argmax U_{ig}(t)\nonumber\\
A(t)=\sum_i a_{i g_i(t)}^{\mu(t)}\\
U_{ig}(t+1)-U_{ig}(t)=\epsilon_i a_{ig}^{\mu(t)}A(t)/N\nonumber
\end{eqnarray}
where as before $\epsilon_i=1$ for trend-followers (or
$i\in\{1,\ldots,fN\}$) and $\epsilon_i=-1$ for fundamentalists (or
$i\in\{fN+1,\ldots,N\}$). The statistical mechanics of this model is
slightly more involved than previous cases. As before, one finds that
the steady state can be characterized in terms of the microscopic
variables $m_i=\avg{\sign(y_i)}$ where
$y_i(t)=\frac{1}{2}[U_{i1}(t)-U_{i2}(t)]$. In particular, the
stationary $m_i$'s for can be obtained by solving the following
problem: \be \max_{\boldsymbol{m}_2}~\min_{\boldsymbol{m}_1}
~H(\boldsymbol{m}_1,\boldsymbol{m}_2) \ee where \be H
(\bsy{m}_1,\bsy{m}_2)=\frac{1}{P}\sum_\mu\l[\Omega^\mu+\sum_i\xi_i^\mu
m_i\r]^2 \ee and $\boldsymbol{m}_1$ (resp. $\boldsymbol{m}_2$) denote
collectively the $m_i$ variables of Minority (resp. Majority) game
players. Hence the mixed game where both minority and majority players
are present at the same time requires a minimization of the
predictability in certain directions (the minority ones) and a
maximization in others (the majority ones). It is possible to tackle
this type of problem by a replica theory \cite{varga}. The idea is to
introduce two `inverse temperatures' $\beta_1$ and $\beta_2$ for
minority and majority players respectively, such that \cite{dgm}
\begin{equation}\label{minimax}
\max_{\boldsymbol{m}_2}~\min_{\boldsymbol{m}_1}
~H(\boldsymbol{m}_1,\boldsymbol{m}_2)=
\lim_{\beta_1,\beta_2\to\infty}
\frac{1}{\beta_2}\davg{\log Z(\beta_1,\beta_2)}
\end{equation}
with the following generalized partition function:
\begin{equation}\fl\label{zeta}
Z(\beta_1,\beta_2)=\int d\boldsymbol{m}_2 ~
e^{\beta_2\l[-\frac{1}{\beta_1}\log\int d\boldsymbol{m}_1~
e^{-\beta_1\mathcal{H}}\r]}=\int d\boldsymbol{m}_2\l[\int
d\boldsymbol{m}_1 ~e^{-\beta_1\mathcal{H}}\r]^{-\gamma}
\end{equation}
where $\gamma=\beta_2/\beta_1>0$. In physical jargon, this describes a
system where: first, the $\boldsymbol{m}_1$ variables are thermalized
at a positive temperature $1/\beta_1$ with Hamiltonian $H$ at fixed
$\boldsymbol{m}_2$; then, the $\boldsymbol{m}_2$ variables are
thermalized at a negative temperature $-1/\beta_2$ with an effective
Hamiltonian $H_{{\rm eff}}$ defined by $-\beta_1 H_{{\rm
eff}}(\boldsymbol{m}_2)=\log\int d\boldsymbol{m}_1~ e^{-\beta_1
H}$. The disorder average can be carried out with the help
of a `nested' replica trick. First, one replicates the minority
variables by treating the exponent $-\gamma$ as a positive integer $R$
(in the end, the limit $R\to-\gamma<0$ must be taken). (\ref{zeta})
thus becomes
\begin{equation} \fl
Z=\int d\boldsymbol{m}_2\l[\int d\boldsymbol{m}_1
~e^{-\beta_1\mathcal{H}}\r]^{R}=\int d\boldsymbol{m}_2
\l[\int e^{-\beta_1\sum_{r}\mathcal{H}(\{\boldsymbol{m}_1^r\},
\boldsymbol{m}_2)}\prod_{r=1,R}d\boldsymbol{m}_1^r\r]
\end{equation} 
Then a second replication is needed, this time on the
$\boldsymbol{m}_2$ variables:
\begin{equation} 
Z^{R'}=\int e^{-\beta_1\sum_{a,r}
\mathcal{H}(\{\boldsymbol{m}_1^{ar}\},\{\boldsymbol{m}_2^a\})}
\prod_{a=1,R'}\prod_{r=1,R}d\boldsymbol{m}_1^{ar}
d\boldsymbol{m}_2^a
\label{zn}
\end{equation} 
At this point we have two replica indexes with different roles: the
replicas labeled $a$ have been introduced to deal with the disorder,
and their number $R'$ will eventually go to zero, as usual; the
replicas labeled $r$ have been introduced to deal with the negative
temperature, and their number $R$ must be set to a negative
value. Majority variables bear just one index, while minority ones
have two. We can interpret this fact by saying that
$\boldsymbol{m}_2^a$ indicates a particular configuration of the
majority variables, i.e. a given manifold in the whole
$\boldsymbol{m}$ space; and $\boldsymbol{m}_1^{ar}$ indicates the
minority coordinates in that particular manifold. Notice that the
$\min$ and $\max$ operations and hence the meaning of coordinates in
the above interpretation can be interchanged. In general, this leads
to different solutions. In our case, however, one can verify that the
main results would not change, though the intermediate steps (e.g. the
definition of $\gamma$) would vary.

\begin{figure}[t]
\centering
\includegraphics[width=8cm]{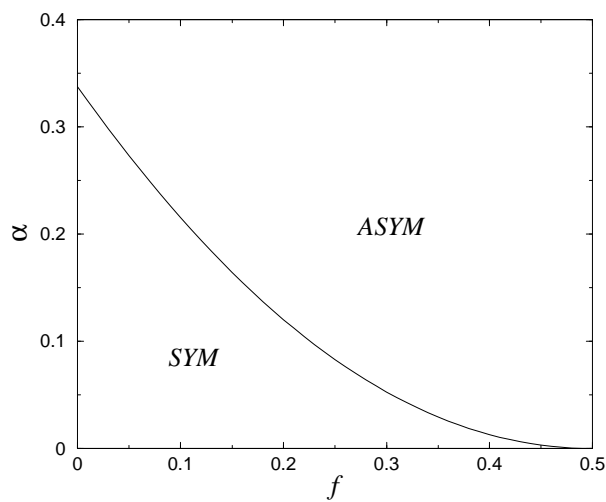}
\caption{Phase diagram of the mixed majority-minority game (from
\cite{dgm}).}
\label{mmmg}
\end{figure}

Following the procedure outlined above it is possible to calculate the
phase diagram of the model (Fig. \ref{mmmg}), namely the line of
critical points $\alpha_c(f)$ for different values of $f$ separating
the asymmetric, information-rich phase ($\alpha>\alpha_c(f)$) from the
symmetric, unpredictable regime ($\alpha<\alpha_c(f)$).
One sees that the efficient regime shrinks as the fraction of
trend-followers increases until, for $f=1/2$ it disappears. Now
trend-followers are the majority group and the market becomes
completely predictable.  The dynamical calculation clarifies the phase
transition further by relating the critical line to the onset of
ergodicity breaking.

While this model captures one of the basic effects of the presence of
trend-followers in the market, namely a decrease in efficiency, it is
clear that the properties of mixed games are to some extent a linear
combination of those of pure games and thus a gross simplification
with respect to a realistic case.  Now it is reasonable to think that
real traders may revise their expectations if they prove wrong or
simply may want to weigh their decisions against other factors than
the expected profit. For instance, in certain market regimes
(e.g. bubbles) a trader could perceive the market as a Majority rather
than Minority Game and consequently switch from a fundamentalist to a
trend-following behavior. Similarly, in situations of high volatility
traders would likely take into account the risk factor when choosing a
trading strategy over another. How would the macroscopic properties of
the Minority Game change if agents were allowed to modify their
behavior and expectations according to the market conditions they
perceive?

\begin{figure}[t]
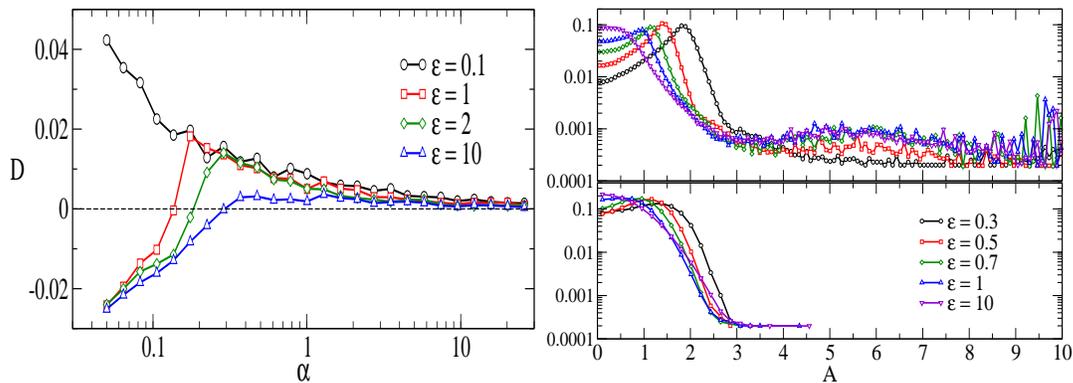

\centering
\includegraphics*[width=7cm,height=5cm]{D.eps}
\includegraphics*[width=7cm,height=5cm]{Pa005a2.eps}
\caption{Normalized return autocorrelation function $D$ as a function
of $\alpha=P/N$ for different values of $\epsilon$ (left) and
probability distributions $P(A)$ of $A>0$ for different values of
$\epsilon$ for $\alpha=0.05$ (top right) and $\alpha=2$ (bottom
right). From \cite{gene1}.}
\label{gmg}
\end{figure}

This issue may be tackled through the introduction of a more general
MG setting with the rationale that traders prefer to adopt a
trend-following attitude, and thus perceive the market as a Majority
Game, when fluctuations are small while they revert to fundamentals,
and hence perceive the market as a Minority Game, when the price
dynamics becomes more chaotic \cite{gene1,gene2}. This mechanism leads
to a surprisingly rich phenomenology which includes the formation and
disruption of trends and the emergence of `heavy tails' in the returns
distribution. The model is defined through
\begin{eqnarray}
g_i(t)=\argmax U_{ig}(t)\nonumber\\
A(t)=\sum_i a_{ig_i(t)}^{\mu(t)}\\
U_{ig}(t+1)-U_{ig}(t)=a_{ig}^{\mu(t)}F_i[A(t)]\nonumber
\end{eqnarray}
where the function $F_i$ embodies the way in which agent $i$ perceives
the performance of his/her $g$-th trading strategy in the market. For
simplicity we shall henceforth assume that $F_i=F$ for all
$i$. Clearly, $F(A)=-A$ for a Minority Game whereas $F(A)=A$ for a
Majority Game. The case we consider is
\begin{equation}
F(A)=A-\epsilon A^3
\end{equation}
with $\epsilon\geq 0$. For $\epsilon=0$ one has a pure Majority Game.
Upon increasing $\epsilon$, the non-linear gains importance, and for
$\epsilon\to\infty$ one obtains a Minority Game with $F(A)\propto
-A^3$. A couple of remarks are in order.
\begin{enumerate}
\item This mechanism is expected to induce a feed-back in the dynamics
  of the excess demand: when it is small, trend-followers dominate and
  drive it to larger values until fundamentalists eventually take over
  and drive it back to smaller values.
\item It is reasonable to think that $\epsilon$ should fluctuate in
  time and possibly be coupled to the system's performance. A possible
  microscopic mechanism is the following. When $\epsilon$ is large a
  high volatility is to be expected as agents are more likely to
  behave as trend-followers. As a consequence, they should likely
  reduce their threshold since the market is risky; however, for small
  $\epsilon$ fundamentalists are expected to dominate and the game
  should acquire a Minority character. Hence the predictability will
  be smaller and there will be less profit opportunities. Agents may
  then decide to adopt a larger threshold to seek for convenient
  speculations on a wider scale. If these two competing effects are
  appropriately described by an evolution equation for $\epsilon$, the
  system should self-organize around an `optimal' value of the
  parameter. However such a time evolution should take place on
  time-scales much longer than those which the model addresses
  (intra-day/daily trading) and hence it is reasonable to study the
  case of fixed $\epsilon$.
\end{enumerate}
It turns out (see Fig. \ref{gmg}) that while for low enough
(resp. high enough) $\epsilon$ the behavior of a pure Majority
(resp. Minority) game is recovered (with some qualitative differences
due to the unconventional nature of the MG in this case), there exists
a range of values of $\epsilon$ for which the two tendencies coexist
and one can cross over from one to the other by changing $\alpha$
and/or $\epsilon$.
This can be seen from the behavior of the (normalized) autocorrelation
$D=\avg{A(t)A(t+1)}/\sigma^2$ as a function of $\alpha$. The crossover
gets sharper and sharper as $\alpha$ increases and turns into a sharp
threshold for $\alpha\gg 1$. In this case, the threshold can be
estimated analytically. Indeed one has \be
v_i\equiv\avg{y_i(t+1)-y_i(t)}=\ovl{\xi_i^\mu\avg{F(A)|\mu}} \ee As
usual, if $v_i\neq 0$, then $y_i(t) \sim v_i t$ and $s_i(t)$ tends
asymptotically to $\sign(v_i)$: there is a well defined preference
towards one of the two strategies and the agent becomes frozen. For
large $\alpha$, we can approximate $A(t)$ with a Gaussian random
variable with variance $H$. By virtue of Wick's theorem, this implies
that $\avg{A^3|\mu}\simeq 3H\avg{A|\mu}$, so \be v_i\simeq
(1-3\epsilon H)\ovl{\xi_i^\mu\avg{A|\mu}} \ee If $1-3\epsilon H>0$,
the agents' spins will freeze on the Majority-type solution
$s_i=\sign(\ovl{\xi_i^\mu\avg{A|\mu}})$, which is unstable for
$1-3\epsilon H\leq 0$. Given that $H=1$ for large $\alpha$, we see
that the crossover from the Majority- to the Minority-regime takes
place at $\epsilon\simeq 1/3$ for $\alpha\gg 1$, which is
significantly close to the numerical value of $\epsilon_c\simeq 0.37$.

For small $\alpha$, when the contribution of frozen agents is small,
we expect the system to self-organize around a value of $A$ such that
$F(A)=0$: indeed one can see from Fig. \ref{gmg} that the peak of the
distribution moves as $1/\sqrt{\epsilon}$. Besides, as $\epsilon$
increases, large excess demands occur with a finite probability. The
emergence of such `tails' in $P(A)$, while not power-law, is a clear
non-Gaussian signature. The dynamics in this regime is particularly
interesting: while the market is mostly chaotic and dominated by
contrarians, `ordered' periods can arise where the excess demand is
small and trends are formed, signaling that chartists have taken over
the market. These trends, that can be arbitrarily long, eventually
eventually die out restoring the fundamentalist regime. 

In order to understand the full impact of trend-followers it is
however necessary to emply endogenous information
\cite{gene2}. Indeed, one identifies two regimes in an intermittent
market dynamics. Phases with small fluctuations, dominated by
contrarians and in which the information dynamics is roughly ergodic
over the possible patterns, are followed by phases with large
fluctuations dominated by trend-followers, where the information
dynamics is strongly non-ergodic (actually a single information
pattern is dynamically selected).

\subsection{Markets with asymmetric information}

A crucial assumption in all models we have been dealing with so far is
that all agents possess the same information, be it the real price
time series or the bar attendance sequence or a random integer. As
long as all agents process the same information pattern the system can
reach some level of coordination and a more or less complicated phase
structure arises. Unfortunately, it is hard to believe that all agents
in real systems possess the same information. This brings us to the
question: how are the coordination properties affected when the
information is private, i.e. agent-dependent?

This question is indeed of fundamental theoretical importance. A
substantial part of economic theory is based on the assumption that
markets are informationally efficient. Roughly speaking, a market is
efficient with respect to an information set if the public revelation
of that information would not change the prices of the assets. In
other words, this means that all the relevant information is
incorporated into prices. This includes both public and private
information. However, it has been understood [Akerlof] that asymmetric
information may cause inefficiency of the equilibrium, given the
strategic incentive of each agent not to reveal the information he
has. The salvation comes from the system size: in fact this nefarious
effect may vanish in large markets, since the single bits of
information possessed by an individual agent become less significant
the larger is the number of agents. Hence, the common understanding is
that prices reflect information more accurately in large systems.

To conclude our review, we shall now discuss a model in which the
above scenario emerges as a phase transition between an
informationally efficient phase and an informationally inefficient one
\cite{BMRZ}. The control parameter is, as in the MG, the ratio between
the size of the information space and the number of traders.

We consider a market with one asset. The market can find itself in any
of $P$ states of the world, labeled $\mu$. The return of the asset
depends on the state of the world only, and is denoted by $R^\mu$. We
assume that each $R^\mu$ is given by \be
R^\mu=\ovl{R}+\frac{r^\mu}{\sqrt{N}} \ee where the $r^\mu$ are
independent samples of a Gaussian random variable with zero mean and
variance $s$ drawn at time $t=0$ and fixed (quenched disorder). We
further assume that at each time step the state of the market is drawn
randomly and independently from $\{1,\ldots,P\}$ with equal
probability. This process determines the time series of returns
$\{R^{\mu(t)}\}_{t\geq 0}$ completely.

$N$ traders act in this market. They have no information concerning
the state of the world but rather they observe a coarse-grained signal
on the information space $\{1,\ldots,P\}$. We denote it as a vector
\be \bsy{k}_i:\{1,\ldots,P\}\ni \mu\to k_i^\mu\in\{-1,1\} \ee in which
every state of the market is associated to a particular value of a
binary variable (in other words, an agent cannot tell which state the
market is in but only knows whether it is an ``up state'' or a ``down
state''). Different agents receive different signals, as each
component $k_i^\mu$ of every vector $\bsy{k}_i$ is taken to be drawn
randomly and independently from $\{-1,1\}$ with equal probability for
all $i$ and $\mu$. This defines the private information structure.
Note that if an agent knew simultaneously the partial information of
all agents he would be able to know the state $\mu$, with probability
one, for $N\to\infty$.

At each time step, traders $i$ has to decide an investment. Let
$z_i(t)$ denote the amount of money he decides to invest (buying or
selling) at time $t$. We assume that the price at time $t$, $p(t)$ is
fixed by a market clearing condition, in which the demand of the asset
is determined by the aggregate money invested and the supply is fixed
at $N$: \be \frac{1}{N}\sum_i z_i(t)=p(t) \ee We further assume that
$z_i(t)$ depends on whether his information $k_i^{\mu(t)}$ about the
state is ``up'' or ``down'':
$z_i(t)=\sum_{m\in\{-1,1\}}z_i^m(t)\delta_{k_i^{\mu(t)},m}$. In this
way the price depends on the state since the amount invested by each
agent depends on the state: $p(t)=p^{\mu(t)}$.

At the end of each period $t$, each unit of asset pays a monetary
amount $R^{\mu(t)}$. If agent $i$ has invested $z_i(t)$ units of
money, he will hold $z_i(t)/p(t)$ units of asset, so his payoff will
be $z_i(t)(\frac{R^{\mu(t)}}{p(t)}-1)$. It follows that the expected
payoff is given by \be \pi_i=\frac{1}{P}\sum_\mu
\sum_{m\in\{-1,1\}}\delta_{k_i^\mu, m} z^m_i \left(
\frac{R^\mu}{p^\mu}-1\right)=\sum_{m\in \{-1,1\}}\ovl{\delta_{k_i, m}
z^m_i \left(\frac{R}{p}-1\right)} \ee Every agent aims at choosing the
$z_i^m$'s so as to maximize their expected payoff. We consider
inductive agents who repeatedly trade in the market. Each agent $i$
has a propensity to invest $U_i^m(t)$ for each of the signals
$m\in\{-1,1\}$. His investment $z_i^m=\chi_i(U_i^m)$ at time $t$ is an
increasing function of $U_i^m(t)$ ($\chi_i: \mathbb{R}\to
\mathbb{R}^+_0$) with $\chi_i(x)\to 0$ if $x\to -\infty$ and
$\chi_i(x)\to \infty$ if $x\to\infty$ (a convenient choice for
numerical experiments is $z_i^m=U_i^m\theta(U_i^m)$). After each
period agents update $U_i^m(t)$ according to the marginal success of
the investment:
\begin{equation}
U_i^m(t+1)=U_i^m(t)+\Gamma
\delta_{k_i^{\mu(t)},m}\left[R(t)-p(t)-\eta\frac{z_i(t)}{N}\right]
\label{learn}
\end{equation}
The idea is that if the return is larger than the price, the agent's
propensity to invest in that signal increases, otherwise it decreases.
The $\eta$ term provides the distinction between na\"\i ve (or
price-taking) agents ($\eta=0$), who are unaware of their market
impact, and ``sophisticated'' traders ($\eta=1$) who instead are able
to disentangle their contribution to the price exactly. $\Gamma>0$ is
a parameter (In \cite{BMRZ} the dynamics \req{learn} is obtained from
a more properly justified process involving the marginal utility of a
certain investment.).

As a measure of coordination we employ the distance between prices and
returns in the steady state: \be
H=|\bsy{R}-\bsy{p}|^2\equiv\sum_\mu\l(R^\mu-p^\mu\r)^2 \ee Clearly, if
$H=0$ prices follow returns and hence incorporate the information
about the states of the world, so that the market is informationally
efficient.

\begin{figure}[t]
\centering
\includegraphics[width=.45\textwidth]{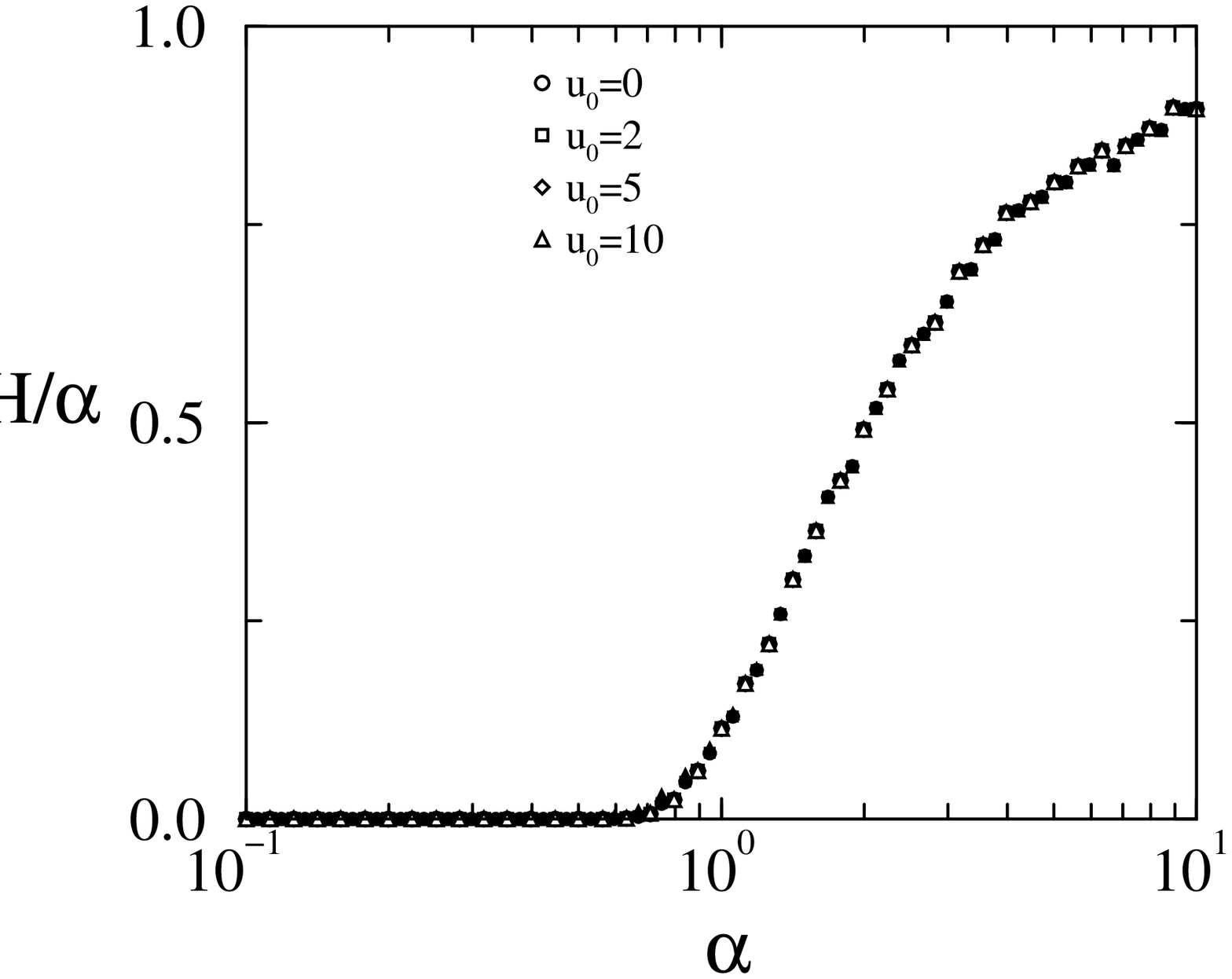}
\includegraphics[width=.4\textwidth]{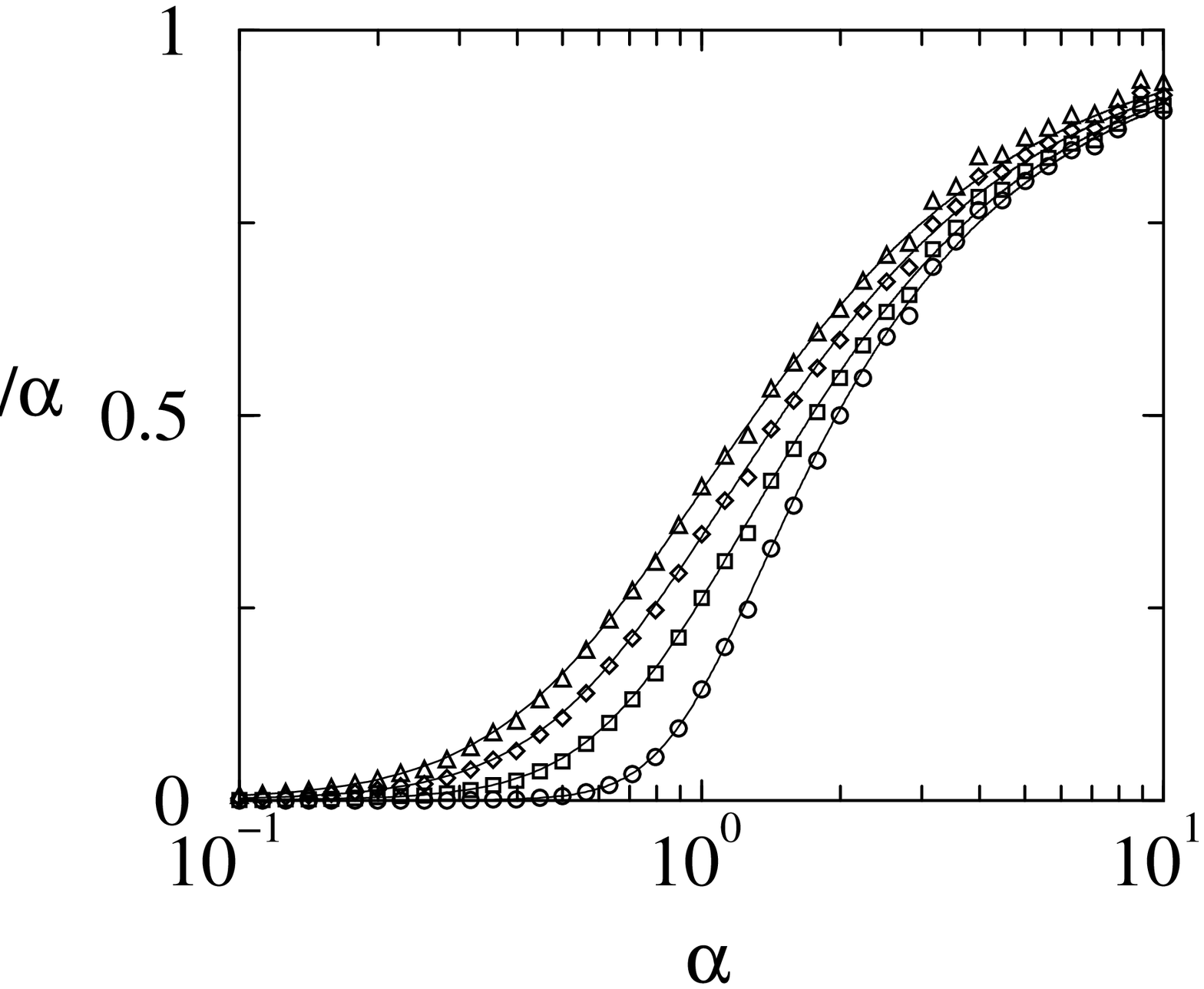}
\caption{Behaviour of $H/\alpha$ versus $\alpha$ for $\eta=0$ (left;
$u(0)$ is the initial bias in the score functions) and various
$\eta>0$ (right: $\eta=0.05$ (circles), $\eta=0.25$ (squares),
$\eta=0.5$ (diamonds) and $\eta=0.75$ (triangles). From \cite{DG}.}
\label{toy}
\end{figure}

Numerical results for the stationary $H$ as a function of $\alpha=P/N$
for $\eta=0$ and $\eta=1$ (and $\Gamma$ small enough) are given in
Fig. \ref{toy}.

Let us start from na\"\i ve traders ($\eta=0$). As the number of
agents increases, i.e. as $\alpha=P/N$ decreases, agents are
collectively more efficient in driving prices close to returns. Indeed
the distance $H$ decreases as $\alpha$ decreases. The price-return
distance vanishes at a critical point $\alpha_c$ which turns out to
mark a {\em second order phase transition} in the statistical
mechanics approach. The value of $\alpha_c$ depends on the intensity
$s$ of fluctuations of returns. The region $\alpha<\alpha_c$ is
characterized by the condition $H=0$, which means $p^\mu=R^\mu$ for
all $\mu$. This means that the market efficiently aggregates the
information dispersed across agents into the price. It can be shown
that the efficient phase, where $H=0$, shrinks as $s$ increases. This
is reasonable because as the fluctuations in $R^\mu$ increase, it
becomes harder and harder for the agents to incorporate them into
prices. This behavior can be understood analytically as usual by
constructing the continuous-time limit of \req{learn}. It turns out
that $H$ is a Lyapunov function of the dynamics: price takers
cooperate to make the market as informationally efficient as possible.
From the agent's point of view the steady states in the efficient
phase ($\alpha<\alpha_c$) are not unique and the state in which agents
will end up depends on the initial conditions $\{U_i^m(t=0)\}$
(prices, of course, do not depend on the initial condition, because
$p^\mu=R^\mu$ for all $\mu$). It can also be shown that these steady
states in which $H$ is minimum correspond to competitive equilibria,
namely configurations obtained when agents choose their investments
$z_i^m$ a priori by solving \be \max_{x \geq 0} ~x
\ovl{\delta_{k_i,m}\left( \frac{R}{p}-1\right)} \ee for
$m\in\{-1,1\}$, namely by maximizing their expected profits.

Turning to sophisticated agents ($\eta=1$), one sees that the phase
transition disappears: the distance between prices and returns
smoothly decreases as $\alpha$ decreases and it vanishes only in the
limit $\alpha\to 0$.  Moreover, the steady state is unique in both
prices and investment for all $\alpha>0$: the asymptotic behavior of
learning dynamics does not depend on initial conditions. It can be
shown that the steady state in this case is a Nash equilibrium, that
is it corresponds to all agents choosing their investments by solving
\be \max_{x \geq 0}~x \sum_\mu \delta_{k_i^\mu,m} \left(
  \frac{R^{\omega}}{p_{-i}^{\omega}+x/N}-1\right) \ee for
$m\in\{-1,1\}$, where
$p_{-i}^{\omega}=p^\omega-\sum_{m\in\{-1,1\}}\delta_{k_i^\omega,m}
z_i^m/N$ is the contribution of all other agents to the price (in
other words, each trader disentangles his contribution from the price
and optimizes the response to all other traders).

\begin{figure}[t]
\centering
\includegraphics[width=7cm]{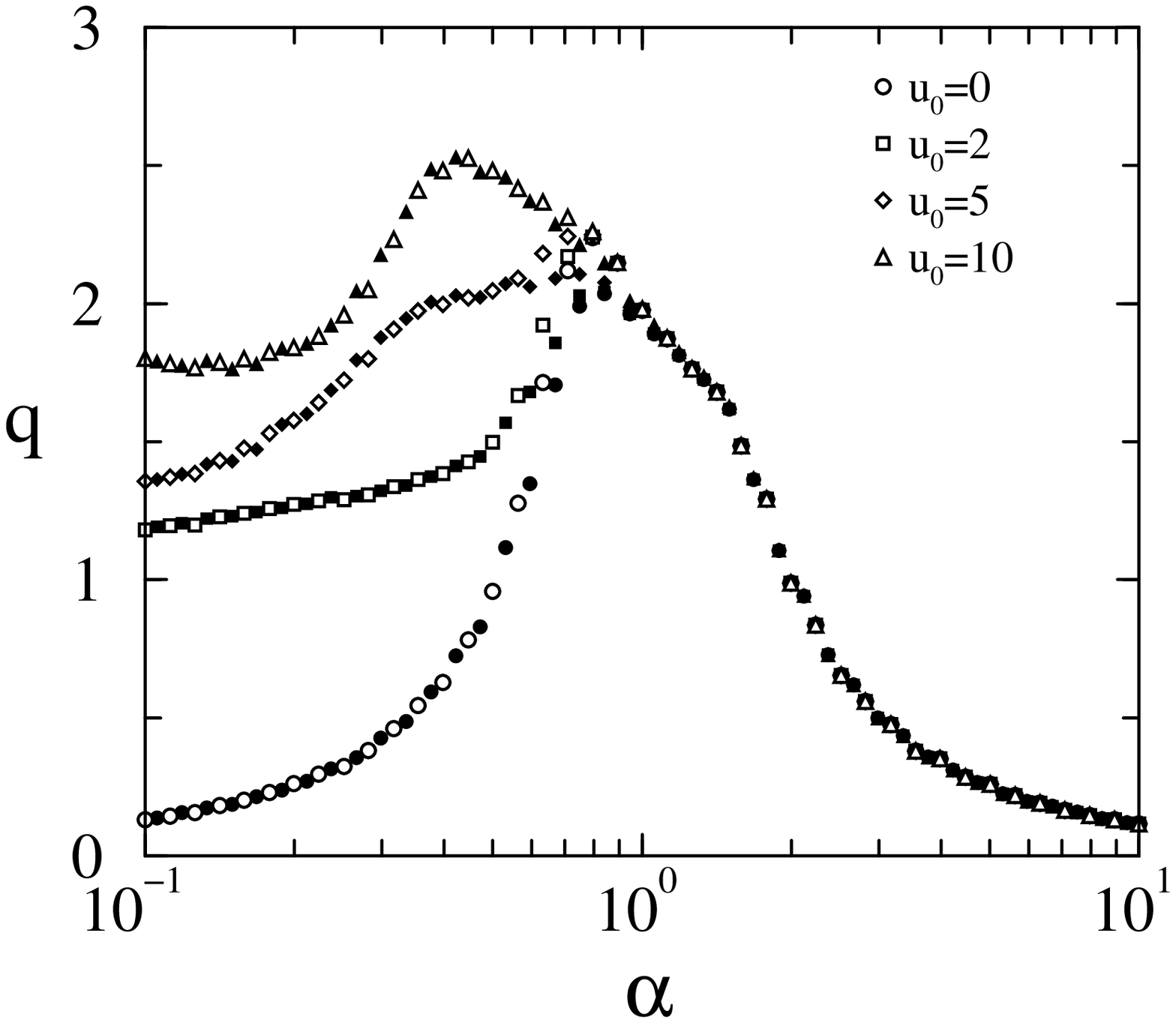}
\includegraphics[width=7cm]{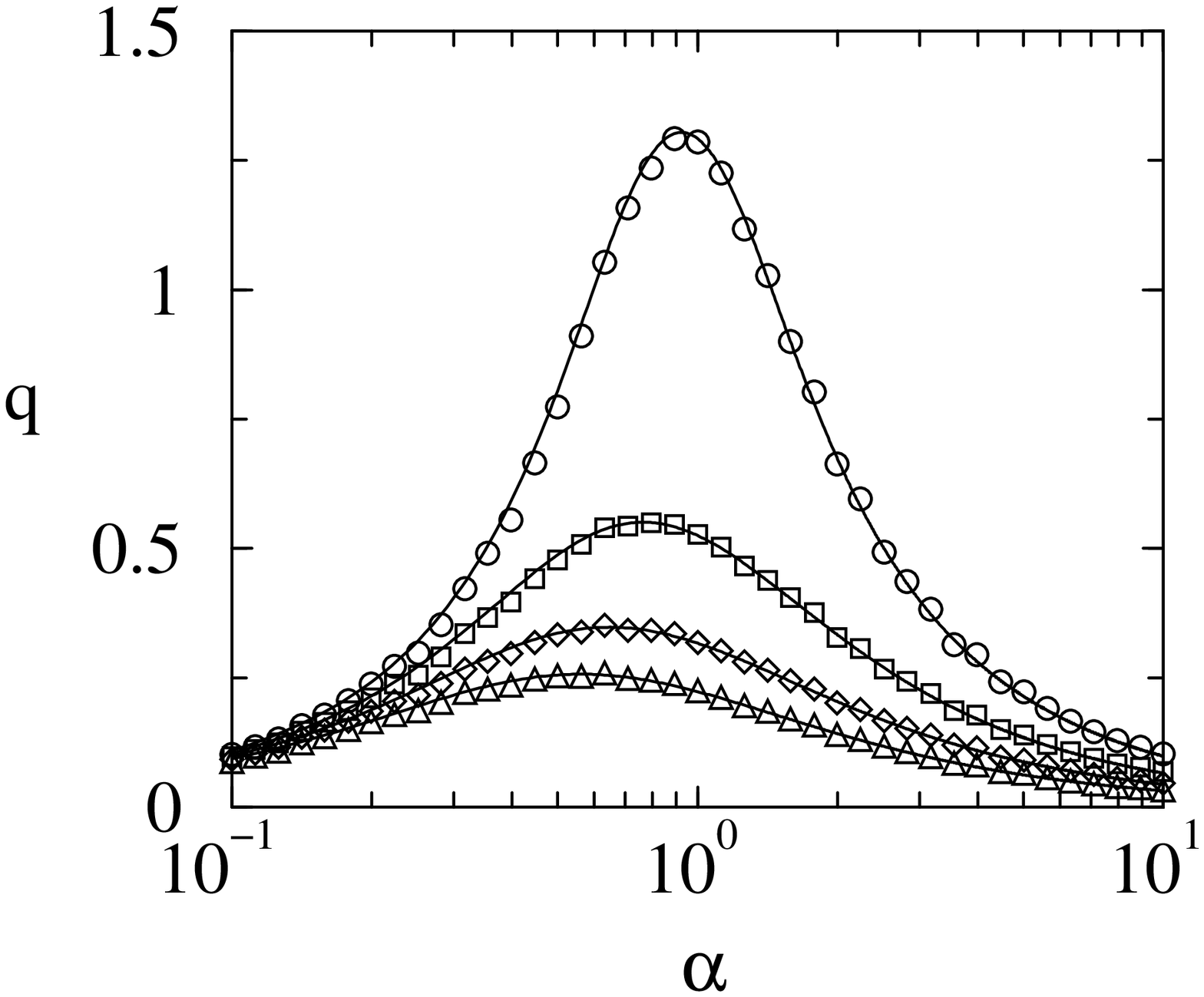}
\caption{Behaviour of $q$ versus $\alpha$ for $\eta=0$ (left; $u(0)$
is the initial bias in the score functions) and various $\eta>0$
(right: $\eta=0.05$ (circles), $\eta=0.25$ (squares), $\eta=0.5$
(diamonds) and $\eta=0.75$ (triangles). From \cite{DG}.}
\label{toyo}
\end{figure}

These findings defy the intuition that Nash equilibria behave
similarly to competitive equilibria when $N\to\infty$. Another
striking proof of the difference between the two equilibrium concepts
is given by the quantity \be q=\frac{1}{N}\sum_{i=1}^N
\left(\frac{z^+_i-z_i^-}{2}\right)^2 \ee which measures how
differently agents invest under the two signals, i.e. how much they
use the information they possess (Fig. \ref{toyo}). 
Price takers exploit their signals much more than sophisticated
traders, who invest very similar amounts of money in the two states
they distinguish. Note that for $\eta=0$ the steady state depends on
initial conditions below $\alpha_c$. The efficient/inefficient
transition may then be characterized also dynamically trhough
transition via path-integral methods \cite{DG}. 

There are several other aspects of the model that deserve attention,
starting with the dependence of fluctuations on $\Gamma$. We refer the
interested reader to \cite{BMRZ,DG} for a more detailed discussion.


\section{Conclusions}

Compared to reality, the models discussed in this review have a
marked theoretical nature. The aim of these models is not that of
providing quantitative predictions but rather to understand under
what conditions the rich variety of behaviors, ranging from
anomalous fluctuations to spontaneous coordination, may emerge in
a simplified controllable setting. This is a complementary
approach to that of empirical analysis, which has been dominating
the scene of interdisciplinary ventures of statistical physicists
into economics and finance. Indeed, a proper understanding of how
interaction propagates from the micro to the macro scale, is
crucial in many cases in order to infer what empirical analysis
should focus on.

Here we have reviewed a number of models with $N$ heterogeneous
interacting agents -- be they firms, species, drivers or traders --
who compete for the exploitations of a number $P$ of resources.  The
collective behavior of all these systems belongs to the same generic
phenomenology, as discussed in Secion 2.1. A key parameter is the
ratio ($\alpha=P/N$) between the number of resources and the number of
agents, and the central quantities of interest are the (in)efficiency
$\sigma^2$, which is related to the amount of unexploited resources,
and the unevenness $H$ with which resources are exploited.

The collective behavior depends strongly on whether agents account
or not for their impact on the resources. This is somewhat
surprising, as one would expect that in the limit $N\to\infty$,
the contribution of each agent to the exploitation of each
resource is vanishing. For the ease of exposition, we distinguish
between the two extreme case of competitive equilibria (CE) and
Nash equilibria (NE), where agents fully neglect or account
exactly for their impact, respectively. The stationary state of
the learning dynamics which converges to these equilibria, in
Minority Game type models markedly differ in the following
respects:
\begin{description}
  \item[Equilibrium condition] In CE resources are exploited, on average, as
  evenly as possible, i.e. $H$ is minimal. In NE fluctuations or
  wastes are as small as possible (i.e. $\sigma^2$ is minimal).
  \item[Phase transition] A phase transition occurs in CE when the
  number of agents exceeds a critical one, i.e. when
  $\alpha<\alpha_c$. This separates an asymmetric ($H>0$ for $\alpha>
  \alpha_c$) from a symmetric ($H=0$ for $\alpha\le\alpha_c$)
  phase. No phase transition takes place in NE (i.e. $H>0$ for all
  $\alpha>0$)
  \item[Degeneracy] The stationary state is unique in CE for
  $\alpha>\alpha_c$ and it is degenerate on a continuous set for
  $\alpha\le\alpha_c$. There is an exponential number of disjoint
  NE.
  \item[Initial conditions] The stationary state does not depend
  on initial conditions for CE and $\alpha>\alpha_c$ and it
  depends continuously on initial conditions for
  $\alpha\le\alpha_c$. The NE to which the system converges
  depends discontinuously on initial conditions.
  \item[Fluctuations] Agents' behavior is stochastic in CE (i.e. $\sigma^2>H$)
  whereas it is deterministic ($\sigma^2=H$) in NE. Put
  differently, in NE agents always play a single strategy, whereas
  in CE agents switch between different strategies.
  \item[Number of choices] Giving more strategies to agents
  improves coordination in NE but it can make agents worse off in
  CE (typically when $\alpha$ is small).
  \item[Convergence] Agents converge fast to CE whereas agents may
  fail to learn to coordinate on NE \cite{mmrz}
\end{description}
Not all these conclusions apply to the asset market model with private
information of Sec.  5.6, though even there CE and NE differ
substantially \cite{BMRZ}.

There still remain interesting theoretical challenges in this
field. Some of these are:
\begin{itemize}
  \item The MG is a prototype model of a systems where the collective
  fluctuations which agents produce feed back into their dynamics.
  Still, there are no analytical tools which allows us to characterize this
  feedback in precise terms in the symmetric phase
  of the MG, i.e. to compute the
  volatility $\sigma^2$ as a function of $\Gamma$.
  \item MG based models of financial markets show that anomalous fluctuations
  similar to the stylized facts observed in real markets arise
  close to the phase transition line. Still the critical
  properties at this phase transition have not yet been
  characterized. Detailed numerical studies of critical properties or
  analytic approaches based on renormalization group techniques
  would be very important to shed light on this issue.
  \item The MG suggests that real markets operate close to a phase
  transition but it does not explicitly describe a mechanism of how
  markets would ``self-organize'' to such a state. Though some
  arguments have been put forward \cite{gcmg}, these have not yet
  been formalized in a definite model.
  \item The extensions to cases where firms behave strategically, as in
  Cournot games \cite{GameTheory}, of the model of economic equilibria may
  prove interesting. The conjecture is that, even in the limit $N\to\infty$
  if the number of commodities (or markets) also diverges, the NE may be
  markedly different from a CE.
\end{itemize}

As a concluding remark, we observe that socio-economic phenomena
have features which are markedly different from those addressed in
natural sciences. Above all, the economy and society change at a
rate which is probably much faster than that at which we
understand them. For example, many of the things which are traded
nowadays in financial markets did not exist few decades ago, not
to speak of internet communities. In addition, we face a situation
in which the density and range of interactions are steadily
increasing, thus making theoretical concepts based on effective
non-interacting theories inadequate.

Definitely, socio-economic systems provide several interesting
theoretical challenges. Our hope is that these effort will help refine
our understanding of how individual behavior, interaction and
randomness may conspire in shaping collective phenomena, which,
broadly speaking, is the aim of statistical physics.

\ack This review has greatly benefited from the interactions we had
with many colleagues over the last few years. It is our pleasure to
thank in particular G Bianconi, D Challet, S Cocco, ACC Coolen, JD
Farmer, FF Ferreira, S Franz, T Galla, I Giardina, JAF Heimel, E
Marinari, R Monasson, R Mulet, G Mosetti, I Perez Castillo, F Ricci
Tersenghi, A Tedeschi, MA Virasoro, R Zecchina and YC Zhang. We
acknowledge financial support from the EU grant HPRN-CT-2002-00319
(STIPCO), the EU-NEST project COMPLEXMARKETS, the MIUR strategic
project ``Dinamica di altissima frequenza dei mercati finanziari'',
and from EVERGROW, integrated project no. 1935 in the complex systems
initiative of the Future and Emerging Technologies directorate of the
IST Priority, EU Sixth Framework.


\section*{References}

\begin{thebibliography}{99}

\bibitem{Wigner1}Wigner E 1955 {\it Ann. of Math} {\bf 62} 548

\bibitem{GameTheory}Vega Redondo F 2003 Game theory and economic
applications (Cambridge University Press, Cambridge)

\bibitem{MantegnaStanley} Mantegna RN and Stanley HE 2000
An introduction to econophysics (Cambridge University Press, Cambridge)

\bibitem{Dacorogna}Dacorogna MM, Gencay R, M\"uller UA,
Olsen RB and Pictet OV 2001
An introduction to high-frequency finance
(Academic Press, San Diego, CA)

\bibitem{BouchaudPotters}Bouchaud JP and 
Potters M 2003 
Theory of financial risk and derivative pricing: from statistical
  physics to risk management
(Cambridge University Press, Cambridge)

\bibitem{NFJBOOK}Johnson NF, Jefferies P and
Hui PM 2003 Financial market complexity (Oxford, University Press, Oxford)

\bibitem{Voit}Voit J 2005 The statistical mechanics of financial markets
(Springer-Verlag, Berlin)

\bibitem{Til}Tilman D 1982 Resource competition and community
structure (Princeton University Press, Princeton)

\bibitem{Rieger}Rieger H 1989 {\it J. Phys. A} {\bf 22} 3447

\bibitem{May}May RM 1973 Stability and complexity in model ecosystems
(Princeton University Press, Princeton)

\bibitem{mitra}Sengupta AM and Mitra PP 1999 {\it Phys. Rev. E} {\bf
60} 3389
  
\bibitem{arth}Arthur WB 1994 {\it Am. Econ. Rev. Pap. Proc.} {\bf 84}
406

\bibitem{elfa}Challet D, Marsili M and Ottino G 2004 {\it Physica A}
{\bf 332} 469

\bibitem{Deb}Weisstein EW
http://mathworld.wolfram.com/deBruijnGraph.html

\bibitem{CMmem}Challet D and Marsili M 2000 {\it Phys. Rev. E} {\bf
62} 1862

\bibitem{CTL}Marsili M and Challet D 2001 {\it Phys. Rev. E} {\bf 64}
056138

\bibitem{weis}Weisbuch G, Kirman A and Herreiner D 2001 {\it Economic
J.}  {\bf 110} 411

\bibitem{spie}De Martino A and Marsili M 2005 {\it Proceedings of
SPIE} {\bf 5848} 165

\bibitem{tra1} Selten R et al. 2004 {\it Experimental Investigation of
Day-to-Day Route Choice-Behaviour and Simulation of Autobahn Traffic
in NRW}. In: Traffic and Human Behaviour (Selten R and Schreckenberg
M, eds) (Springer, Heidelberg)

\bibitem{tra2}Helbing D, Schoenhof M and Kern D 2002 {\it New
J. Phys.} {\bf 4} 33

\bibitem{urb}De Martino A, Marsili M and Mulet R 2004 {\it
Europhys. Lett.} {\bf 65} 283

\bibitem{Masco}Mas-Colell A, Whinston MD and Green JR 1995 
Microeconomic theory (Oxford University Press, Oxford)

\bibitem{Lancaster}Lancaster K 1987 Mathematical economics (Dover, New York)

\bibitem{gale}Gale D 1960 The theory of linear economic models (The
University of Chicago Press, Chicago)

\bibitem{spdsa}De Martino A 2005 {\it Prog. Theor. Phys. Suppl.} {\bf
157} 308

\bibitem{Knaps1}Korutcheva E, Opper M and Lopez B 1994 {\it J. Phys. A}
{\bf 27} L645

\bibitem{Knaps2}Inoue J 1997 {\it J. Phys. A} {\bf 30} 1047

\bibitem{Nishi}Nishimori H 2001 Statistical physics of spin-glasses
and information processing: an introduction (Oxford University Press,
Oxford)

\bibitem{Kirman}Kirman AP 1992 {\it J. Econ. Persp.} {\bf 6} 117

\bibitem{Wigner2}Wigner E 1958 {\it Ann. of Math.} {\bf 67} 325

\bibitem{geneq1}De Martino A, Marsili M and Perez Castillo I 2004 {\it
JSTAT} P04002

\bibitem{geneq2}De Martino A, Marsili M and Perez Castillo I 2006 {\it
Macroeconomic Dynamics} (to appear)

\bibitem{Takayasu}Okuyama K, Takayasu M and Takayasu H 1999 {\it
Physica A} 269 125

\bibitem{john}Von Neumann J 1937 Ergebn. eines Math. Kolloq. {\bf
8}. English translation: Von Neumann J 1945 {\it Rev. Econ. Studies}
{\bf 13} 1

\bibitem{turnpike}McKenzie LW 1986 {\it Optimal Economic Growth,
Turnpike Theorems and Comparative Dynamics}, in Arrow KJ and
Intriligator MD (eds), Handbook of Mathematical Economics, Vol. III
(North-Holland, Amsterdam)

\bibitem{VN}De Martino A and Marsili M 2005 JSTAT L09003

\bibitem{Gardner}Gardner E 1988 {\it J. Phys. A: Math. Gen.} {\bf 21}
257

\bibitem{Romer} Romer P 1990 {\it J. Pol. Econ.} {\bf 98} S72

\bibitem{CZ}Challet D and Zhang YC 1997 {\it Physica A} {\bf 246} 407

\bibitem{Fama}Fama EF 1965 {\it J. Business} {\bf 36} 420

\bibitem{Stylized}Pagan A 1999 {\it J. Empirical Finance} {\bf 3} 15

\bibitem{mgbook}Challet D, Marsili M and Zhang YC 2005 Minority Games
(Oxford University Press, Oxford)

\bibitem{coolen}Coolen ACC 2005 The mathematical theory of Minority
Games (Oxford University Press, Oxford)

\bibitem{Mats}Marsili M 2001 {\it Physica A} {\bf 299} 93

\bibitem{Farmer1}Farmer JD 1999 {\it SFI Technical Report} 98-12-117

\bibitem{Luxm}Lux T and Marchesi M 1999 {\it Nature} {\bf 397} 498

\bibitem{ChM}Challet D and Marsili M 1999 {\it Phys. Rev. E} {\bf 60}
R6271

\bibitem{Cavagna}Cavagna A 1999 {\it Phys. Rev. E} {\bf 59} R3783

\bibitem{Savit}Savit R, Manuca R and Riolo R 1999 {\it
Phys. Rev. Lett.} {\bf 82} 2203

\bibitem{cac}Hart M, Jefferies P, Hui PM and Johnson NF 2001 {\it
Eur. Phys. J. B} {\bf 20} 547

\bibitem{thermal}Cavagna A, Garrahan JP, Giardina I and Sherrington D
1999 {\it Phys. Rev. Lett.} {\bf 83} 4429

\bibitem{physa}Marsili M, Challet D and Zecchina R 2000 {\it Physica
A} {\bf 280} 522

\bibitem{multnoise}Coolen ACC, Heimel JAF  and  Sherrington D 2001 
{\it Phys. Rev. E} {\bf 65} 016126

\bibitem{andemar}De Martino A and Marsili M 2001 {\it J. Phys. A} {\bf
34} 2525

\bibitem{relev}Challet D and Marsili M 2000 {\it Phys. Rev. E} {\bf
62} 1862

\bibitem{coolong}Coolen ACC 2005 {\it J. Phys. A} {\bf 38} 2311

\bibitem{gcmg}Challet D and Marsili M 2003 {\it Phys. Rev. E} {\bf 68}
036132

\bibitem{jasymm}Johnson NF, Hui PM, Zheng D and Tai CW 1999 {\it
Physica A} {\bf 269} 493

\bibitem{CMZha}Challet D, Marsili M and Zhang YC 2000 {\it Physica A}
{\bf 276} 284

\bibitem{ccmz}Challet D, Chessa A, Marsili M and Zhang YC 2000 {\it
Quant. Finance} {\bf 1} 168

\bibitem{Rodgers} D'Hulst R and Rodgers GJ 1999 Preprint
adap-org/9904003

\bibitem{Chau} Chow FK and Chau HF 2003 {\it Physica A} {\bf 319} 601

\bibitem{new}Bianconi G, De Martino A, Ferreira FF and Marsili M 2006
Preprint physics/0603152

\bibitem{MSR}Martin PC, Siggia ED and Rose HA 1973 {\it Phys. Rev. A} {\bf 8} 423

\bibitem{dedo}De Dominicis C 1978 {\it Phys. Rev. B} {\bf 18} 4913

\bibitem{HeimCool}Heimel JAF and Coolen ACC 2001 {\it Phys. Rev. E} {\bf 63} 056121

\bibitem{Kozlo}Kozlowski P and Marsili M 2003 {\it J. Phys. A} {\bf
36} 11725

\bibitem{dgm}De Martino A, Giardina I and Mosetti G 2003 {\it
J. Phys. A} {\bf 36} 8935

\bibitem{varga}Varga P 1998 {\it Phys. Rev. E} {\bf 57} 6487

\bibitem{gene1}De Martino A, Giardina I, Marsili M and Tedeschi A 2004
{\it Phys. Rev. E} {\bf 70} 025104(R)

\bibitem{gene2}Tedeschi A, De Martino A and Giardina I 2005 {\it
Physica A} {\bf 358} 529

\bibitem{BMRZ}Berg J, Marsili M, Rustichini A and Zecchina R 2001 {\it
Quant. Finance} {\bf 1} 203

\bibitem{DG}De Martino A and Galla T 2005 {\it JSTAT} P08008

\bibitem{mmrz} Marsili M, Mulet RG, Ricci-Tersenghi F and
Zecchina R 2001 {\it Phys. Rev. Lett.} {\bf87} {208701}

\end{thebibliography}
\end{document}